\documentclass[largeformat]{interact}

\usepackage{epstopdf}
\usepackage{subfigure}

\usepackage{natbib}
\bibpunct[, ]{(}{)}{;}{a}{}{,}
\renewcommand\bibfont{\fontsize{10}{12}\selectfont}
\allowdisplaybreaks[4]

\usepackage{hyperref}
\usepackage{bm}
\def\btheta{\bm{\theta}}
\def\bbeta{\mbox{\boldmath{$\beta$}}}

\DeclareMathOperator*{\argmin}{argmin}

\def\w{\textbf{w}}

\def\mH{\mathcal{H}}

\theoremstyle{plain}
\newtheorem{theorem}{Theorem}[section]

\theoremstyle{definition}

\theoremstyle{remark}

\usepackage{cleveref}
\usepackage{xcolor}
\newcommand{\conditionname}{C}
\newcounter{condition}
\renewcommand{\thecondition}{\conditionname\arabic{condition}}
\crefname{condition}{condition}{conditions}
\Crefname{condition}{Condition}{Conditions}
\newcommand{\condition}[1]{%
	\refstepcounter{condition}%
	\label{#1}%
	\textbf{\thecondition}%
}

\begin{document}


\title{Spatial weights matrix selection and model averaging for multivariate spatial autoregressive models}

\author{
\name{Xin Miao\textsuperscript{a}, Fang Fang\textsuperscript{a,*}\thanks{* Corresponding author. Email: ffang@sfs.ecnu.edu.cn}, Xuening Zhu\textsuperscript{b} and Hansheng Wang\textsuperscript{c}}
\affil{\textit{\textsuperscript{a}School of Statistic, East China Normal University, Shanghai, China; \textsuperscript{b}School of Data Science, Fudan University, Shanghai, China; \textsuperscript{c}Guanghua School of Management, Peking University, Beijing, China}
}}

\maketitle

\begin{abstract}
In this paper, we focus on the model specification problem in multivariate spatial econometric models when a candidate set for the spatial weights matrix is available. We propose a model selection method for the multivariate spatial autoregressive model, when the true spatial weights matrix may not be in the candidates. We show that the selected estimator is asymptotically optimal in the sense of minimizing the squared loss. If the candidate set contains the true spatial weights matrix, the method has selection consistency.  We further propose a model averaging estimator that combines a set of candidate models and show its asymptotic optimality. Monte Carlo simulation results indicate that the proposed model selection and model averaging estimators perform quite well in finite samples. The proposed methods are applied to a Sina Weibo data to reveal how the user's posting behavior is influenced by the users that he follows.  The analysis results indicate that
the influence tends to be uniformly distributed among the user¡¯s followee, or linearly correlated
with the number of followers of the followee.
\end{abstract}


\newpage
\setcounter{equation}{0}

\section{Introduction}\label{sec1}

Spatial autoregressive (SAR) model proposed by \cite{cliff1973} has received a lot of attention for analyzing spatial data \citep{anselin1988, Banerjee2014} and social network data \citep{chen2013, liu2014, cohen2018multivariate}. The SAR model assumes that the observation in a region or node is influenced not only by its own factors but also by its neighboring regions or nodes. \cite{anselin1988} summarized early development in estimation and testing for SAR
models, while \cite{kelejian1998generalized} proposed the two stage least squares method. Based on these fundamental research, the quasi-maximum likelihood \citep{lee2004asymptotic} and the generalized method of moments estimation method \citep{lee2007gmm} are investigated and widely used.

Although univariate SAR models have been well developed, multivariate responses are commonly encountered in practice. Consider social networks such as Sina Weibo  (the largest Twitter-type social media in China). Since the user's posts can be classified  by contents (e.g., Finance, Economics), we can treat the number of user's posts in a single classification as a univariate response and all classifications naturally constitute a multivariate response \citep{zhu2020multivariate}. Similar situations exist in other examples as well. \cite{jeanty2010estimation} investigated the importance of accounting for spatial interdependencies between population change and housing values and modeled a spatial simultaneous equation  to identify  the local interactions between housing prices and population migration. \cite{cohen2018multivariate}  proposed a simultaneous-equation network model and discussed the identification of social interaction effects. Using the Add Health data, they found that a student's academic performance is not only affected
by academic performance of his peers but also affected by screen-related activities of his peers. To model the data with multivariate responses and spatial interactions, \cite{yang2017identification} investigated a multivariate spatial autoregressive (MSAR) model and associated it with the simultaneous equations SAR models \citep{Kelejian2004, Baltagi2011, Graaff2012, liu2014, cohen2018multivariate}. A quasi-maximum likelihood estimator (QMLE) was proposed to estimate the MSAR model. To reduce the computational complexity of QMLE, \cite{zhu2020multivariate} developed a least squares method and investigated its identification conditions and asymptotic properties.

 It is remarkable that these estimation methods (including SAR and MSAR models) treat the spatial weights matrix as given and implicitly assume that the specification of the spatial weights matrix is correct. However, for characterizing the spatial dependence, spatial weights matrix is usually constructed from geographical or economic information, which means the spatial weights matrix may have many candidates \citep{pace2000method, pace2002semiparametric}. In addition, it is also very difficult to determine whether the true weight matrix is included in the candidates or not.
To select the true spatial weights matrix from a set of non-nested spatial effects, \cite{kelejian2008spatial} proposed a J-test for SAR model specifications. The method is related to whether or not the predictive value of an alternative model adds significance to the null model. Several improvements have been made to the J-test afterward \citep{burridge2010bootstrap, kelejian2011extension}. The main shortcoming of the J-test is that if we reject the null model, there is no formal method to select alternatives. \cite{zhang2018spatial} formally proposed a model selection method using a Mallows type function \citep{mallows1973some}, and showed that the method is asymptotically optimal if the true weights matrix is not in the candidates and it is consistent when the true weights matrix is in the candidates. For the matrix exponential spatial specification \citep[MESS,][]{lesage2007matrix}, an alternative to the SAR type models, \cite{yang2022model} proposed a model selection method and showed that the selection estimator is asymptotically optimal in the sense that it is as efficient as the infeasible estimator that uses the best candidate spatial weights matrix.

Compared to model selection, model averaging can usually reduce the risk of model misspecification. \cite{zhang2018spatial} also studied model averaging to reduce the squared loss for estimating SAR model with a set of candidate spatial weights matrices. They proposed a model averaging method based on a Mallows type criterion function,  and showed that the MA method is asymptotically optimal under certain conditions. \cite{yang2022model} also proposed a MA method for MESS. \cite{debarsy2022bayesian} investigated a Bayesian model averaging method focusing on convex combinations of weight matrices that results in a single weight matrix reflecting multiple types of connectivity, where coefficients from the convex combination can be used for inference regarding the relative importance of each type of connectivity in the global cross-sectional dependence scheme.

All the above mentioned model selection and model averaging methods for spatial weights matrix specification are only applicable to univariate response. In this paper, we focus on model selection (MS) and model averaging (MA) from a set of candidates for the multivariate spatial autoregressive (MSAR) model. Our main contributions are summarized as follows.

\begin{enumerate}
\item[(1)] We propose a model selection method based a Mallows type selection criterion. The selected estimator is asymptotically optimal in the sense that it is as efficient as the infeasible estimator that uses the best candidate spatial weights matrix. In addition, the model selection method selects the true spatial weights matrix with probability approaching one in large samples when the true model is in the candidate set.
\item[(2)] Further, without putting all our inferential eggs in one unevenly woven basket \citep{Longford2005}, we propose a model averaging method based on an unbiased estimator of the squared loss of the model averaging estimator. The proposed estimator is asymptotically optimal over all the weighted models, and it's more robust to model misspecification  compared to model selection.
\item[(3)] Simulation studies confirm the theoretical results and show the superiorities of the model selection and model averaging methods compared to MSAR with each single candidate model. It's also revealed that the MS and MA methods for MSAR  perform much better than MS and MA of univariate SAR model by treating each element of the response alone, especially when the elements are correlated each other.
\item[(4)] When there are multiple spatial weights matrices available, another option is the high-order MSAR model which extends the classical high-order SAR model \citep{anselin1988, 2010EFFICIENT, 2015Inference} to the multivariate response case. Our simulation results show that the high-order MSAR model only performs better than our proposed methods when the true model is high-order MSAR and the candidate spatial weights matrices used for model fitting are exactly the same as the true ones, which is almost impossible in practice.
\item[(5)] The proposed methods are applied to a Sina Weibo data to investigate how the posting behavior of a user is influenced by the followee, i.e., the users that he follows. Analysis results show that considering a bivariate response is helpful in improving the prediction performance, compared to models with each univariate response alone. The influence tends to be uniformly distributed among the user's followee, or linearly correlated with the number of followers of the followee.
\end{enumerate}

The rest of the paper is organized as follows. Section \ref{sec2} introduces the MSAR model and the estimation method under each candidate spatial weights matrix. Section \ref{sec3} and Section \ref{sec4} propose the model selection and model averaging methods, respectively, and establish their asymptotic properties. Section \ref{sec5} and Section \ref{sec6} present the results of simulation studies and a real example, respectively. Some concluding remarks are provided in Section \ref{sec7}. Some techincal details and all the proofs are in the Appendix.

\section{Multivariate Spatial Autoregressive Model} \label{sec2}
Consider a cross-sectional MSAR model \citep{zhu2020multivariate}
\begin{equation} \label{MSARmodel}
    \bm{Y} = \bm{W} \bm{Y} \bm{D} + \bm{X} \bm{B} + \bm{E},
\end{equation}
where $\bm{Y} = (\bm{y}_1,\cdots,\bm{y}_q)\in\mathcal{R}^{n\times q}$ is a response matrix which consists of $q$ endogenous variables, $n$ is the number of observations, $\bm{W}\in\mathcal{R}^{n \times n}$ is a nonstochastic row-normalized spatial weights matrix which summaries the neighborhood relationships, {\color{black}the diagonal elements of $\bm{W}$ are zeros,} $\bm{X}\in\mathcal{R}^{n \times p}$ represents the matrix of all observations on $p$ nonstochastic exogenous variables, $\bm{E} = (\bm{\varepsilon}_1,\cdots,\bm{\varepsilon}_q)\in\mathcal{R}^{n \times q}$ is the disturbance matrix, and $\bm{D}\in\mathcal{R}^{q \times q}$ and $\bm{B}\in\mathcal{R}^{p \times q}$ are two parameter matrices. In addition, each row of $\bm{E}$ is identically and independently distributed {\color{black} with zero mean and covariance matrix $\bm{\Sigma_e}\in\mathcal{R}^{q \times q}$, which} also needs to be estimated. We assume $p$ and $q$ are fixed.

By taking vectorization, model (\ref{MSARmodel}) can be rewritten as
\begin{equation} \label{vecMSAR}
    \bm{y} = (\bm{D}^T \otimes \bm{W}) \bm{y} + \widetilde{\bm{X}} \bm{\beta} + \bm{\varepsilon},
\end{equation}
where $\bm{y} = \text{vec} (\bm{Y}) = (\bm{y}_1^T,\cdots,\bm{y}_q^T)^T \in \mathcal{R}^{nq}$, $ \widetilde{\bm{X}} = \bm{I}_q \otimes \bm{X}\in\mathcal{R}^{nq\times pq} $, $ \bm{\varepsilon} = \text{vec} (\bm{E}) = (\bm{\varepsilon}_1^T,\cdots,\bm{\varepsilon}_q^T)^T \in \mathcal{R}^{nq} $, $\bm{\beta} = \text{vec} (\bm{B}) \in \mathcal{R}^{pq}$, and $\otimes$ is the Kronecker product. Here, $\bm{\varepsilon}$ is {\color{black} a random vector} with zero mean and {\color{black}covariance matrix} $\text{cov} (\bm{\varepsilon}) =  \bm{\Sigma} = \bm{\Sigma_e} \otimes \bm{I}_n$.

The reduced form of model (\ref{vecMSAR}) is
\begin{equation} \label{reduceVecMSAR}
    \bm{y} = (\bm{I}_{nq} - \bm{D}^T \otimes \bm{W})^{-1}( \bm{\widetilde{X}} \bm{\beta} + \bm{\varepsilon}).
\end{equation}

\noindent In order to ensure the matrix $(\bm{I}_{nq} - \bm{D}^T \otimes \bm{W})$ to be invertible, {\color{black} since $\bm{W}$ is row-normalized}, according to Lemma 1 of \cite{zhu2020multivariate}, we just need to assume that the maximum absolute eigenvalue of $\bm{D}$ is less than 1 throughout the paper.
From (\ref{reduceVecMSAR}), we know that
$\bm{\mu} = E(\bm{y}) = (\bm{I}_{nq} - \bm{D}^T \otimes \bm{W})^{-1} \bm{\widetilde{X}} \bm{\beta}$ and $\bm{\Omega} = \text{cov}(\bm{y}) = (\bm{I}_{nq} - \bm{D}^T \otimes \bm{W})^{-1} \bm{\Sigma} (\bm{I}_{nq} - \bm{D} \otimes \bm{W}^T)^{-1}$.  Our main target is to predict $\bm{\mu}$ as best as possible.

Denote $\bm{d} = \text{vec} (\bm{D}) \in \mathcal{R}^{q^2}$, $\bm{\xi} = \text{vec}^* (\bm{\Sigma_e}^{-1}) \in \mathcal{R}^{q(q+1) / 2}$, where $ \text{vec}^* (\bm{\Sigma_e}^{-1})$ selects only up-triangle parameters of $\bm{\Sigma_e}^{-1}$ since the covariance matrix $\bm{\Sigma_e}$ is symmetric. Let $\btheta = (\bm{d}^T, \bm{\beta}^T, \bm{\xi}^T)^T \in \mathcal{R}^{n_{pq}}$ collect the parameters to be estimated, where $n_{pq} = q^2 + pq + q (q+1) / 2$. Define $ \bm{S} = \bm{I}_{nq} - \bm{D}^T \otimes \bm{W} $ and $ \bm{M} = \text{diag}^{-1} (\bm{\Omega}^{-1}) = \{ \text{diag} (\bm{\Sigma_e}^{-1}) \otimes \bm{I}_n + \text{diag} (\bm{D} \bm{\Sigma_e}^{-1} \bm{D}^T ) \otimes \text{diag} (\bm{W}^T \bm{W}) \}^{-1}$. Based on \cite{zhu2020multivariate}, a least squares type objective function can be constructed as
\begin{equation}\label{loss}
    Q(\btheta) = \left \Vert \bm{M} \bm{S}^T (\bm{\Sigma_e}^{-1} \otimes \bm{I}_n) (\bm{S} \bm{y} - \bm{\widetilde{X}} \bm{\beta})   \right \Vert^2.
\end{equation}

\vspace{0.4cm}
{\color{black}\noindent{\bf Remark 1.} Based on \eqref{reduceVecMSAR}, the MSAR model can be estimated by the QMLE method \citep{yang2017identification} by treating $\bm{\varepsilon}\sim N(\bm{0},\bm{\Sigma})$.
However, as shown by \cite{zhu2020multivariate}, the QMLE objective function involves the determinant of a high dimensional matrix $\bm{S} = \bm{I}_{nq} - \bm{D}^T \otimes \bm{W}$, whose computation is time consuming especially for large scale networks.
Therefore, they design a least squares type objective function as
$\sum_{i=1}^n\sum_{j=1}^q\{Y_{ij} - E(Y_{ij}| \bm{Y}_{-(ij)})\}^2$, where $Y_{ij}$ is the $(i,j)$-th element of $\bm{Y}$ and $\bm{Y}_{-(ij)} =
\{Y_{kl}:(k,l)\ne (i,j), 1\le k\le n, 1\le l\le q\}$.
This leads to the objective function in \eqref{loss}.
It is remarkable that in \eqref{loss},
the determinant of a high dimensional matrix is no longer involved, thus it is more computationally efficient.}
\vspace{0.4cm}

Since $\bm{W}$ can be constructed in many ways, assume we have a spatial weights matrix candidate set $\mathcal{W} = \big \{ \bm{W}_1, \cdots, \bm{W}_K \big \}$, where each $\bm{W}_k$ is a plausible $n\times n$ row-normalized spatial weights matrix, $k=1,\cdots,K$. We allow $K$ to increase with the sample size $n$, although in practice a limited number of candidate models are used. Under the $k$-th candidate model, let $\widehat\btheta_k$ be the least squares type estimator of $\btheta$ by minimizing (\ref{loss}) with $\bm{W}$ being replaced by $\bm{W}_k$. We estimate $\bbeta$ by $\widetilde{\bm{\beta}}_k=(\bm{\widetilde{X}}^T \bm{\widetilde{X}})^{-1} \bm{\widetilde{X}}^T\bm{S}_k \bm{y}$, where $\bm{S}_k = \bm{I}_{nq} - \bm{\widehat{D}}_k^T \otimes \bm{W}_k$ and $\widehat{\bm{D}}_k$ is the estimator of $\bm{D}$ in $\widehat\btheta_k$. Thus, the estimator of the expected value $\bm{\mu}$ is $ \widetilde{\bm{\mu}}_k = \bm{S}_k^{-1} \bm{\widetilde{X}} \widetilde{\bm{\beta}}_k = \bm{S}_k^{-1} \bm{P} \bm{S}_k \bm{y} = \widetilde{\bm{P}}_k \bm{y} $, where $\bm{P} = \bm{\widetilde{X}} (\bm{\widetilde{X}}^T \bm{\widetilde{X}})^{-1} \bm{\widetilde{X}}^T$ is the projection matrix and $\widetilde{\bm{P}}_k = \bm{S}_k^{-1} \bm{P} \bm{S}_k=( \bm{I}_{nq} - \bm{\widehat{D}}_k^T \otimes \bm{W}_k)^{-1}\bm{P} (\bm{I}_{nq} - \bm{\widehat{\bm{D}}}_k^T \otimes \bm{W}_k)$ is a function of $\bm{\widehat{D}}_k$. If $\bm{W}_k$ is the true spatial weights matrix, $\widehat{\bm{D}}_k$ is a consistent estimator of $\bm{D}$ {\color{black} under some regularity conditions} by \cite{zhu2020multivariate} and hence $\widetilde{\bm{\mu}}_k$ is approximately unbiased for $\bm{\mu}$. Therefore, we can use $\widetilde{\bm{\mu}}_k$ to evaluate the performance of the $k$th model by the difference of $\widetilde{\bm{\mu}}_k$ and the true $\bm{\mu}$.

\vspace{0.4cm}
{\color{black}\noindent{\bf Remark 2.} Note that we do not estimate $\bbeta$ by $\widehat{\bm{\beta}}_k$ in $\widehat\btheta_k$. Instead, we use the estimator $\widetilde{\bm{\beta}}_k=(\bm{\widetilde{X}}^T \bm{\widetilde{X}})^{-1} \bm{\widetilde{X}}^T\bm{S}_k \bm{y}$. Note that $\widetilde{\bm{\beta}}_k$ depends on $\widehat{\bm{D}}_k$ in $\widehat\btheta_k$. By doing so, $\widetilde{\bm{\mu}}_k$ can be written as $\widetilde{\bm{P}}_k \bm{y}$ and $\widetilde{\bm{P}}_k$ is a function of $\bm{\widehat{D}}_k$, which are critical for the derivation of the model selection criterion in (\ref{Rk_}) and (\ref{Ck}).}
\vspace{0.4cm}

\section{Model Selection of Spatial Weights Matrix} \label{sec3}

In this section we construct a Mallows type model selection criterion for the spatial weights matrix selection, and establish the asymptotic properties of the selected matrix. Let $ L_k = \Vert \widetilde{\bm{\mu}}_k - \bm{\mu} \Vert^2 $ be the squared loss of the $k$-th model for predicting $\bm{\mu}$. We try to find the $k$ that minimizes $ L_k $. Denote the associated risk $ R_k = E(L_k) = E \Vert \widetilde{\bm{\mu}}_k - \bm{\mu} \Vert^2 $.

The key part is to find an appropriate estimator of $R_k$ for building the model selection criterion. We follow the classical Mallows type techniques in the literature \citep{mallows1973some, hansen2007, zhang2018spatial, yang2022model}. {\color{black} To motivate our proposed method, here we temporally assume that $\bm{\varepsilon}\sim N(\bm{0},\bm{\Sigma})$. But the theories later are established under a much weaker condition than the normality.}
Note that
\begin{eqnarray} \label{Rk_}
& R_k & = E \Vert \widetilde{\bm{\mu}}_k - \bm{\mu} \Vert^2 = E \Vert \widetilde{\bm{\mu}}_k - \bm{y} + \bm{y} - \bm{\mu} \Vert^2  \nonumber\\
&& =  E \Vert \widetilde{\bm{\mu}}_k - \bm{y} \Vert^2 + \text{tr} (\bm{\Omega}) + 2 E \big\{ ( \widetilde{\bm{\mu}}_k - \bm{y} )^T (\bm{y} - \bm{\mu}) \big \} \nonumber\\
&& =  E \Vert \widetilde{\bm{\mu}}_k - \bm{y} \Vert^2 - \text{tr} (\bm{\Omega}) + 2 E \big  \{ ( \widetilde{\bm{P}}_k \bm{y})^T (\bm{y} - \bm{\mu}) \big  \}.
\end{eqnarray}
\noindent Define $\bm{z} = \bm{\Omega}^{-1/2} \bm{y}$. From the normality of $\bm{\varepsilon}$ and Stein's Lemma \citep{stein1981estimation}, we have
\begin{eqnarray} \label{Ck}
&& E  \big \{ (\widetilde{\bm{P}}_k \bm{y} )^T  \left(\bm{y} - \bm{\mu} \right) \big  \} \nonumber\\
& = &  E \big  \{ (\bm{\Omega}^{1/2} \widetilde{\bm{P}}_k \bm{\Omega}^{1/2} \bm{z} )^T (\bm{z} - \bm{\Omega}^{-1/2} \bm{\mu} ) \big  \} \nonumber\\
& = & E \Bigg \{ \text{tr} \Bigg( \frac{\partial( \bm{\Omega}^{1/2} \widetilde{\bm{P}}_k \bm{\Omega}^{1/2} \bm{z} )}{\partial \bm{z}^T} \Bigg)  \Bigg \} \nonumber\\
& = & E \big  \{ \text{tr} (\bm{\Omega}^{1/2} \widetilde{\bm{P}}_k \bm{\Omega}^{1/2}) \big  \} +  E \Bigg[ \text{tr} \Bigg \{ \frac{\partial (\bm{\Omega}^{1/2} \widetilde{\bm{P}}_k \bm{\Omega}^{1/2} \bm{z})}{\partial \text{vec} (\widehat{\bm{D}}_k)} \frac{\partial \text{vec} (\widehat{\bm{D}}_k)}{\partial \bm{z}^T} \Bigg \} \Bigg] \nonumber\\
& = & E \big  \{ \text{tr} (\widetilde{\bm{P}}_k \bm{\Omega}) \big  \} +  E \Bigg[ \text{tr} \Bigg( \frac{\partial \text{vec}(\widehat{\bm{D}}_k)}{\partial \bm{y}^T} (\bm{y}^T \otimes \bm{\Omega}) \frac{\partial \text{vec}(\widetilde{\bm{P}}_k)}{\partial \text{vec}(\widehat{\bm{D}}_k)^T} \Bigg) \Bigg].
\end{eqnarray}

\noindent Combing (\ref{Rk_}) and (\ref{Ck}), we have
\begin{equation} \label{expectedSelectionCri}
    E ( C_k \big ) = R_k + \text{tr} (\bm{\Omega}),
\end{equation}
where
\begin{equation} \label{selectionCri_k}
C_k = \big \Vert \widetilde{\bm{P}}_k \bm{y} - \bm{y} \big \Vert^2 + 2 \left \{ \text{tr}(\widetilde{\bm{P}}_k \bm{\Omega}) + \text{tr} \left( \frac{\partial \text{vec} (\widehat{\bm{D}}_k)}{\partial \bm{y}^T} (\bm{y}^T \otimes \bm{\Omega}) \frac{\partial \text{vec} (\widetilde{\bm{P}}_k)}{\partial \text{vec} (\widehat{\bm{D}}_k)^T} \right) \right \}.
\end{equation}

\noindent The first term of $C_k$ is the sum of squared errors of the $k$th model. The second term  of $C_k$ is an unbiased estimator of $2\times df_k$ by the derivation of (\ref{Ck}),  where $df_k = $ cov$\big ( \widetilde{\bm{\mu}}_k, \bm{y} \big )$ is the degrees of freedom of the $k$-th model following \cite{efron2004estimation} and \cite{zou2007degrees}.

Based on (\ref{expectedSelectionCri}), we may select the best model by minimizing $C_k$ in (\ref{selectionCri_k}) over the candidate set. However, the evaluation of $\frac{\partial \text{vec} (\widetilde{\bm{P}}_k)}{\partial \text{vec} (\widehat{\bm{D}}_k)^T} $, $\frac{\partial \text{vec} (\widehat{\bm{D}}_k)}{\partial \bm{y}^T}$ and the estimation of $\bm{\Omega}$ need to be addressed to obtain $C_k$. Let $d_{ij}^k$ be the
$(i,j)$-th element of $\widehat{\bm{D}}_k$. Since $\frac{\partial \bm{S}_k}{\partial d_{ij}^k} = - \bm{I}_{ji} \otimes \bm{W}_k$ and $\frac{\partial \bm{S}_k^{-1}}{\partial d_{ij}^k} = - \bm{S}_k^{-1} \frac{\partial \bm{S}_k}{\partial d_{ij}^k} \bm{S}_k^{-1} $, where $\bm{I}_{ji}$ is a zero matrix with only the $(j, i)$-th element being $1$, we have
\begin{eqnarray*}
\frac{\partial \widetilde{\bm{P}}_k}{\partial d_{ij}^k} & =&  \frac{\partial \bm{S}_k^{-1}}{\partial d_{ij}^k} \bm{P} \bm{S}_k + \bm{S}_k^{-1} \bm{P} \frac{\partial \bm{S}_k}{\partial d_{ij}^k}\\
&=&- \bm{S}_k^{-1} \frac{\partial \bm{S}_k}{\partial d_{ij}^k} \bm{S}_k^{-1}\bm{P} \bm{S}_k +\bm{S}_k^{-1} \bm{P} \frac{\partial \bm{S}_k}{\partial d_{ij}^k}\\
&=&\bm{S}_k^{-1}( \bm{I}_{ji} \otimes \bm{W}_k)\bm{S}_k^{-1}\bm{P} \bm{S}_k -\bm{S}_k^{-1} \bm{P}( \bm{I}_{ji} \otimes \bm{W}_k).
\end{eqnarray*}
By vectorization, we can get $\frac{\partial \text{vec}(\widetilde{\bm{P}}_k)}{\partial \text{vec}(\widehat{\bm{D}}_k)^T}$. The evaluation of $\frac{\partial \text{vec} (\widehat{\bm{D}}_k)}{\partial \bm{y}^T}$ involves the first-order condition of (\ref{loss}) and is quite complicated. We leave the details to Appendix A.

\vspace{0.4cm}
{\color{black}\noindent{\bf Remark 3.} The evaluation of $\frac{\partial \text{vec} (\widehat{\bm{D}}_k)}{\partial \bm{y}^T}$ is much more complicated than its counterpart in the univariate SAR model selection method \citep{zhang2018spatial}, and it is one of the major challenges for the MSAR model selection. The exact form of $\frac{\partial \text{vec} (\widehat{\bm{D}}_k)}{\partial \bm{y}^T}$ is very difficult to derive. So we propose to ignore the randomness of $\widehat{\bm{\Sigma}}_k$, the estimator of $\bm{\Sigma}$ in $\widehat\btheta_k$, and derive an approximated form of $\frac{\partial \text{vec} (\widehat{\bm{D}}_k)}{\partial \bm{y}^T}$ in Appendix A. Fortunately, the theoretical results do not require a precise evaluation of $\frac{\partial \text{vec} (\widehat{\bm{D}}_k)}{\partial \bm{y}^T}$. To be specific, among the technical conditions that will be presented later, only conditions (\ref{C5}), (\ref{C8}) and (\ref{C11}) involve $\frac{\partial \text{vec} (\widehat{\bm{D}}_k)}{\partial \bm{y}^T}$. These conditions are very similar and only require the penalty term $\text{tr} \left( \frac{\partial \text{vec}(\widehat{\bm{D}}_k)}{\partial \bm{y}^T} (\bm{y}^T \otimes \widehat{\bm{\Omega}}) \frac{\partial \text{vec}(\widetilde{\bm{P}}_k)}{\partial \text{vec}(\widehat{\bm{D}}_k)^T} \right)$, where $\widehat{\bm{\Omega}}$ is an estimator of $\bm{\Omega}$, to have certain convergence rate. As long as the approximated form of $\frac{\partial \text{vec} (\widehat{\bm{D}}_k)}{\partial \bm{y}^T}$ satisfies the convergence rate, the theoretical results still hold. This phenomenon is quite typical in the model selection methods. The most well-known example is that, the penalty term $\log(n)$ in BIC can be replaced by any factor $\phi_n$ that satisfies $\phi_n\rightarrow\infty$ and $\phi_n=o(n)$, without changing the model selection consistency. Similar results are also presented in the model averaging literature, for example, \cite{fangET2023}. The more critical issue is whether the approximation form affects the empirical performance of the proposed method. Actually, it works very well as we will see in the simulation studies and real data analysis.}
\vspace{0.4cm}

Let $\widehat{\bm{\Omega}}$ be an estimator of $\bm{\Omega}$. We can define a feasible model selection criterion as
\begin{equation*} \label{selectionCri}
\widehat{C}_k = \big \Vert \widetilde{\bm{P}}_k \bm{y} - \bm{y} \big \Vert^2 + 2 \left \{ \text{tr}(\widetilde{\bm{P}}_k \widehat{\bm{\Omega}}) + \text{tr} \left( \frac{\partial \text{vec}(\widehat{\bm{D}}_k)}{\partial \bm{y}^T} (\bm{y}^T \otimes \widehat{\bm{\Omega}}) \frac{\partial \text{vec}(\widetilde{\bm{P}}_k)}{\partial \text{vec}(\widehat{\bm{D}}_k)^T} \right) \right \}
\end{equation*}
and define
\begin{eqnarray}\label{khat}
\widehat{k} = \underset{k \in \{1,\cdots,K\}}{\argmin} \widehat{C}_k
\end{eqnarray}
as the selected model.

Note that the consistency of $\widehat{\bm{\Omega}}$ is not necessary by the theoretical results that we will establish soon, which means the model used to estimate $\bm{\Omega}$ can be misspecified. In the model selection and model averaging literature, this is a quite common phenomenon and people usually use the estimator from the ``largest" model. However, for the spatial weights matrix, the largest model is hard to define. In the simulation studies, we use a relatively dense candidate matrix to estimate $\bm{\Omega}$ and it works quite well no matter the matrix is correctly specified or not. 
Actually, based on our simulation experience, the results are not sensitive to the choice of  $\widehat{\bm{\Omega}}$, which is consistent to the findings of \cite{zhang2018spatial}.

Now we establish the asymptotic results for the proposed model selection method.  Let $\bm{W}^*$ be the true spatial weights matrix that generates the response $\bm{Y}$. We first assume that $\mathcal{W}$ does not include $\bm{W}^*$, and establish the asymptotic optimality of the selected model. Assume $\widehat{\bm{D}}_k$ has a limit $\bm{D}_k^*$ as $n$ goes to infinity. Let $\overline{\bm{P}}_k = \widetilde{\bm{P}}_k |_{\widehat{\bm{D}}_k = \bm{D}_k^*}$, $R_k^* = E \Vert\overline{\bm{P}}_k\bm{y} - \bm{\mu} \Vert^2$ and $\xi_n = \inf_k R_k^*$. Let $\lambda_{\max} (\bm{\Pi})$ and $\lambda_{\min} (\bm{\Pi})$ be the maximum and minimum eigenvalue of any square matrix $\bm{\Pi}$, respectively. We need the following technical conditions.

\begin{description}

{\color{black}\item[(\condition{C00})] The elements of $\bm{\Sigma}^{-\frac{1}{2}}\bm{\varepsilon}\in\mathcal{R}^{nq}$ are independent from each other. }
{\color{black}\item[(\condition{C0})] The conditions (C1)-(C6) in \cite{zhu2020multivariate} are satisfied. }

	\item[(\condition{C1})]
 There exists a positive integer $G$ such that $ \sum_{k=1}^K (R_k^*)^{-G} = o(1)$.

    \item[(\condition{C2})]
$\Vert \bm{\mu} \Vert^2 = O(n)$, $\lambda_{\max} (\bm{\Omega}) = O(1)$ and $\lambda_{\max} (\widehat{\bm{\Omega}}) = O_p (1)$.

\item[(\condition{C3})]
          $ c_1 \leq \lambda_{\min} (\widetilde{\bm{X}}^T \widetilde{\bm{X}} / n) \leq \lambda_{\max} (\widetilde{\bm{X}}^T \widetilde{\bm{X}} / n) \leq c_2
        $
    and
    $c_3 \leq \lambda_{\min} \big \{ \widetilde{\bm{X}}^T \bm{W}_k (\bm{I}_{nq} - \bm{D}^T \otimes \bm{W}^*)^{-1} \widetilde{\bm{X}} / n \big \} \leq \lambda_{\max} \big \{ \widetilde{\bm{X}}^T \bm{W}_k (\bm{I}_{nq} - \bm{D}^T \otimes \bm{W}^*)^{-1} \widetilde{\bm{X}} / n \big \} \leq c_4
$      for all $k \in \{ 1,\cdots, K \}$, where $c_1,c_2,c_3,c_4$ are positive constants.

    \item[(\condition{C4})]
    $\sup_k \lambda_{\max} (\bm{D}_k^{*T} \otimes \bm{W}_k) = O(1)$ and $\sup_k \lambda_{\max} \big \{ (\bm{I}_{nq}- \bm{D}_k^{*T} \otimes \bm{W}_k)^{-1} \big \} = O(1)$.

    \item[(\condition{C5})]
    $ \xi_n^{-1} \sup_k \left| \text{tr} \left( \frac{\partial \text{vec} (\widehat{\bm{D}}_k)}{\partial \bm{y}^T} (\bm{y}^T \otimes \widehat{\bm{\Omega}}) \frac{\partial \text{vec} (\widetilde{\bm{P}}_k)}{\partial \text{vec}(\widehat{\bm{D}}_k)^T} \right)  \right| = o_p (1)$.

    \item[(\condition{C6})]
    $\xi_n^{-1} = o(1)$ and $n\xi_n^{-1} \sup_k \lambda_{\max} (\widetilde{\bm{P}}_k - \overline{\bm{P}}_k) = o_p(1)$.
\end{description}

Most of these conditions are commonly use in the model selection and model averaging literature such as \cite{li1987}, \cite{shao1997asymptotic}, \cite{andoandli2014}, \cite{zhang2018spatial} and \cite{yang2022model}.
{\color{black} Condition (\ref{C00}) is used to replace the assumption of $\bm{\varepsilon}\sim N(\bm{0},\bm{\Sigma})$, and it is much weaker than the normality. Note that (\ref{C00}) is naturally satisfied if $q=1$. Condition (\ref{C0}) is required since our method is based on the least square estimation of \cite{zhu2020multivariate}. It imposes some regular conditions of the network structure, moment of the disturbance matrix, covariates and law of large numbers. Interested readers can check \cite{zhu2020multivariate} for more details.}
 Condition (\ref{C1}) indicates that $\xi_n\rightarrow\infty$, which requires that all the candidate models are misspecified. In fact, if model $k$ is correct, then $\bm{D}_k^*$ equals true $\bm{D}$ by \cite{zhu2020multivariate}, $\overline{\bm{P}}_k\bm{y}=\bm{S}^{-1}\bm{P}\bm{S}\bm{y}=\bm{S}^{-1}\bm{P}(\widetilde{\bm{X}}\bbeta+\bm{\varepsilon})= \bm{S}^{-1}\widetilde{\bm{X}}\bbeta+\bm{S}^{-1}\bm{P}\bm{\varepsilon}=\bm{\mu}+\bm{S}^{-1}\bm{P}\bm{\varepsilon}$ and
\vspace{-0.5cm}
\begin{eqnarray}
\xi_n\leq R_k^*&=&E \Vert\overline{\bm{P}}_k\bm{y} - \bm{\mu} \Vert^2=E\Vert\bm{S}^{-1}\bm{P}\bm{\varepsilon}\Vert^2=E\big\{\bm{\varepsilon}^T\bm{P}(\bm{S}^{-1})^T\bm{S}^{-1}\bm{P}\bm{\varepsilon}\big\}\nonumber\\
&=&\text{tr}\big\{\bm{\widetilde{X}} (\bm{\widetilde{X}}^T \bm{\widetilde{X}})^{-1} \bm{\widetilde{X}}^T  (\bm{S}^{-1})^T \bm{S}^{-1}\bm{\widetilde{X}} (\bm{\widetilde{X}}^T \bm{\widetilde{X}})^{-1} \bm{\widetilde{X}}^T\bm{\Sigma}\big\}\nonumber\\
&\leq&\lambda_{\max}\big\{(\bm{\widetilde{X}}^T \bm{\widetilde{X}})^{-1} \bm{\widetilde{X}}^T  (\bm{S}^{-1})^T \bm{S}^{-1}\bm{\widetilde{X}} (\bm{\widetilde{X}}^T \bm{\widetilde{X}})^{-1}\big\}\text{tr}(\bm{\widetilde{X}}^T\bm{\Sigma}\bm{\widetilde{X}})\nonumber\\
&\leq&\lambda_{\max}^2\Big\{\Big(\frac{\bm{\widetilde{X}}^T \bm{\widetilde{X}}}{nq}\Big)^{-1}\Big\}\lambda_{\max}\Big\{\frac{\bm{\widetilde{X}}^T (\bm{S}^{-1})^T \bm{S}^{-1}  \bm{\widetilde{X}}}{nq}\Big\} \frac{1}{nq}\text{tr}(\bm{\widetilde{X}}\bm{\widetilde{X}}^T\bm{\Sigma})\nonumber,
\end{eqnarray}
which is $O(1)$ under conditions (\ref{C3}), (\ref{C4}) and a common assumption that the $nq$ diagonal elements of $\bm{\widetilde{X}}\bm{\widetilde{X}}^T\bm{\Sigma}$ are uniformly upper-bounded. Then $\xi_n\rightarrow\infty$ can not be satisfied.

The first part of (\ref{C2}) is quite common concerning the squared sum of $nq$ elements of $\bm{\mu}$, see for example, \cite{shao1997asymptotic}, \cite{liang2011optimal} and \cite{andoandli2014}. The other parts of (\ref{C2}) and (\ref{C3})-(\ref{C4}) require the minimum and maximum eigenvalues of some matrices to be bounded by constants and away from 0. \cite{zhang2018spatial} discussed some elementary assumptions to justify these conditions. Actually, the first part of (\ref{C3}) is common for regressor matrices, while the second part of  (\ref{C3}) and  (\ref{C4}) require $\sup_k \lambda_{\max} (\bm{W}_k) = O(1)$, $\lambda_{\max} (\bm{W}^*) = O(1)$, $\lambda_{\max} \{(\bm{I}_{nq} - \bm{D}^T \otimes \bm{W}^*)^{-1}\} = O(1)$ and $\sup_k \lambda_{\max} \{(\bm{I}_{nq} - \bm{D}_k^{*T} \otimes \bm{W}_k)^{-1} \}= O(1)$.

 Condition (\ref{C5}) is a high-level condition. Similarly to the discussion of \cite{zhang2018spatial}, with the approximated form of $\frac{\partial \text{vec} (\widehat{\bm{D}}_k)}{\partial \bm{y}^T}$, (\ref{C5}) requires that $\Vert \bm{y} \Vert^2$ has order $n$, and uniformly for $k \in \{ 1,\cdots, K \}$, $\lambda_{\min} (\bm{W}_k^T \bm{W_k})$ is bounded away from zero. In addition, it also requires $\lambda_{\max} \{(\bm{I}_{nq} - \widehat{\bm{D}}_k^T \otimes \bm{W}_k)^{-1}\}$ are bounded uniformly.  
 Condition (\ref{C6}) requires that $\widehat{\bm{D}}_k$ converge to $\bm{D}_k^*$ at a rate such that $\xi_n^{-1} \sup_k \lambda_{\max} (\widetilde{\bm{P}}_k - \overline{\bm{P}}_k)$ converges to $0$ at a rate faster than $n\to \infty$. It's similar to condition (A5) of \cite{zhang2014}, Assumption 6 of \cite{zhang2018spatial} and Assumption 8 of \cite{yang2022model}.

\medskip
\begin{theorem}\label{Th1} Under Conditions (\ref{C00})-(\ref{C6}), we have
	\begin{equation*}
	    \frac{L_{\widehat{k}}}{\inf_{k \in \{ 1,\cdots,K \}} L_k}\rightarrow_p1\quad \mbox{ as } n \mbox{ goes to infinity},
	\end{equation*}
 where $\widehat k$ is defined in (\ref{khat}).
\end{theorem}
\medskip

The proof is provided in Appendix B.  Theorem \ref{Th1} indicates that the squared loss of our selected $\widetilde{\bm{\mu}}_{\hat{k}}$ estimator is asymptotically identical to the squared loss of the infeasible estimator that uses the best candidate spatial weights matrix. That is, the selected model is asymptotically optimal among all the candidate models.

Next, we assume that $\mathcal{W}$ includes the true weights matrix $\bm{W}^*$ and show that our selection method has model selection consistency. Let $\mathcal{S}$ be the index set of spatial weights matrices excluding the true $\bm{W}^*$, i.e., $\mathcal{S} = \{1\leq k \leq K: \bm{W}_k \neq \bm{W}^* \}$. Define $\xi_n^* = \inf_{k \in \mathcal{S}} R_k^*$. To show the selection consistency, we need to modify the previous conditions (\ref{C1}), (\ref{C5}) and (\ref{C6}) to the following three conditions, respectively.

\begin{description}
\item[(\condition{C7})] There exists a positive integer $G$ such that $ \sum_{k\in\mathcal{S}} (R_k^*)^{-G} = o(1)$.
	\item[(\condition{C8})]
 $ \xi_n^{*-1}\sup_k \left| \text{tr} \left( \frac{\partial \text{vec} (\widehat{\bm{D}}_k)}{\partial \bm{y}^T} (\bm{y}^T \otimes \widehat{\bm{\Omega}}) \frac{\partial \text{vec} (\widetilde{\bm{P}}_k)}{\partial \text{vec}(\widehat{\bm{D}}_k)^T} \right)  \right|  = o_p (1)$.
 \item[(\condition{C9})] $n \xi_n^{*-1}\sup_k \lambda_{\max} (\widetilde{\bm{P}}_k - \overline{\bm{P}}_k)  = o_p(1)$ and $\xi_n^{* -2} n = o(1)$.
 \end{description}

The main difference of these three conditions to their previous analogues is that $\xi_n$ is replaced by $\xi_n^*$ and the summation in (\ref{C7}) is taken over $\mathcal{S}$ instead of $\{1,\cdots,K\}$.  The last part of (\ref{C9}) requires that $\xi_n^*$ increases at a rate faster than $n^{1/2}$.  These rates play important roles in bounding certain terms in the proof of the following theorem.

\medskip
\begin{theorem}\label{Th2} If the true weights matrix is included in the candidates set, under Conditions (\ref{C00})-(\ref{C0}), (\ref{C2})-(\ref{C4}) and (\ref{C7})-(\ref{C9}), we have
\begin{equation*}
P \left( \bm{W}_{\widehat{k}} = \bm{W}^* \right) \rightarrow_p 1 \quad \mbox{ as } n \mbox{ goes to infinity},
\end{equation*}
where $\widehat k$ is defined in (\ref{khat}).
\end{theorem}
\medskip

The proof is provided in Appendix C. Theorem \ref{Th2} indicates that our selection method has the selection consistency in the sense that it chooses the true weights matrix with probability approaching one in large samples if $\mathcal{W}$ includes the true weights matrix $\bm{W}^*$.

\vspace{0.4cm}
{\color{black}\noindent{\bf Remark 4.} It is well-known that, in variable selection, Mallows $C_p$ is an AIC-type criterion, which has asymptotic  optimality but not selection consistency. In Theorem \ref{Th2} we shows the selection consistency of our method. It is understandable since here we focus on the selection of the spatial weights matrix but not the selection of the covariates. }
\vspace{0.2cm}

\section{Model Averaging of Spatial Weights Matrix} \label{sec4}
Instead of ``putting all our inferential eggs in one unevenly woven basket" \citep{Longford2005} as model selection, model averaging aims to assign weights to candidate models based on their importance, and tends to achieve better prediction performance than model selection in finite samples.  In this section, we propose a model averaging method for the spatial weights matrix, which has the potential of reducing squared loss.

Let $\widetilde{\bm{P}}(\w) = \sum_{k = 1}^K w_k \widetilde{\bm{P}}_k $, where $\w = (w_1, \cdots, w_K)^T$ is a weight vector belonging to the set $\mathcal{H} = \left \{ \w \in [0,1]^K : \sum_{k = 1}^K w_k = 1 \right \}$. The model averaging estimator of $\bm{\mu}$ is
\begin{equation} \label{estimator}
\widetilde{\bm{\mu}}(\w) = \sum_{k=1}^K w_k \widetilde{\bm{\mu}}_k = \sum_{k=1}^K w_k \widetilde{\bm{P}}_k \bm{y} = \widetilde{\bm{P}} (\w) \bm{y}.
\end{equation}

\noindent Define $L(\w) = \Vert \widetilde{\bm{\mu}}(\w) - \bm{\mu} \Vert^2$ and $ R (\w) = E(L(\w)) = E \Vert \widetilde{\bm{\mu}}(\w) - \bm{\mu} \Vert^2 $.

To determine the optimal weight $\w$ that minimizes the squared loss, we define a model averaging analogue of (\ref{selectionCri_k}) as
 \begin{equation*} \label{cri}
C (\w) \!=\! \big  \Vert \widetilde{\bm{P}} (\w) \bm{y} -  \bm{y} \big  \Vert^2 \!+\! 2 \left \{ \text{tr}(\widetilde{\bm{P}}(\w) \bm{\Omega}) \!+\!  \sum_{k=1}^K w_k \text{tr} \left( \frac{\partial \text{vec}(\widehat{\bm{D}}_k)}{\partial \bm{y}^T} (\bm{y}^T \otimes \bm{\Omega}) \frac{\partial \text{vec}(\widetilde{\bm{P}}_k)}{\partial \text{vec}(\widehat{\bm{D}}_k)^T} \right) \right \}.
\end{equation*}

\noindent Similar to the arguments in the previous section for model selection, {\color{black} if $\bm{\varepsilon}\sim N(\bm{0},\bm{\Sigma})$,} we have
\begin{equation*} \label{expectedCri}
    E \big \{ C (\w) \big \} = R (\w) + \text{tr} (\bm{\Omega}).
\end{equation*}


\noindent {\color{black} Note that the normality assumption is only used to derive the weight selection criterion but not required for the theoretical results.} Let $\widehat{\bm{\Omega}}$ is an estimator of $\bm{\Omega}$, which again is not necessarily to be consistent. We can define a feasible weight choice criterion as
\begin{equation*} \label{averagingCri}
\widehat{C} (\w) = \big \Vert \widetilde{\bm{P}} (\w) \bm{y} -  \bm{y} \big \Vert^2 + 2 \left \{ \text{tr}(\widetilde{\bm{P}}(\w) \bm{\widehat{\Omega}}) +  \sum_{k=1}^K w_k \text{tr} \left( \frac{\partial \text{vec}(\widehat{\bm{D}}_k)}{\partial \bm{y}^T} (\bm{y}^T \otimes \widehat{\bm{\Omega}}) \frac{\partial \text{vec}(\widetilde{\bm{P}}_k)}{\partial \text{vec}(\widehat{\bm{D}}_k)^T} \right) \right \}.
\end{equation*}

\noindent The optimal weight is selected by $\widehat{\w} = \underset{\w \in \mathcal{H}}{\text{argmin}} \big\{\widehat{C}(\w)\big\}$.

Define $\bm{H} = (\widetilde{\bm{P}}_1 \bm{y} - \bm{y}, \cdots, \widetilde{\bm{P}}_K \bm{y} - \bm{y})$ and
$$
\bm{h} =
\begin{bmatrix}
& \text{tr} (\widetilde{\bm{P}}_1 \widehat{\bm{\Omega}}) + \text{tr} \left( \frac{\partial \text{vec}(\widehat{\bm{D}}_1)}{\partial \bm{y}^T}  (\bm{y}^T \otimes \widehat{\bm{\Omega}}) \frac{\partial \text{vec}(\widetilde{\bm{P}}_1)}{\partial \text{vec}(\widehat{\bm{D}}_1)^T} \right) \\
& \vdots \\
& \text{tr} (\widetilde{\bm{P}}_K \widehat{\bm{\Omega}}) + \text{tr} \left( \frac{\partial \text{vec}(\widehat{\bm{D}}_K)}{\partial \bm{y}^T} (\bm{y}^T \otimes \widehat{\bm{\Omega}}) \frac{\partial \text{vec}(\widetilde{\bm{P}}_K)}{\partial \text{vec}(\widehat{\bm{D}}_K)^T} \right)
\end{bmatrix}.
$$

\noindent It is straightforward to show that $\widehat{C} (\w) = \w^T \bm{H}^T \bm{H} \w + 2 \w^T \bm{h}$, which is a quadratic function of $\w$. The optimization problem of a quadratic function with linear constraints has been well-studied, and numerical solvers are available in popular statistical softwares such as Python, R
and Matlab.

{\color{black} Let $\hat w_k$ be the $k$-th element of $\widehat{\w}$. The estimated spatial weights matrix is $\sum_{k=1}^K\hat w_k\bm{W}_k$. By taking average over $K$ candidate spatial weights matrices, it covers a large range of spatial weights matrix in the convex hull of $\{\sum_{k=1}^Kw_k\bm{W}_k:0\leq w_k\leq 1, \sum_{k=1}^Kw_k=1\}$. Also, by checking the estimated $\hat w_k$, we can understand which spatial weighting mechanisms play more important roles, as illustrated in the real data analysis. }

Next, we show that our model averaging estimator is asymptotically optimal in the sense that its associated squared loss is asymptotically equivalent to the smallest squared loss, if all the candidate models are misspecified. Define $R^*(\w) = E \Vert \sum_{k=1}^K w_k \widetilde{\bm{\mu}}_k |_{\widehat{\bm{D}}_k = \bm{D}_k^*} - \bm{\mu} \Vert^2$ and  $\widetilde{\xi}_n = \inf_{\w \in \mH} R^* (\w)$. We need to modify the model selection conditions (\ref{C1}), (\ref{C5}) and (\ref{C6}) to the following three conditions, respectively.

\begin{description}
\item[(\condition{C10})] There exists a positive integer $G$ such that $K \widetilde{\xi_n}^{-2G} \sum_{k=1}^K (R_k^*)^G = o(1)$.
	\item[(\condition{C11})]
 $ \widetilde{\xi_n}^{-1} \sup_k \left| \text{tr} \left( \frac{\partial \text{vec} (\widehat{\bm{D}}_k)}{\partial \bm{y}^T} (\bm{y}^T \otimes \widehat{\bm{\Omega}}) \frac{\partial \text{vec} (\widetilde{\bm{P}}_k)}{\partial \text{vec}(\widehat{\bm{D}}_k)^T} \right)  \right|= o_p (1)$.
 \item[(\condition{C12})] $\widetilde{\xi}_n^{-1} = o(1)$ and $n \widetilde{\xi}_n^{-1} \sup_k \lambda_{\max} (\widetilde{\bm{P}}_k - \overline{\bm{P}}_k) = o_p(1)$.
 \end{description}

The main difference of these three conditions to their previous analogues is that $\xi_n$ is replaced by $\widetilde{\xi}_n$. Condition (\ref{C10}) states the convergence of squared loss in each candidate model, which is common in the model averaging literature \citep{wan2010least, liu2013heteroscedasticity, zhang2018spatial}. It allows that $K \sum_{k=1}^K (R_k^*)^G$ diverges at a rate slower than $\widetilde{\xi}_n^{2G}$ as $n \to \infty$. It's obvious that $\widetilde{\xi}_n \to \infty$ is a necessary condition for it to hold, which indicates that the candidate set can not include the true spatial weights matrix by similar discussions of Condition (\ref{C1}).

\medskip
\begin{theorem}\label{Th3} Under Conditions  (\ref{C00})-(\ref{C0}), (\ref{C2})-(\ref{C4}) and (\ref{C10})-(\ref{C12}), we have
\begin{equation*}
\frac{L(\widehat{\w})}{\inf_{\w\in\mH} L(\w)}\rightarrow_p1\quad \mbox{ as } n \mbox{ goes to infinity}.
\end{equation*}
\end{theorem}
\medskip

The proof is provided in Appendix D. Theorem \ref{Th3} means the selected weight vector yields a squared loss which is asymptotically identical to that of the infeasible optimal weight vector. This implies the model averaging estimator $\widetilde{\bm{\mu}} (\widehat{\w})$ is asymptotically optimal in the class of model averaging estimators defined in (\ref{estimator}). In addition, since $\inf_{\w \in \mathcal{H}} \Vert \sum_{k=1}^K w_k (\widetilde{\bm{\mu}}_k - \bm{\mu}) \Vert \leq \inf_{\w \in \mathcal{H}} \sum_{k=1}^K w_k \Vert \widetilde{\bm{\mu}}_k - \bm{\mu} \Vert \leq \inf_k \Vert \widetilde{\bm{\mu}}_k - \bm{\mu} \Vert $, it is expected that model averaging can perform better than any candidate model in the sense of squared loss.

\section{Simulation Studies} \label{sec5}

In this section, we conduct simulation studies to evaluate the finite sample performances of the proposed model selection (MS) and model averaging (MA) methods for the MSAR model. We are mainly interested in the following questions:
\begin{itemize}
\item[(i)] Will MSAR with MS and MA be better than MSAR model based on each single candidate $\bm{W}_k$, except in the case where $\bm{W}^*$ belongs to candidate set $\mathcal{W}$?
\item[(ii)] When $\bm{Y}$ is multivariate, will MSAR with MS and MA be better than univariate SAR with MS and MA for each single component of $\bm{Y}$ alone?
\item[(iii)] If the true model is included in the candidate set $\mathcal{W}$, will MS have a high probability of selecting the true model, and MA concentrate the weights on the true model?
\item[(iv)] How is MA compared to MS?
\end{itemize}

In addition, we compare our proposed MS and MA methods to the high-order MSAR model which extends the classical high-order SAR model \citep{anselin1988, 2010EFFICIENT, 2015Inference} to the multivariate response case, although it's less addressed in the literature and in practice.

{\color{black}\subsection{MS and MA with normal error}\label{exDesign}}

Consider $p=q=2$, i.e., $\bm{Y}=(\bm{Y_1},\bm{Y_2})$ and $\bm{X}$ both belong to $\mathcal{R}^{n\times 2}$. Each row of $\bm{X}$ follows a bivariate normal distribution with mean $\boldsymbol{0}$ and a covariance matrix of $(1,0.5;0.5,1)$. Let $\bm{B} = (-0.5, 1;1.3, 0.3)$. {\color{black} Each row of $\bm{E}$ is identically and independently distributed as $N(\bm{0},\bm{\Sigma_e})$.} Two cases of different $\bm{\Sigma}_e$ and $\bm{D}$ are considered. In Case 1: $\bm{\Sigma}_e = (0.5,0.3;0.3,0.8)$, $\bm{D} = (0.3,-0.3;0.5,0.4)$; In Case 2: $\bm{\Sigma}_e = (0.5,0.1;0.1,0.8)$, $\bm{D} = (0.3,-0.1;0.1,0.4)$. In Case 2, the correlation between $\bm{Y_1}$ and $\bm{Y_2}$ is relatively weak. The sample size is $n = 300$. The simulation round is {\color{black} 500}.

We consider $K=4$ candidate spatial weights matrices, by the definitions in \cite{zhang2018spatial}. $\bm{W}_1$ is a square tessellation in which each unit interacts only with its left neighbor (with the left edge unit having the right edge unit as its neighbor), referred to as the ``left" matrix. In $\bm{W}_2$, each unit interacts only with its left and right units, referred to as the ``left-right" matrix. In $\bm{W}_3$, each unit interacts only with its top, bottom, left, and right units, referred to as the ``rook" matrix. $\bm{W}_4$ is the ``queen" matrix, where each unit interacts with the surrounding 8 neighboring units. Note that the spatial weights matrices become denser from $\bm{W}_1$ to $\bm{W}_4$. All weights matrices are row-normalized. We use $\bm{W}_1$, $\bm{W}_4$ or $(\bm{W}_1+\bm{W}_4)/2$ as the true spatial weights matrix to generate the response data $\bm{Y}$.

We consider the following estimation methods: (i) Direct MSAR estimation based on each candidate model $\bm{W}_k$. (ii) The proposed MS and MA methods for MSAR. We use $\bm{W}_4$ to obtain $\widehat{\bm{\Omega}}$ for our proposed method, which is not necessarily consistent. (iii) Direct SAR estimation based solely on $\bm{Y}_1$ or $\bm{Y}_2$ with each candidate model $\bm{W}_k$. (iv) MS and MA methods for SAR \citep{zhang2018spatial} based solely on $\bm{Y}_1$ or $\bm{Y}_2$.

The methods are compared based on the following criteria. (i) The main criterion is the prediction error of $\bm{\mu}$. We calculate the mean squared error (MSE) of $\bm{\mu}_1$ and $\bm{\mu}_2$ (i.e., the means of $\bm{Y}_1$ and $\bm{Y}_2$) separately, for the convenience to compare with the univariate SAR model based methods. (ii) We use the Frobenius norm of matrix to evaluate the estimation error of $\bm{D}$, $\bm{B}$, and $\bm{W}$. For example, the estimation error for $\bm{D}$ is $ \Vert \widehat{\bm{D}} - \bm{D} \Vert_F$. The weighted estimators of  $\bm{D}$ and $\bm{B}$  for MA have no explicit explanations. So their estimation errors are not reported for MA. (iii) We are also interested in the averaged frequency (denoted as ``MS accuracy") of selecting each candidate model for model selection, as well as the averaged weights (denoted as ``MA weights") assigned to each candidate model for model averaging.


Tables \ref{Tab1}, \ref{Tab2} and \ref{Tab3}  present the simulation results for the situations that the true spatial weights matrix is $\bm{W}_1$, $\bm{W}_4$  and $(\bm{W}_1+\bm{W}_4)/2$, respectively.  The simulation results can be summarized as follows.

First, when the true model is included in the candidate models (Tables \ref{Tab1} and \ref{Tab2}), MSAR estimations based on the true model are the best among all the candidates, while the results based on wrong models could be very poor in terms of large estimation errors. Our proposed MS and MA methods based on MSAR model perform quite well. The MSEs of $(\widehat{\bm{\mu}}_1,\widehat{\bm{\mu}}_2)$ are even smaller than the true model in many cases. When the true model is not included in the candidate models (Table \ref{Tab3}), results based on MS are close to the MSAR with the single candidate model that performs the best among all the candidates. MA performs  better than MSAR based on single candidate models.

Second, methods utilizing only single $\bm{Y}_1$ or $\bm{Y}_2$ (direct SAR estimation based on each single candidate model, MS and MA) perform much worse than the methods using multivariate $\bm{Y}$. The advantage of MSAR is weakened when the correlation between $\bm{Y}_1$ and $\bm{Y}_2$ is weaker (case 2). In addition, MA methods based on single $\bm{Y}_1$ or $\bm{Y}_2$ are also better than univariate SAR based on single candidate models.

Third, when the true model is included in the candidate models (Tables \ref{Tab1} and \ref{Tab2}), our MS method of MSAR model can select the true model at most of the times, while the MA method tends to assign the largest weights to the true model. MA performs worse than MS in terms of choosing the correct spatial weights matrix. But when the true model is not included in the candidate models (Table \ref{Tab3}), this disadvantage of MA becomes an advantage over MS, because MS can only select among the wrong models,  while MA can assign weights across different candidate models to further improve the prediction performance of $\bm{\mu}$.

Fourth, comparing MS and MA methods of MSAR, when the true model is included in the candidate models (Tables \ref{Tab1} and \ref{Tab2}), the MS method usually provides better estimates for $\bm{W}$ {\color{black} in terms of $\Vert \widehat{\bm{W}} - \bm{W} \Vert_F$}, but results may be worse for estimating $\bm{\mu}$ (MA has better performance in most cases). {\color{black} The reason is that MS can only select one single model and it often selects the true model, which is not necessarily the best model for predicting $\bm{\mu}$ in finite samples. On the other hand, MA can assign weights across different models to achieve better estimation of $\bm{\mu}$, but may lose estimation accuracy of $\bm{W}$ (note that the model averaging criterion is established based on $\bm{\mu}$ but not $\bm{W}$).} However, when the true model is not included in the candidate models (Table \ref{Tab3}), undoubtedly, MA outperforms MS method for the estimation of {\color{black}$\bm{W}$ and} $\bm{\mu}$.

{\color{black}
\begin{table}[t]
\scriptsize
\begin{center}
\caption{$\bm{W}_1$ is true and error terms are multivariate normal\label{Tab1}}\smallskip
\resizebox{\linewidth}{!}{
{\color{black}
\begin{tabular}{cccccccc}
\multicolumn{8}{c}{{\normalsize Case 1: $\bm{\Sigma}_e = (0.5,0.3;0.3,0.8)$, $\bm{D} = (0.3,-0.3;0.5,0.4)$}}\\ \hline
\multicolumn{2}{c}{}          & $\bm{W}_1$ & $\bm{W}_2$ & $\bm{W}_3$ & $\bm{W}_4$ & MS & MA  \\ \hline
 MSAR & $\Vert \widehat{\bm{D}} - \bm{D} \Vert_F$ & 0.126  & 0.264  & 0.315  & 0.325 & 0.124  \\
Based $\bm{Y} = (\bm{Y}_1,\bm{Y}_2)$    & $\Vert \widehat{\bm{B}} - \bm{B} \Vert_F$          & 0.213  & 0.308  & 0.193  & 0.153  &  0.213    \\
     & MSE of $(\widehat{\bm{\mu}}_1,\widehat{\bm{\mu}}_2)$                 & (0.030, 0.065)  & (0.417, 0.185)  & (0.626, 0.273)  & (0.667, 0.281) & (0.032, 0.060)     & (0.028, 0.038) \\
   & $\Vert \widehat{\bm{W}} - \bm{W} \Vert_F$   &        &        &        &    & 0.163     & 1.805 \\
   & MS accuracy         & 0.986  & 0.014  & 0.000  & 0.000 \\
      & MA weights         & 0.864  & 0.060  & 0.028  & 0.048 \\
 \hline
 SAR     & MSE of $\widehat{\bm{\mu}}_1$                   & 0.810  & 2.195  & 1.784  & 0.714  & 0.705   & 0.589 \\
 Based $\bm{Y}_1$   & $\Vert \widehat{\bm{W}} - \bm{W} \Vert_F$ &        &        &        &    & 11.130     & 9.079 \\
   & MS accuracy         & 0.298  & 0.008  & 0.072  & 0.622 \\
      & MA weights            & 0.423  & 0.012  & 0.045  & 0.520 \\
 \hline
 SAR     & MSE of $\widehat{\bm{\mu}}_2$                  & 0.327  & 0.330  & 0.301  & 0.293  & 0.289  & 0.248 \\
Based $\bm{Y}_2$     & $\Vert \widehat{\bm{W}} - \bm{W} \Vert_F$ &        &        &        &   & 11.107    & 9.381 \\
    & MS accuracy         & 0.260  & 0.122  & 0.176  & 0.442 \\
      & MA weights           & 0.370  & 0.092  & 0.078  & 0.460 \\
 \hline
 \\
 \multicolumn{8}{c}{{\normalsize Case 2: $\bm{\Sigma}_e = (0.5,0.1;0.1,0.8)$, $\bm{D} = (0.3,-0.1;0.1,0.4)$}}\\ \hline
\multicolumn{2}{c}{}          & $\bm{W}_1$ & $\bm{W}_2$ & $\bm{W}_3$ & $\bm{W}_4$ &  MS & MA \\ \hline
 MSAR & $\Vert \widehat{\bm{D}} - \bm{D} \Vert_F$   & 0.061  & 0.087  & 0.177  & 0.263  & 0.061 \\
Based $\bm{Y} = (\bm{Y}_1,\bm{Y}_2)$   & $\Vert \widehat{\bm{B}} - \bm{B} \Vert_F$            & 0.095  & 0.121  & 0.111  & 0.109  & 0.095     \\
   &  MSE of $(\widehat{\bm{\mu}}_1,\widehat{\bm{\mu}}_2)$                  & (0.004, 0.014)  & (0.093, 0.127)  & (0.156, 0.211)  & (0.168, 0.222) &  (0.004, 0.014)  & (0.005, 0.014) \\
  & $\Vert \widehat{\bm{W}} - \bm{W} \Vert_F$&        &        &        &    & 0.000      & 1.025 \\
  & MS accuracy         & 1.000  & 0.000  & 0.000  & 0.000 \\
      & MA weights           & 0.926  & 0.030  & 0.017  & 0.027 \\
 \hline
  SAR    & MSE of $\widehat{\bm{\mu}}_1$                   & 0.026  & 0.103  & 0.153  & 0.169  & 0.026 & 0.026 \\
Based $\bm{Y}_2$   & $\Vert \widehat{\bm{W}} - \bm{W} \Vert_F$  &        &        &        &    & 0.000     & 1.003 \\
  & MS accuracy         & 1.000  & 0.000  & 0.000  & 0.000 \\
      & MA weights          & 0.927  & 0.028  & 0.019  & 0.025 \\
\hline
  SAR    & MSE of $\widehat{\bm{\mu}}_2$                   & 0.032  & 0.134  & 0.205  & 0.219  & 0.032 & 0.032 \\
Based $\bm{Y}_2$   & $\Vert \widehat{\bm{W}} - \bm{W} \Vert_F$  &        &        &        &    & 0.076     & 2.098 \\
  & MS accuracy         & 0.994  & 0.004  & 0.002  & 0.000 \\
      & MA weights          & 0.848  & 0.059  & 0.033  & 0.059 \\
      \hline
\multicolumn{8}{l}{$\bm{W}_k$: methods based on $\bm{W}_k$; MS: model selection; MA: Model averaging.}
\end{tabular}}}
\end{center}
\end{table}
}

{\color{black}
\begin{table}[h]
\scriptsize
\begin{center}
\caption{$\bm{W}_4$  is true and error terms are multivariate normal\label{Tab2}}\smallskip
\resizebox{\linewidth}{!}{
{\color{black}
\begin{tabular}{cccccccc}
\multicolumn{8}{c}{{\normalsize Case 1: $\bm{\Sigma}_e = (0.5,0.3;0.3,0.8)$, $\bm{D} = (0.3,-0.3;0.5,0.4)$}}\\ \hline
\multicolumn{2}{c}{}          & $\bm{W}_1$ & $\bm{W}_2$ & $\bm{W}_3$ & $\bm{W}_4$ & MS & MA  \\ \hline
 MSAR & $\Vert \widehat{\bm{D}} - \bm{D} \Vert_F$ & 0.697  & 0.536  & 0.431  & 0.214 & 0.220  \\
Based $\bm{Y} = (\bm{Y}_1,\bm{Y}_2)$    & $\Vert \widehat{\bm{B}} - \bm{B} \Vert_F$          & 0.120  & 0.115  & 0.109  & 0.108  & 0.108   \\
     & MSE of $(\widehat{\bm{\mu}}_1,\widehat{\bm{\mu}}_2)$                 & (0.128, 0.050)  & (0.098, 0.038)  & (0.055, 0.026)  & (0.014, 0.013)  & (0.015, 0.013)   & (0.013, 0.011) \\
   & $\Vert \widehat{\bm{W}} - \bm{W} \Vert_F$   &        &        &        &    & 0.464     & 1.609 \\
   & MS accuracy         & 0.000  & 0.004  & 0.068  & 0.928 \\
      & MA weights         & 0.051  & 0.030  & 0.087  & 0.831 \\
 \hline
  SAR    & MSE of $\widehat{\bm{\mu}}_1$                   & 0.130  & 0.119  & 0.095  & 0.153   & 0.098  & 0.094 \\
Based $\bm{Y}_1$    & $\Vert \widehat{\bm{W}} - \bm{W} \Vert_F$ &        &        &        &    & 5.187    & 5.455 \\
   & MS accuracy         & 0.024  & 0.042  & 0.702  & 0.232 \\
      & MA weights            & 0.131  & 0.088  & 0.481  & 0.300 \\
 \hline
  SAR    & MSE of $\widehat{\bm{\mu}}_2$                  & 0.043  & 0.041  & 0.038  & 0.038  & 0.038   & 0.037 \\
Based $\bm{Y}_2$     & $\Vert \widehat{\bm{W}} - \bm{W} \Vert_F$ &        &        &        &   & 5.593   & 5.882 \\
    & MS accuracy         & 0.192  & 0.118  & 0.196  & 0.494 \\
      & MA weights           & 0.249  & 0.100  & 0.158  & 0.492 \\
 \hline
 \\
 \multicolumn{8}{c}{{\normalsize Case 2: $\bm{\Sigma}_e = (0.5,0.1;0.1,0.8)$, $\bm{D} = (0.3,-0.1;0.1,0.4)$}}\\ \hline
\multicolumn{2}{c}{}          & $\bm{W}_1$ & $\bm{W}_2$ & $\bm{W}_3$ & $\bm{W}_4$ &  MS & MA  \\ \hline
 MSAR & $\Vert \widehat{\bm{D}} - \bm{D} \Vert_F$   & 0.382  & 0.332  & 0.266  & 0.144 & 0.149  \\
Based $\bm{Y} = (\bm{Y}_1,\bm{Y}_2)$   & $\Vert \widehat{\bm{B}} - \bm{B} \Vert_F$            & 0.091  & 0.091  & 0.086  & 0.082  & 0.083     \\
   &  MSE of $(\widehat{\bm{\mu}}_1,\widehat{\bm{\mu}}_2)$                  & (0.032, 0.049)  & (0.027, 0.041)  & (0.016, 0.026)  & (0.004, 0.014)  & (0.005, 0.014)   & (0.004, 0.012) \\
  & $\Vert \widehat{\bm{W}} - \bm{W} \Vert_F$&        &        &        &      & 0.615    & 1.816 \\
  & MS accuracy         & 0.002  & 0.008  & 0.080  & 0.910 \\
      & MA weights           & 0.050  & 0.038  & 0.114  & 0.798 \\
 \hline
   SAR   & MSE of $\widehat{\bm{\mu}}_1$                    & 0.032  & 0.028  & 0.017  & 0.007  & 0.007  & 0.007 \\
Based $\bm{Y}_1$   & $\Vert \widehat{\bm{W}} - \bm{W} \Vert_F$ &        &        &        &      & 0.304     & 1.423 \\
  & MS accuracy         & 0.002  & 0.000  & 0.044  & 0.954 \\
      & MA weights         & 0.043  & 0.033  & 0.068  & 0.856 \\
\hline
   SAR   & MSE of $\widehat{\bm{\mu}}_2$                   & 0.048  & 0.039  & 0.024  & 0.013 & 0.016  & 0.013 \\
Based $\bm{Y}_2$   & $\Vert \widehat{\bm{W}} - \bm{W} \Vert_F$  &        &        &        &    & 1.880     & 2.638 \\
  & MS accuracy         & 0.024  & 0.038  & 0.174  & 0.764 \\
      & MA weights          & 0.078  & 0.059  & 0.141  & 0.722 \\
      \hline
\multicolumn{8}{l}{$\bm{W}_k$: methods based on $\bm{W}_k$; MS: model selection; MA: Model averaging.}
\end{tabular}}}
\end{center}
\end{table}
}


\begin{table}[h]
\scriptsize
\begin{center}
\caption{$(\bm{W}_1 + \bm{W}_4) / 2$  is true and error terms are multivariate normal\label{Tab3}}\smallskip
\resizebox{\linewidth}{!}{
{\color{black}\begin{tabular}{cccccccc}
\multicolumn{8}{c}{{\normalsize Case 1: $\bm{\Sigma}_e = (0.5,0.3;0.3,0.8)$, $\bm{D} = (0.3,-0.3;0.5,0.4)$}}\\ \hline
\multicolumn{2}{c}{}          & $\bm{W}_1$ & $\bm{W}_2$ & $\bm{W}_3$ & $\bm{W}_4$ & MS  & MA  \\ \hline
 MSAR & $\Vert \widehat{\bm{D}} - \bm{D} \Vert_F$ & 0.325  & 0.287  & 0.255  & 0.237 & 0.324  \\
 Based $\bm{Y} = (\bm{Y}_1,\bm{Y}_2)$   & $\Vert \widehat{\bm{B}} - \bm{B} \Vert_F$          & 0.147  & 0.154  & 0.130  & 0.113  & 0.146   \\
     & MSE of $(\widehat{\bm{\mu}}_1,\widehat{\bm{\mu}}_2)$                 & (0.045, 0.033)  & (0.121, 0.056)  & (0.132, 0.063)  & (0.142, 0.070)  & (0.046, 0.033)   & (0.026, 0.023) \\
   & $\Vert \widehat{\bm{W}} - \bm{W} \Vert_F$   &        &        &        &     & 8.016    & 3.533 \\
   & MS accuracy         & 0.988  & 0.002  & 0.006  & 0.004 \\
      & MA weights         & 0.691  & 0.018  & 0.080  & 0.211 \\
 \hline

   SAR   & MSE of $\widehat{\bm{\mu}}_1$                   & 0.187  & 0.240  & 0.228  & 0.249  & 0.188   & 0.165 \\
 Based $\bm{Y}_1$   & $\Vert \widehat{\bm{W}} - \bm{W} \Vert_F$ &        &        &        &     & 8.016   & 2.891 \\
   & MS accuracy         & 0.892  & 0.008  & 0.068  & 0.032 \\
      & MA weights            & 0.620  & 0.029  & 0.125  & 0.227 \\
 \hline

  SAR    & MSE of $\widehat{\bm{\mu}}_2$                  & 0.078  & 0.084  & 0.088  & 0.093  & 0.082   & 0.071 \\
  Based $\bm{Y}_2$    & $\Vert \widehat{\bm{W}} - \bm{W} \Vert_F$ &        &        &        &   & 8.016   & 4.221 \\
    & MS accuracy         & 0.560  & 0.190  & 0.134  & 0.116 \\
      & MA weights           & 0.500  & 0.164  & 0.126  & 0.210 \\
 \hline
 \\
 \multicolumn{8}{c}{{\normalsize Case 2: $\bm{\Sigma}_e = (0.5,0.1;0.1,0.8)$, $\bm{D} = (0.3,-0.1;0.1,0.4)$}}\\ \hline
\multicolumn{2}{c}{}          & $\bm{W}_1$ & $\bm{W}_2$ & $\bm{W}_3$ & $\bm{W}_4$  &  MS & MA  \\ \hline
MSAR & $\Vert \widehat{\bm{D}} - \bm{D} \Vert_F$   & 0.203  & 0.187  & 0.163  & 0.161  & 0.199 \\
Based $\bm{Y} = (\bm{Y}_1,\bm{Y}_2)$    & $\Vert \widehat{\bm{B}} - \bm{B} \Vert_F$            & 0.090  & 0.089  & 0.089  & 0.089  & 0.090     \\
   &  MSE of $(\widehat{\bm{\mu}}_1,\widehat{\bm{\mu}}_2)$                  & (0.013, 0.024)  & (0.032, 0.047)  & (0.037, 0.053)  & (0.039, 0.058)  & (0.015, 0.026)   & (0.009, 0.016) \\
  & $\Vert \widehat{\bm{W}} - \bm{W} \Vert_F$&        &        &        &      & 8.016    & 3.275 \\
  & MS accuracy         & 0.918  & 0.006  & 0.028  & 0.048 \\
      & MA weights           & 0.659  & 0.028  & 0.077  & 0.236 \\
 \hline

   SAR  & MSE of $\widehat{\bm{\mu}}_1$                    & 0.018  & 0.035  & 0.039  & 0.042  & 0.020   & 0.014 \\
 Based $\bm{Y}_1$   & $\Vert \widehat{\bm{W}} - \bm{W} \Vert_F$ &        &        &        &      & 8.016     & 3.707 \\
  & MS accuracy         & 0.930  & 0.014  & 0.018  & 0.038 \\
      & MA weights         & 0.688  & 0.034  & 0.065  & 0.213 \\
\hline

   SAR   & MSE of $\widehat{\bm{\mu}}_2$                   & 0.027  & 0.049  & 0.052  & 0.056 & 0.033  & 0.020 \\
 Based $\bm{Y}_2$   & $\Vert \widehat{\bm{W}} - \bm{W} \Vert_F$  &        &        &        &    & 8.016     & 3.687 \\
  & MS accuracy         & 0.758  & 0.062  & 0.070  & 0.110 \\
      & MA weights          & 0.611  & 0.069  & 0.089  & 0.230 \\
      \hline
\multicolumn{8}{l}{$\bm{W}_k$: methods based on $\bm{W}_k$; MS: model selection; MA: Model averaging.}
\end{tabular}}}
\end{center}
\end{table}

\vspace{0.5cm}
{\color{black}
\subsection{MS and MA with non-normal error}
In this subsection, we further conduct simulation studies for non-normal distributed errors. Each row of $\bm{E}$ is identically and independently distributed as a multivariate t-distribution with location parameter $\bm{0}_q$, scale matrix $\bm{\Sigma}_e$, and degree of freedom 5. All the other simulation settings are the same as subsection \ref{exDesign}. Tables \ref{tTab1}, \ref{tTab2} and \ref{tTab3}  present the simulation results for the situations that the true spatial weights matrix is $\bm{W}_1$, $\bm{W}_4$  and $(\bm{W}_1+\bm{W}_4)/2$, respectively. The simulation conclusions are quite similar to subsection \ref{exDesign}. The proposed method still performs well even when the errors are not normally distributed.
}

\begin{table}[t]
\scriptsize
\begin{center}
\caption{$\bm{W}_1$ is true and  error terms are  multivariate t-distribution \label{tTab1}}\smallskip
\resizebox{\linewidth}{!}{
{\color{black}
\begin{tabular}{cccccccc}
\multicolumn{8}{c}{{\normalsize Case 1: $\bm{\Sigma}_e = (0.5,0.3;0.3,0.8)$, $\bm{D} = (0.3,-0.3;0.5,0.4)$}}\\ \hline
\multicolumn{2}{c}{}          & $\bm{W}_1$ & $\bm{W}_2$ & $\bm{W}_3$ & $\bm{W}_4$ & MS & MA  \\ \hline
 MSAR & $\Vert \widehat{\bm{D}} - \bm{D} \Vert_F$ & 0.091  & 0.261  &  0.372  & 0.371 & 0.091  \\
Based $\bm{Y} = (\bm{Y}_1,\bm{Y}_2)$    & $\Vert \widehat{\bm{B}} - \bm{B} \Vert_F$          & 0.168  & 0.263  & 0.214  & 0.184  &  0.168    \\
     & MSE of $(\widehat{\bm{\mu}}_1,\widehat{\bm{\mu}}_2)$                 & (0.029, 0.026)  & (0.423, 0.184)  & (0.657, 0.275)  & (0.694, 0.284) & (0.029, 0.026)     & (0.029, 0.022) \\
   & $\Vert \widehat{\bm{W}} - \bm{W} \Vert_F$   &        &        &        &    & 0.000     & 1.342 \\
   & MS accuracy         & 1.000  & 0.000  & 0.000  & 0.000 \\
      & MA weights         & 0.901  & 0.042  & 0.022  & 0.034 \\
 \hline

 SAR     & MSE of $\widehat{\bm{\mu}}_1$                   & 0.937  & 1.435  & 1.649  & 0.736  & 0.742   & 0.622 \\
 Based $\bm{Y}_1$   & $\Vert \widehat{\bm{W}} - \bm{W} \Vert_F$ &        &        &        &    & 11.874     & 9.643 \\
   & MS accuracy         & 0.246  & 0.030  & 0.066  & 0.658 \\
      & MA weights            & 0.376  & 0.041  & 0.051  & 0.532 \\
 \hline

 SAR     & MSE of $\widehat{\bm{\mu}}_2$                  & 0.275  & 0.287  & 0.289  & 0.295  & 0.269   & 0.242 \\
Based $\bm{Y}_2$     & $\Vert \widehat{\bm{W}} - \bm{W} \Vert_F$ &        &        &        &   & 7.268    & 7.278 \\
    & MS accuracy         & 0.488  & 0.154  & 0.210  & 0.148 \\
      & MA weights           & 0.491  & 0.123  & 0.130  & 0.257  \\
 \hline
 \\
 \multicolumn{8}{c}{{\normalsize Case 2: $\bm{\Sigma}_e = (0.5,0.1;0.1,0.8)$, $\bm{D} = (0.3,-0.1;0.1,0.4)$}}\\ \hline
\multicolumn{2}{c}{}          & $\bm{W}_1$ & $\bm{W}_2$ & $\bm{W}_3$ & $\bm{W}_4$ &  MS & MA \\ \hline
 MSAR & $\Vert \widehat{\bm{D}} - \bm{D} \Vert_F$   & 0.081  & 0.114  & 0.235  & 0.295  & 0.081 \\
Based $\bm{Y} = (\bm{Y}_1,\bm{Y}_2)$   & $\Vert \widehat{\bm{B}} - \bm{B} \Vert_F$            & 0.144  & 0.166  & 0.158  & 0.155  & 0.144     \\
   &  MSE of $(\widehat{\bm{\mu}}_1,\widehat{\bm{\mu}}_2)$                  & (0.013, 0.023)  & (0.100, 0.136)  & (0.165, 0.220)  & (0.174, 0.233) &  (0.013, 0.023)  & (0.014, 0.024) \\
  & $\Vert \widehat{\bm{W}} - \bm{W} \Vert_F$&        &        &        &    & 0.023      & 1.649 \\
  & MS accuracy         & 0.998  & 0.002  & 0.000  & 0.000 \\
      & MA weights           & 0.879  & 0.050  & 0.031  & 0.039 \\
 \hline

  SAR    & MSE of $\widehat{\bm{\mu}}_1$                    & 0.032  & 0.108  & 0.157  & 0.173  & 0.033  & 0.034 \\
Based $\bm{Y}_1$   & $\Vert \widehat{\bm{W}} - \bm{W} \Vert_F$ &        &        &        &     & 0.148      & 1.713 \\
  & MS accuracy         & 0.988  & 0.010  & 0.000  & 0.002 \\
      & MA weights         & 0.875  & 0.054  & 0.030  & 0.041 \\
\hline

  SAR    & MSE of $\widehat{\bm{\mu}}_2$                   & 0.039  & 0.138  & 0.209  & 0.225  & 0.042 & 0.040 \\
Based $\bm{Y}_2$   & $\Vert \widehat{\bm{W}} - \bm{W} \Vert_F$  &        &        &        &    & 0.383     & 2.319 \\
  & MS accuracy         & 0.970  & 0.020  & 0.008  & 0.002 \\
      & MA weights          & 0.832  & 0.065  & 0.040  & 0.062 \\
      \hline
\multicolumn{8}{l}{$\bm{W}_k$: methods based on $\bm{W}_k$; MS: model selection; MA: Model averaging.}
\end{tabular}}}
\end{center}
\end{table}

\begin{table}[t]
\scriptsize
\begin{center}
\caption{$\bm{W}_4$ is true and  error terms are  multivariate t-distribution \label{tTab2}}\smallskip
\resizebox{\linewidth}{!}{
{\color{black}
\begin{tabular}{cccccccc}
\multicolumn{8}{c}{{\normalsize Case 1: $\bm{\Sigma}_e = (0.5,0.3;0.3,0.8)$, $\bm{D} = (0.3,-0.3;0.5,0.4)$}}\\ \hline
\multicolumn{2}{c}{}          & $\bm{W}_1$ & $\bm{W}_2$ & $\bm{W}_3$ & $\bm{W}_4$ & MS & MA  \\ \hline
 MSAR & $\Vert \widehat{\bm{D}} - \bm{D} \Vert_F$ & 0.720  & 0.540  &  0.447  & 0.258 & 0.261  \\
Based $\bm{Y} = (\bm{Y}_1,\bm{Y}_2)$    & $\Vert \widehat{\bm{B}} - \bm{B} \Vert_F$          & 0.146  & 0.140  & 0.137  & 0.134  &  0.132    \\
     & MSE of $(\widehat{\bm{\mu}}_1,\widehat{\bm{\mu}}_2)$                 & (0.140, 0.059)  & (0.103, 0.042)  & (0.059, 0.032)  & (0.021, 0.019) & (0.023, 0.019)     & (0.018, 0.016) \\
   & $\Vert \widehat{\bm{W}} - \bm{W} \Vert_F$   &        &        &        &    & 0.674    & 1.988 \\
   & MS accuracy         & 0.000  & 0.004  & 0.102  & 0.894 \\
      & MA weights         & 0.056  & 0.041  & 0.123  & 0.780 \\
 \hline

 SAR     & MSE of $\widehat{\bm{\mu}}_1$                   & 0.137  & 0.122  & 0.097  & 0.225  & 0.101   & 0.097 \\
 Based $\bm{Y}_1$   & $\Vert \widehat{\bm{W}} - \bm{W} \Vert_F$ &        &        &        &    & 5.582     & 5.626 \\
   & MS accuracy         & 0.038  & 0.072  & 0.676  & 0.214 \\
      & MA weights            & 0.132  & 0.101  & 0.412  & 0.295 \\
 \hline

 SAR     & MSE of $\widehat{\bm{\mu}}_2$                  & 0.046  & 0.043  & 0.041  & 0.040  & 0.040   & 0.039 \\
Based $\bm{Y}_2$     & $\Vert \widehat{\bm{W}} - \bm{W} \Vert_F$ &        &        &        &   & 5.882    & 5.903 \\
    & MS accuracy         & 0.198  & 0.112  & 0.238  & 0.452 \\
      & MA weights           & 0.221  & 0.112  & 0.213  & 0.453  \\
 \hline
 \\
 \multicolumn{8}{c}{{\normalsize Case 2: $\bm{\Sigma}_e = (0.5,0.1;0.1,0.8)$, $\bm{D} = (0.3,-0.1;0.1,0.4)$}}\\ \hline
\multicolumn{2}{c}{}          & $\bm{W}_1$ & $\bm{W}_2$ & $\bm{W}_3$ & $\bm{W}_4$ &  MS & MA \\ \hline
 MSAR & $\Vert \widehat{\bm{D}} - \bm{D} \Vert_F$   & 0.383  & 0.353  & 0.297  & 0.208  & 0.231 \\
Based $\bm{Y} = (\bm{Y}_1,\bm{Y}_2)$   & $\Vert \widehat{\bm{B}} - \bm{B} \Vert_F$            & 0.137  & 0.138  & 0.136  & 0.131  & 0.132     \\
   &  MSE of $(\widehat{\bm{\mu}}_1,\widehat{\bm{\mu}}_2)$                  & (0.039, 0.055)  & (0.034, 0.048)  & (0.023, 0.034)  & (0.013, 0.024) &  (0.016, 0.027)  & (0.014, 0.022) \\
  & $\Vert \widehat{\bm{W}} - \bm{W} \Vert_F$&        &        &        &    & 1.950      & 3.017 \\
  & MS accuracy         & 0.020  & 0.024  & 0.198  & 0.758 \\
      & MA weights           & 0.078  & 0.066  & 0.157  & 0.699 \\
 \hline

  SAR    & MSE of $\widehat{\bm{\mu}}_1$                    & 0.038  & 0.032  & 0.022  & 0.012  & 0.016  & 0.015 \\
Based $\bm{Y}_1$   & $\Vert \widehat{\bm{W}} - \bm{W} \Vert_F$ &        &        &        &     & 2.284      & 2.948 \\
  & MS accuracy         & 0.046  & 0.046  & 0.168  & 0.740 \\
      & MA weights         & 0.091  & 0.069  & 0.136  & 0.704 \\
\hline

  SAR    & MSE of $\widehat{\bm{\mu}}_2$                   & 0.053  & 0.044  & 0.029  & 0.018  & 0.023 & 0.021 \\
Based $\bm{Y}_2$   & $\Vert \widehat{\bm{W}} - \bm{W} \Vert_F$  &        &        &        &    & 3.092     & 3.420 \\
  & MS accuracy         & 0.064  & 0.074  & 0.202  & 0.660 \\
      & MA weights          & 0.102  & 0.081  & 0.166  & 0.650 \\
      \hline
\multicolumn{8}{l}{$\bm{W}_k$: methods based on $\bm{W}_k$; MS: model selection; MA: Model averaging.}
\end{tabular}}}
\end{center}
\end{table}

\begin{table}[t]
\scriptsize
\begin{center}
\caption{$(\bm{W}_1+\bm{W}_4)/2$ is true and  error terms are  multivariate t-distribution \label{tTab3}}\smallskip
\resizebox{\linewidth}{!}{
{\color{black}
\begin{tabular}{cccccccc}
\multicolumn{8}{c}{{\normalsize Case 1: $\bm{\Sigma}_e = (0.5,0.3;0.3,0.8)$, $\bm{D} = (0.3,-0.3;0.5,0.4)$}}\\ \hline
\multicolumn{2}{c}{}          & $\bm{W}_1$ & $\bm{W}_2$ & $\bm{W}_3$ & $\bm{W}_4$ & MS & MA  \\ \hline
 MSAR & $\Vert \widehat{\bm{D}} - \bm{D} \Vert_F$ & 0.347  & 0.269  &  0.252  & 0.269 & 0.328  \\
Based $\bm{Y} = (\bm{Y}_1,\bm{Y}_2)$    & $\Vert \widehat{\bm{B}} - \bm{B} \Vert_F$          & 0.158  & 0.159  & 0.155  & 0.151  &  0.154    \\
     & MSE of $(\widehat{\bm{\mu}}_1,\widehat{\bm{\mu}}_2)$                 & (0.062, 0.033)  & (0.129, 0.057)  & (0.141, 0.065)  & (0.155, 0.070) & (0.070, 0.036)     & (0.038, 0.027) \\
   & $\Vert \widehat{\bm{W}} - \bm{W} \Vert_F$   &        &        &        &    & 8.016     & 3.470 \\
   & MS accuracy         & 0.868  & 0.034  & 0.032  & 0.066 \\
      & MA weights         & 0.646  & 0.054  & 0.082  & 0.218 \\
 \hline

 SAR     & MSE of $\widehat{\bm{\mu}}_1$                   & 0.216  & 0.247  & 0.237  & 0.248  & 0.218   & 0.180 \\
 Based $\bm{Y}_1$   & $\Vert \widehat{\bm{W}} - \bm{W} \Vert_F$ &        &        &        &    & 8.016      & 3.097 \\
   & MS accuracy         & 0.542  & 0.108  & 0.190  & 0.160 \\
      & MA weights            & 0.504  & 0.077  & 0.145  & 0.234 \\
 \hline

 SAR     & MSE of $\widehat{\bm{\mu}}_2$                  & 0.072  & 0.080  & 0.087  & 0.091  & 0.077   & 0.072 \\
Based $\bm{Y}_2$     & $\Vert \widehat{\bm{W}} - \bm{W} \Vert_F$ &        &        &        &   & 8.016    & 5.325 \\
    & MS accuracy         & 0.604  & 0.186  & 0.108  & 0.102 \\
      & MA weights           & 0.576  & 0.159  & 0.111  & 0.154  \\
 \hline
 \\
 \multicolumn{8}{c}{{\normalsize Case 2: $\bm{\Sigma}_e = (0.5,0.1;0.1,0.8)$, $\bm{D} = (0.3,-0.1;0.1,0.4)$}}\\ \hline
\multicolumn{2}{c}{}          & $\bm{W}_1$ & $\bm{W}_2$ & $\bm{W}_3$ & $\bm{W}_4$ &  MS & MA \\ \hline
 MSAR & $\Vert \widehat{\bm{D}} - \bm{D} \Vert_F$   & 0.209  & 0.209  & 0.200  & 0.220  & 0.208 \\
Based $\bm{Y} = (\bm{Y}_1,\bm{Y}_2)$   & $\Vert \widehat{\bm{B}} - \bm{B} \Vert_F$            & 0.135  & 0.135  & 0.133  & 0.135  & 0.134     \\
   &  MSE of $(\widehat{\bm{\mu}}_1,\widehat{\bm{\mu}}_2)$                  & (0.020, 0.031)  & (0.040, 0.055)  & (0.044, 0.062)  & (0.049, 0.067) &  (0.026, 0.039)  & (0.016, 0.025) \\
  & $\Vert \widehat{\bm{W}} - \bm{W} \Vert_F$&        &        &        &    & 8.016      & 3.543 \\
  & MS accuracy         & 0.756  & 0.060  & 0.060  & 0.124 \\
      & MA weights           & 0.606  & 0.066  & 0.098  & 0.229 \\
 \hline

  SAR    & MSE of $\widehat{\bm{\mu}}_1$                    & 0.024  & 0.040  & 0.044  & 0.047  & 0.030  & 0.022 \\
Based $\bm{Y}_1$   & $\Vert \widehat{\bm{W}} - \bm{W} \Vert_F$ &        &        &        &     & 8.016      & 4.267 \\
  & MS accuracy         & 0.690  & 0.072  & 0.088  & 0.150 \\
      & MA weights         & 0.617  & 0.068  & 0.088  & 0.228 \\
\hline

  SAR    & MSE of $\widehat{\bm{\mu}}_2$                   & 0.033  & 0.053  & 0.057  & 0.060  & 0.042 & 0.028 \\
Based $\bm{Y}_2$   & $\Vert \widehat{\bm{W}} - \bm{W} \Vert_F$  &        &        &        &    & 8.016     & 4.330 \\
  & MS accuracy         & 0.630  & 0.120  & 0.096  & 0.154 \\
      & MA weights          & 0.581  & 0.101  & 0.099  & 0.219 \\
      \hline
\multicolumn{8}{l}{$\bm{W}_k$: methods based on $\bm{W}_k$; MS: model selection; MA: Model averaging.}
\end{tabular}}}
\end{center}
\end{table}

{\color{black}
\subsection{Simulation with weights matrices from geographic distance}

In this subsection, we conduct a simulation that the candidate weights matrices are derived from geographic distance. The simulation settings are the same as subsection 5.1 but the sample size and the candidate weights matrices. We consider the $n=473$ counties in five provinces in China (Zhejiang, Jiangsu, Fujian, Anhui, Jiangxi) and $K = 6$  candidate spatial weights matrices. $\bm{W}_1$ is the connection matrix, representing the adjacency relationship among the samples. Since distance matrix $[dis_{ij}]_{i,j=1}^{473}$ is available, we can construct one- and two-window distance bands or exponential specifications. $\bm{W}_2$ is the two-window distance band, where $w_{ij}^{(2)} = 1$ if $dis_{ij} \leq 50$, $w_{ij}^{(2)} = 0.5$ if $50 < dis_{ij} \leq 100$ and $w_{ij}^{(2)} = 0$ if $dis_{ij} > 100$. The one-window distance bands are (1) $\bm{W}_3$, where the counties are neighbors if the $dis_{ij} \leq 100$; and (2) $\bm{W}_4$, where counties are neighbors if the $dis_{ij} \leq 50$. For the exponential specification, we specify $w_{ij} = \exp \{\theta_d * Dis_{ij}\}$ with $\theta_d = -0.1$ for $\bm{W}_5$ and $\theta_d = -0.8$ for $\bm{W}_6$, where $Dis_{ij} = \frac{dis_{ij}}{80}$.

 Tables \ref{GeoTab1}, \ref{GeoTab2}, \ref{GeoTab3} present the simulation results when the true weights matrix is $\bm{W}_1, \bm{W}_4, (\bm{W}_1+\bm{W}_4) / 2$, respectively. The simulation conclusions are quite similar to subsection \ref{exDesign}.}

\renewcommand{\arraystretch}{1.1}
\begin{table}[h]
\scriptsize
\begin{center}
\caption{$\bm{W}_1$ is true with weights matrices from geographic distance \label{GeoTab1}}\smallskip
\resizebox{\linewidth}{!}{
{\color{black}
\begin{tabular}{cccccccccc}
\multicolumn{10}{c}{{\normalsize Case 1: $\bm{\Sigma}_e = (0.5,0.3;0.3,0.8)$, $\bm{D} = (0.3,-0.3;0.5,0.4)$}}\\ \hline
\multicolumn{2}{c}{}          & $\bm{W}_1$ & $\bm{W}_2$ & $\bm{W}_3$ & $\bm{W}_4$ & $\bm{W}_5$ & $\bm{W}_6$ & MS & MA  \\ \hline
 MSAR & $\Vert \widehat{\bm{D}} - \bm{D} \Vert_F$ & 0.127  & 0.292  &  0.344  & 0.446  &  4.238  &  1.528 & 0.569 \\
Based $\bm{Y} = (\bm{Y}_1,\bm{Y}_2)$    & $\Vert \widehat{\bm{B}} - \bm{B} \Vert_F$          & 0.088  & 0.122  & 0.129  & 0.117 &  0.381 & 0.123  &  0.125  \\
     & MSE of $(\widehat{\bm{\mu}}_1,\widehat{\bm{\mu}}_2)$    & (0.009, 0.008)  & (0.162, 0.061)  & (0.160, 0.057)  & (0.147, 0.054) & (0.232, 0.093) & (0.938, 0.156)  & (0.041, 0.016)     & (0.033, 0.014) \\
   & $\Vert \widehat{\bm{W}} - \bm{W} \Vert_F$   &        &        &    & &    &    & 0.874     & 1.761 \\
   & MS accuracy         & 0.920  & 0.000  & 0.000  & 0.000 & 0.070 & 0.010 \\
      & MA weights         & 0.834  & 0.016  & 0.021  & 0.007 & 0.096 & 0.026 \\
 \hline

 SAR     & MSE of $\widehat{\bm{\mu}}_1$                   & 0.232  & 0.266  & 0.195  & 0.179 & 0.264  & 2.664  & 0.179  & 0.146 \\
 Based $\bm{Y}_1$   & $\Vert \widehat{\bm{W}} - \bm{W} \Vert_F$ &        &        &    & &     &    & 8.302     & 6.205 \\
   & MS accuracy         & 0.240  & 0.020  & 0.160  & 0.410 & 0.170 & 0.000 \\
      & MA weights            & 0.388  & 0.003  & 0.120  &  0.189  & 0.276 & 0.023 \\
 \hline

 SAR     & MSE of $\widehat{\bm{\mu}}_2$                  & 0.055  & 0.063  & 0.063  & 0.061 & 0.075  & 0.994  & 0.057   & 0.052 \\
Based $\bm{Y}_2$     & $\Vert \widehat{\bm{W}} - \bm{W} \Vert_F$ &        &      &    & &    &   & 3.504    & 4.363 \\
    & MS accuracy         & 0.670  & 0.060  & 0.040  & 0.090 & 0.090 & 0.050  \\
      & MA weights           & 0.584  & 0.025  & 0.049  & 0.096 & 0.182 & 0.063 \\
 \hline
 \\
 \multicolumn{10}{c}{{\normalsize Case 2: $\bm{\Sigma}_e = (0.5,0.1;0.1,0.8)$, $\bm{D} = (0.3,-0.1;0.1,0.4)$}}\\ \hline
\multicolumn{2}{c}{}          & $\bm{W}_1$ & $\bm{W}_2$ & $\bm{W}_3$ & $\bm{W}_4$ & $\bm{W}_5$ & $\bm{W}_6$ &  MS & MA \\ \hline
 MSAR & $\Vert \widehat{\bm{D}} - \bm{D} \Vert_F$   & 0.105  & 0.224  & 0.244 & 0.266 & 3.820  & 1.460  & 0.417 \\
Based $\bm{Y} = (\bm{Y}_1,\bm{Y}_2)$   & $\Vert \widehat{\bm{B}} - \bm{B} \Vert_F$            & 0.083  & 0.096  & 0.099 & 0.095 &  0.179 & 0.101  & 0.095     \\
   &  MSE of $(\widehat{\bm{\mu}}_1,\widehat{\bm{\mu}}_2)$     & (0.005, 0.010)  & (0.046, 0.063)  & (0.049, 0.067)  & (0.045, 0.060) & (0.101, 0.144) & (0.498, 0.592) &  (0.008, 0.014)  & (0.010, 0.017) \\
  & $\Vert \widehat{\bm{W}} - \bm{W} \Vert_F$&        &        &   & &     &    & 0.440      & 1.958 \\
  & MS accuracy         & 0.960  & 0.000  & 0.000  & 0.000 & 0.040  & 0.000 \\
      & MA weights          & 0.813  & 0.025  & 0.022  & 0.028 & 0.084 & 0.028 \\
 \hline

  SAR    & MSE of $\widehat{\bm{\mu}}_1$                    & 0.011  & 0.045  & 0.048  & 0.046 & 0.557 & 4.089 & 0.012  & 0.012 \\
Based $\bm{Y}_1$   & $\Vert \widehat{\bm{W}} - \bm{W} \Vert_F$ &        &    & &    &        &     & 0.419     & 1.693 \\
  & MS accuracy         & 0.960  & 0.010  & 0.010  & 0.010 & 0.010 & 0.000 \\
      & MA weights      & 0.837  & 0.038  & 0.020  & 0.032 & 0.058 & 0.015 \\
\hline

  SAR    & MSE of $\widehat{\bm{\mu}}_2$                   & 0.012  & 0.059  & 0.063  & 0.060 & 1.224 & 9.136 & 0.014 & 0.015 \\
Based $\bm{Y}_2$   & $\Vert \widehat{\bm{W}} - \bm{W} \Vert_F$  &        &   & &     &        &    & 0.437     & 1.683 \\
  & MS accuracy         & 0.960  & 0.000  & 0.010  & 0.020 & 0.010 & 0.000 \\
      & MA weights       & 0.840  & 0.026  & 0.015  & 0.037 & 0.064 & 0.019 \\
      \hline
\multicolumn{10}{l}{$\bm{W}_k$: methods based on $\bm{W}_k$; MS: model selection; MA: Model averaging.}
\end{tabular}}}
\end{center}
\end{table}

\renewcommand{\arraystretch}{1.1}
\begin{table}[h]
\scriptsize
\begin{center}
\caption{$\bm{W}_4$ is true  with weights matrices from geographic distance \label{GeoTab2}}\smallskip
\resizebox{\linewidth}{!}{
{\color{black}
\begin{tabular}{cccccccccc}
\multicolumn{10}{c}{{\normalsize Case 1: $\bm{\Sigma}_e = (0.5,0.3;0.3,0.8)$, $\bm{D} = (0.3,-0.3;0.5,0.4)$}}\\ \hline
\multicolumn{2}{c}{}          & $\bm{W}_1$ & $\bm{W}_2$ & $\bm{W}_3$ & $\bm{W}_4$ & $\bm{W}_5$ & $\bm{W}_6$ & MS & MA  \\ \hline
 MSAR & $\Vert \widehat{\bm{D}} - \bm{D} \Vert_F$ & 0.446  & 0.453  &  0.362  & 0.134 & 4.157  &  1.669  &  0.349 \\
Based $\bm{Y} = (\bm{Y}_1,\bm{Y}_2)$    & $\Vert \widehat{\bm{B}} - \bm{B} \Vert_F$          & 0.108  & 0.098  & 0.120  & 0.085 &  0.302 & 0.115  &  0.097  \\
     & MSE of $(\widehat{\bm{\mu}}_1,\widehat{\bm{\mu}}_2)$    & (0.155, 0.053)  & (0.159, 0.044)  & (0.156, 0.054)  & (0.009, 0.007) & (0.265, 0.094) & (1.044, 0.157)  & (0.026, 0.011)     & (0.023, 0.010) \\
   & $\Vert \widehat{\bm{W}} - \bm{W} \Vert_F$   &        &        &    & &    &    & 0.756     & 2.040 \\
   & MS accuracy         & 0.000  & 0.010  & 0.000  & 0.930 & 0.060 & 0.000 \\
      & MA weights         & 0.010  & 0.048  & 0.030  & 0.800 & 0.096 & 0.016 \\
 \hline

 SAR     & MSE of $\widehat{\bm{\mu}}_1$                   & 0.188  & 1.780  & 0.188  & 0.231 & 0.275  & 2.686  & 0.185   & 0.155 \\
 Based $\bm{Y}_1$   & $\Vert \widehat{\bm{W}} - \bm{W} \Vert_F$ &        &        &    & &     &    & 7.801     & 6.203 \\
   & MS accuracy         & 0.260  & 0.010  & 0.320  & 0.270 & 0.140 & 0.000 \\
      & MA weights            & 0.158  & 0.021  & 0.161  &  0.385  & 0.237 & 0.037 \\
 \hline

 SAR     & MSE of $\widehat{\bm{\mu}}_2$                  & 0.062  & 0.066  & 0.064 & 0.057 & 0.070  & 1.127  & 0.058   & 0.054 \\
Based $\bm{Y}_2$     & $\Vert \widehat{\bm{W}} - \bm{W} \Vert_F$ &        &      &    & &    &   & 4.753    & 5.012 \\
    & MS accuracy         & 0.080  & 0.020  & 0.080  & 0.560 & 0.220 & 0.040  \\
      & MA weights           & 0.092  & 0.022  & 0.056  & 0.530 & 0.254 & 0.046 \\
 \hline
 \\
 \multicolumn{10}{c}{{\normalsize Case 2: $\bm{\Sigma}_e = (0.5,0.1;0.1,0.8)$, $\bm{D} = (0.3,-0.1;0.1,0.4)$}}\\ \hline
\multicolumn{2}{c}{}          & $\bm{W}_1$ & $\bm{W}_2$ & $\bm{W}_3$ & $\bm{W}_4$ & $\bm{W}_5$ & $\bm{W}_6$ &  MS & MA \\ \hline
 MSAR & $\Vert \widehat{\bm{D}} - \bm{D} \Vert_F$   & 0.242  & 0.350  & 0.224 & 0.107 & 3.472  & 1.765  & 0.443 \\
Based $\bm{Y} = (\bm{Y}_1,\bm{Y}_2)$   & $\Vert \widehat{\bm{B}} - \bm{B} \Vert_F$            & 0.098  & 0.094  & 0.097 & 0.089 &  0.165 & 0.100  & 0.096     \\
   &  MSE of $(\widehat{\bm{\mu}}_1,\widehat{\bm{\mu}}_2)$     & (0.048, 0.063)  & (0.049, 0.076)  & (0.049, 0.065)  & (0.006, 0.011) & (0.115, 0.163) & (0.530, 0.639) &  (0.022, 0.026)  & (0.018, 0.022) \\
  & $\Vert \widehat{\bm{W}} - \bm{W} \Vert_F$&        &        &   & &     &    & 0.551      & 2.284 \\
  & MS accuracy         & 0.000  & 0.000  & 0.000  & 0.950 & 0.040  & 0.010 \\
      & MA weights          & 0.024  & 0.060  & 0.033  & 0.772 & 0.083 & 0.028 \\
 \hline

  SAR    & MSE of $\widehat{\bm{\mu}}_1$                    & 0.049  & 0.044  & 0.049  & 0.011 & 0.118 & 9.478 & 0.013  & 0.014 \\
Based $\bm{Y}_1$   & $\Vert \widehat{\bm{W}} - \bm{W} \Vert_F$ &        &    & &    &        &     & 0.496      & 1.608 \\
  & MS accuracy         & 0.010  & 0.020  & 0.010  & 0.950 & 0.010 & 0.000 \\
      & MA weights      & 0.033  & 0.054  & 0.010  & 0.839 & 0.044 & 0.020 \\
\hline

  SAR    & MSE of $\widehat{\bm{\mu}}_2$                   & 0.063  & 0.085  & 0.062  & 0.015 & 0.170 & 8.393 & 0.018 & 0.019 \\
Based $\bm{Y}_2$   & $\Vert \widehat{\bm{W}} - \bm{W} \Vert_F$  &        &   & &     &        &    & 0.938     & 2.415 \\
  & MS accuracy         & 0.030  & 0.010  & 0.040  & 0.910 & 0.010 & 0.000 \\
      & MA weights       & 0.044  & 0.069  & 0.034  & 0.753 & 0.075 & 0.025 \\
      \hline
\multicolumn{10}{l}{$\bm{W}_k$: methods based on $\bm{W}_k$; MS: model selection; MA: Model averaging.}
\end{tabular}}}
\end{center}
\end{table}

\renewcommand{\arraystretch}{1.1}
\begin{table}[h]
\scriptsize
\begin{center}
\caption{$(\bm{W}_1 + \bm{W}_4) / 2$  is true with weights matrices from geographic distance \label{GeoTab3}}\smallskip
\resizebox{\linewidth}{!}{
{\color{black}
\begin{tabular}{cccccccccc}
\multicolumn{10}{c}{{\normalsize Case 1: $\bm{\Sigma}_e = (0.5,0.3;0.3,0.8)$, $\bm{D} = (0.3,-0.3;0.5,0.4)$}}\\ \hline
\multicolumn{2}{c}{}          & $\bm{W}_1$ & $\bm{W}_2$ & $\bm{W}_3$ & $\bm{W}_4$ & $\bm{W}_5$ & $\bm{W}_6$ & MS & MA  \\ \hline
 MSAR & $\Vert \widehat{\bm{D}} - \bm{D} \Vert_F$ & 0.232  & 0.325  &  0.330  & 0.233 & 4.189  &  1.547  &  0.672 \\
Based $\bm{Y} = (\bm{Y}_1,\bm{Y}_2)$    & $\Vert \widehat{\bm{B}} - \bm{B} \Vert_F$          & 0.090  & 0.095  & 0.106  & 0.091 &  0.357 & 0.102  &  0.130  \\
     & MSE of $(\widehat{\bm{\mu}}_1,\widehat{\bm{\mu}}_2)$    & (0.045, 0.020)  & (0.092, 0.033)  & (0.110, 0.041)  & (0.043, 0.020) & (0.196, 0.064) & (0.929, 0.133)  & (0.069, 0.026)     & (0.035, 0.017) \\
   & $\Vert \widehat{\bm{W}} - \bm{W} \Vert_F$   &        &        &    & &    &    & 5.949     & 1.969 \\
   & MS accuracy         & 0.450  & 0.010  & 0.000  & 0.460 & 0.060 & 0.020 \\
      & MA weights         & 0.426  & 0.047  & 0.000  & 0.434 & 0.067 & 0.026 \\
 \hline

 SAR     & MSE of $\widehat{\bm{\mu}}_1$                   & 0.135  & 1.068  & 0.146  & 0.136 & 0.416  & 2.617  & 0.131   & 0.109 \\
 Based $\bm{Y}_1$   & $\Vert \widehat{\bm{W}} - \bm{W} \Vert_F$ &        &        &    & &     &    & 6.383     & 3.075 \\
   & MS accuracy         & 0.430  & 0.000  & 0.130  & 0.410 & 0.100 & 0.000 \\
      & MA weights            & 0.375  & 0.023  & 0.082  &  0.347  & 0.150 & 0.022 \\
 \hline

 SAR     & MSE of $\widehat{\bm{\mu}}_2$                  & 0.042 & 0.049  & 0.049 & 0.043 & 0.098  & 0.764  & 0.044   & 0.041 \\
Based $\bm{Y}_2$     & $\Vert \widehat{\bm{W}} - \bm{W} \Vert_F$ &        &      &    & &    &   & 6.178    & 4.118 \\
    & MS accuracy         & 0.450  & 0.020  & 0.010  & 0.390 & 0.100 & 0.030  \\
      & MA weights           & 0.404  & 0.023  & 0.009  & 0.368 & 0.144 & 0.053 \\
 \hline
 \\
 \multicolumn{10}{c}{{\normalsize Case 2: $\bm{\Sigma}_e = (0.5,0.1;0.1,0.8)$, $\bm{D} = (0.3,-0.1;0.1,0.4)$}}\\ \hline
\multicolumn{2}{c}{}          & $\bm{W}_1$ & $\bm{W}_2$ & $\bm{W}_3$ & $\bm{W}_4$ & $\bm{W}_5$ & $\bm{W}_6$ &  MS & MA \\ \hline
 MSAR & $\Vert \widehat{\bm{D}} - \bm{D} \Vert_F$   &  0.152  & 0.269 & 0.214 & 0.161  & 3.894  & 1.675 & 0.347 \\
Based $\bm{Y} = (\bm{Y}_1,\bm{Y}_2)$   & $\Vert \widehat{\bm{B}} - \bm{B} \Vert_F$            & 0.085  & 0.087  & 0.090 & 0.082 &  0.180 & 0.090  & 0.099     \\
   &  MSE of $(\widehat{\bm{\mu}}_1,\widehat{\bm{\mu}}_2)$     & (0.015, 0.022)  & (0.032, 0.044)  & (0.033, 0.044)  & (0.014, 0.021) & (0.099, 0.113) & (0.471, 0.771) &  (0.018, 0.025)  & (0.011, 0.016) \\
  & $\Vert \widehat{\bm{W}} - \bm{W} \Vert_F$&        &        &   & &     &    & 5.852      & 2.216 \\
  & MS accuracy         & 0.520  & 0.040  & 0.000  & 0.400 & 0.040  & 0.000 \\
      & MA weights          & 0.431  & 0.075  & 0.009  & 0.409 & 0.052 & 0.024 \\
 \hline

  SAR    & MSE of $\widehat{\bm{\mu}}_1$                    & 0.017  & 0.028  & 0.033  & 0.017 & 0.977 & 5.730 & 0.018  & 0.023 \\
Based $\bm{Y}_1$   & $\Vert \widehat{\bm{W}} - \bm{W} \Vert_F$ &        &    & &    &        &     & 5.797      & 2.373 \\
  & MS accuracy         & 0.490  & 0.090  & 0.010  & 0.410 & 0.000 & 0.000 \\
      & MA weights      & 0.488  & 0.067  & 0.003  & 0.413 & 0.018 & 0.011 \\
\hline

  SAR    & MSE of $\widehat{\bm{\mu}}_2$                   & 0.023  & 0.040  & 0.043  & 0.022 & 0.081 & 11.219 & 0.024 & 0.023 \\
Based $\bm{Y}_2$   & $\Vert \widehat{\bm{W}} - \bm{W} \Vert_F$  &        &   & &     &        &    & 5.806     & 2.820 \\
  & MS accuracy         & 0.420  & 0.070  & 0.010  & 0.490 & 0.010 & 0.000 \\
      & MA weights       & 0.425  & 0.101  & 0.000  & 0.410 & 0.046 & 0.017 \\
      \hline
\multicolumn{10}{l}{$\bm{W}_k$: methods based on $\bm{W}_k$; MS: model selection; MA: Model averaging.}
\end{tabular}}}
\end{center}
\end{table}

\subsection{Compare with high-order MSAR model}\label{sec5.4}
 When there are multiple spatial weights matrices available, another option is the high-order SAR model \citep{2010EFFICIENT, 2015Inference}.  The SAR model with high order spatial lags can characterize spatial interdependence based on different types of relationships such as social relation and geographic distance among cross-sectional units. A high-order SAR model can be expressed as $\bm{y} = \sum_{k=1}^K \rho_k \bm{W}_k \bm{y} + {\bm{X}} \bm{\beta} + \bm{\epsilon}$, where $\rho_1, \cdots, \rho_K$ are unknown scalars that need to estimated, $\bm{\beta}$ is the unknown regression vector and $\bm{\epsilon}$ is a disturbance vector. This model can also be extended to multivariate response as the high-order MSAR model.

 In this subsection we compare our proposed MS and MA methods to the high-order MSAR model. We will focus on the following two questions. (i) When the true model is MSAR, including the situations that the true spatial weights matrix is in the candidate set or not, will our methods perform better? (ii) When the true model is the high-order MSAR, which method will perform better?  The simulation round in this subsection is 100 since high-order MSAR is very time-consuming.

We first generate data following the MSAR model settings in the subsection \ref{exDesign} but only consider case 1. Table \ref{Tab4} reports the prediction errors of $\bm{\mu}$ based on MS, MA and high-order MSAR.  Since the estimation results of the high-order MSAR are not stable, we report the median of the losses over the simulation rounds instead of mean.  When the true model is MSAR, our proposed MS and MA methods perform much better than the high-order MSAR model, no matter the true spatial weights matrix is included in the candidate set or not.

\vspace{0.5cm}
\begin{table}[h]
\begin{center}
\caption{True model is MSAR. \label{Tab4}}\smallskip
{\small\begin{tabular}{cccc}
\multicolumn{1}{c}{Methods} & True $\bm{W}$ is $\bm{W}_1$ & True $\bm{W}$ is $\bm{W}_4$ & True $\bm{W}$ is $(\bm{W}_1 + \bm{W}_4)/2$         \\ \hline
MS loss & 0.074   & 0.050  & 0.087 \\
MA loss & 0.062   & 0.036  & 0.044 \\
high-order loss (median) & 0.718   & 2.865  & 1.204\\
\hline
\multicolumn{4}{l}{The candidate set is $[\bm{W}_1, \bm{W}_2, \bm{W}_3, \bm{W}_4]$.}
\end{tabular}}
\end{center}
\end{table}
\vspace{-0.5cm}

We then generate the response data from a high-order MSAR model $\bm{Y} = \sum_k \bm{W}_k \bm{Y} \bm{D}_k + \bm{X} \bm{B} + \bm{E}$, where $\bm{X}, \bm{B}, \bm{E}$ are the same as case 1 in the subsection \ref{exDesign}, $[\bm{W}_1, \bm{W}_4]$ in the subsection \ref{exDesign} are used for data generation with the corresponding $\bm{D}_1 = (0.3, 0.0; -0.2, 0.1)$ and $\bm{D}_4 = (0.4, 0.0; -0.1, 0.2)$. For the estimation of MS, MA and high-order MSAR, we consider four situations of candidate set: (i) $[\bm{W}_1, \bm{W}_4]$; (ii) $[\bm{W}_1, \bm{W}_2, \bm{W}_4]$; (iii) $[\bm{W}_1, \bm{W}_2, \bm{W}_3]$; (iv) $[\bm{W}_2, \bm{W}_3]$. For MS and MA, we use $\bm{W}_4$ to estimate $\bm{\Omega}$.  Table \ref{Tab5} reports the prediction errors of $\bm{\mu}$ based on MS, MA and high-order MSAR. The high-order MSAR model has better performance only when the candidate set used in model fitting happens to be the same as the one used to generate the data.

\vspace{0.5cm}
\begin{table}[h]
\begin{center}
\caption{True model is high-order MSAR. \label{Tab5}}\smallskip
{\small\begin{tabular}{ccccc}
\multicolumn{1}{c}{Candidate set} & $[\bm{W}_1, \bm{W}_4]$ & $[\bm{W}_1, \bm{W}_2, \bm{W}_4]$ & $[\bm{W}_1, \bm{W}_2, \bm{W}_3]$   & $[\bm{W}_2, \bm{W}_3]$         \\ \hline
MS loss & 0.080   & 0.079  & 0.076 & 0.088 \\
MA loss & 0.042   & 0.046  & 0.052 & 0.078 \\
high-order loss (median) & 0.038   & 0.298  & 0.205 & 0.099
\\ \hline
\multicolumn{4}{l}{The  true model in data generation is $[\bm{W}_1, \bm{W}_4]$.}
\end{tabular}}
\end{center}
\end{table}

\vspace{-1cm}

\section{Sina Weibo Data Analysis} \label{sec6}
In this section, we apply the proposed model selection and model averaging methods to a Twitter-type social network dataset, which is collected from Sina Weibo to investigate the user's online posting behaviors related to finance and economics  \citep{zhu2020multivariate}.

The response variables are the log-transformed numbers of characters in the posts related to FINANCE and ECONOMICS topics for each user within 75 days. The posts are tagged as FINANCE and ECONOMICS if corresponding keywords are involved and the keywords are obtained from an online public Chinese dictionary, where the finance dictionary contains mostly keywords of stock markets and products and the economic dictionary contains keywords related to the macroeconomic trend and policies. There are four exogenous nodal covariates: GENDER (which is equal to 1 if the user is male), BEIJING (which is equal to 1 if the user is located in Beijing), SHANGHAI (which is equal to 1 if the user is located in Shanghai), TENURE (which is the time length since the user's registration with Sina Weibo). By the analysis in \cite{zhu2020multivariate}, the spatial effect is significant.

While \cite{zhu2020multivariate} simply uses the adjacent matrix as the spatial weights matrix to fit the MSAR model, our major interest here is to investigate what kind of spatial weights matrix is more plausible, i.e., how the Sina Weibo users influence each other in terms of posting behaviors.  We consider four candidate models (before row normalization) as follows.

\begin{itemize}
\item[(1)]  $\bm{W}_1 = (w_{ij}^{(1)})$: If the $i$th user follows the $j$th user ($i \neq j$), then $w_{ij}^{(1)} = 1$. Otherwise  $w_{ij}^{(1)} = 0$.
\item[(2)]  $\bm{W}_2 = (w_{ij}^{(2)})$: If the $i$th user follows the $j$th user ($i \neq j$), then $w_{ij}^{(2)} = \text{follower}_j$, which is the number of users that follow user $j$. Otherwise  $w_{ij}^{(2)} = 0$.
\item[(3)] $\bm{W}_3 = (w_{ij}^{(3)})$: If the $i$th user follows the $j$th user ($i \neq j$), then $w_{ij}^{(3)}\!\! = \!\exp \left \{ 0.05 \!\times \!\text{follower}_j \right \}$. Otherwise  $w_{ij}^{(3)} = 0$.
\item[(4)] $\bm{W}_4 = (w_{ij}^{(4)})$: If the $i$th user follows the $j$th user ($i \neq j$), then $w_{ij}^{(4)}\!\! = \!\exp \left \{ 0.2 \!\times \!\text{follower}_j \right \}$. Otherwise  $w_{ij}^{(4)} = 0$.
\end{itemize}

\noindent Basically, we believe that one user's posting behavior will be influenced by the posting behaviors of the users that he follows, i.e., his followee. But we do not know whether this influence is uniformly distributed among the followee ($\bm{W}_1$), or it is positively correlated with the number of followers of the followee, either linearly ($\bm{W}_2$) or exponentially ($\bm{W}_3$ and $\bm{W}_4$ with different correlation strength).

\begin{table}[t]
\centering
\scriptsize
\caption{Comparison of SAR and MSAR model for Sina Weibo data. \label{sar_vs_msar_in_real_data}}
{\color{black}
    \begin{tabular}{llcccc}
    \hline
        & & \multicolumn{2}{c}{SAR} & \multicolumn{2}{c}{MSAR} \\
        & & FINANCE & ECONOMICS & (FINANCE & ECONOMICS) \\
        \hline
        MS & Training MSE & 1.142 (0.060) & 1.191 (0.078) & 0.988 (0.000) & 0.991 (0.000) \\
        & Test MSE & 1.146 (0.060) & 1.194 (0.101) & 0.986 (0.000) & 0.992 (0.000) \\
        \hline
        MA & Training MSE & 1.119 (0.048) & 1.148 (0.077) & 0.988 (0.000) & 0.991 (0.000) \\
        & Test MSE & 1.122 (0.047) & 1.141 (0.075) & 0.986 (0.000) & 0.991 (0.000) \\
        \hline
        \multicolumn{6}{l}{MSE is the average over 10 splitting replications (variance in bracket).}
    \end{tabular}}
\end{table}

\begin{table}[t]
\begin{center}
\caption{Comparison of SAR and MSAR on model selection and model averaging results for Sina Weibo data. \label{ms_in_real_data}}
{\color{black}
{\footnotesize\begin{tabular}{cccccc}
\hline
Model \& Method & Response variable $\bm{Y}$ & $\bm{W}_1$ & $\bm{W}_2$ & $\bm{W}_3$ & $\bm{W}_4$         \\ \hline
SAR \& MS  & FINANCE & 0.1 & 0.4 & 0.2 & 0.3
\\
 SAR \& MS   & ECONOMICS & 0.3   & 0.1 & 0.4 & 0.2
\\
MSAR \& MS  & (FINANCE, ECONOMICS) & 0.3   & 0.5 & 0.0 & 0.2
\\ \hline
SAR \& MA & FINANCE & 0.080 & 0.394 & 0.261 & 0.265
\\
SAR \& MA   & ECONOMICS & 0.258   & 0.090 & 0.393 & 0.260
\\
MSAR \& MA & (FINANCE, ECONOMICS) & 0.301   & 0.496 & 0.003 & 0.199\\
\hline
\multicolumn{4}{l}{All values are the average over 10 splitting replications.}
\end{tabular}}}
\end{center}
\end{table}

To eliminate the ``island" effects in the dataset, the users with the lowest in-degrees are deleted and $1211$ users are involved in the analysis. Since the true $\bm{\mu}$ is not available for the real data analysis, we randomly split the users into a training set and a test set with equal sample sizes. The proposed MS and MA methods of MSAR with response (FINANCE, ECONOMICS) are used to the training data. {\color{black} We use $\bm{W}_4$ to estimate $\bm{\Omega}$.} Also, the MS and MA methods of univariate SAR \citep{zhang2018spatial} are used with response FINANCE and ECONOMICS, separately. {\color{black} High-order MSAR including $\bm{W}_1$ to $\bm{W}_4$ as the candidate set is also considered.} Then we apply the fitted models (estimated parameters, selected spatial weights matrix for MS, weight of each candidate model for MA) to the test data to obtain $\widehat{\bm{\mu}}$. Training MSE and test MSE can be calculated. We repeat this splitting method 10 times and report the averaged training MSE and test MSE in Table \ref{sar_vs_msar_in_real_data}.  Compared to the SAR model on single FINANCE or ECONOMICS, MSAR model is more stable and has lower training and test MSE on both FINANCE and ECONOMICS. Compared to MS method, MA of MSAR has better predictive performance and can reduce estimation error. {\color{black} The results of the high-order MSAR are not reported in Table \ref{sar_vs_msar_in_real_data} since it performs very poor with huge mean and median of training and test MSE, which indicates that high-order model may not be a good choice for multivariate SAR.}

Further we report the model selection results and the model averaging weights in Table \ref{ms_in_real_data}. Among the 10 replications, MS of univariate SAR with FINANCE as the response selects $\bm{W}_1$,  $\bm{W}_2$, $\bm{W}_3$ and $\bm{W}_4$ by 1, 4, 2 and 3 times, respectively. Meanwhile, MS of univariate SAR with ECONOMICS as the response selects $\bm{W}_1$,  $\bm{W}_2$, $\bm{W}_3$ and $\bm{W}_4$ by 3, 1, 4 and 2 times, respectively. The model selection results of the two responses are different, which indicates that the influencing mechanisms may be different for different types of post. The MS and MA results of MSAR are quite close: the weights on $\bm{W}_1,  \bm{W}_2, \bm{W}_3$ and $\bm{W}_4$ are roughly 0.3, 0.5, 0.0 and 0.2, respectively. The weighted spatial weights matrix is $0.3\bm{W}_1+0.5\bm{W}_2+0.2\bm{W}_4$. It indicates that the influence tends to be uniformly distributed among the user's followee, or linearly correlated with the number of followers of the followee.

\section{Concluding Remarks}\label{sec7}

This paper focuses on choosing an optimal spatial weights matrix from several candidates for the MSAR model. A Mallows type model selection method is proposed. It can consistently select the true model when the true weights matrix is in the candidates; otherwise, it is asymptotically optimal. We also propose a model averaging method to reduce estimation error. The optimal weight is obtained by minimizing an unbiased estimator of the squared loss of model averaging estimator. The proposed model averaging estimator is asymptotically optimal and more robust to model misspecificaitions. Monte Carlo simulations show that the proposed MS and MA estimators perform quite well in finite samples.



{\color{black} In this paper we assume that the spatial weights matrix $\bm{W}_k$ to be row-normalized. This assumption might be restrictive in some empirical applications, although it is a common practice in literature  \citep{lee2004asymptotic,lee2009spatial,yang2017identification,zhu2020multivariate}. It is a sufficient assumption for guaranteeing estimation consistency in combination with other sufficient conditions, as assumed in \cite{zhu2020multivariate}. However, this specification can be relaxed without the restriction of the row-normalization for ensuring  corresponding theoretical properties; see \cite{dou2016generalized,liu2014endogenous,lewbel2023social} for detailed assumptions and discussions therein. It is an interesting future research topic to consider MS and MA with non-row-normalized weights matrices and study corresponding theoretical properties. In practice, we find the row-normalized weights matrices can yield more numerically stable estimation results, thus we keep the row-normalized $\bm{W}_k$ in the empirical studies.}

{\color{black} Further, we would like to discuss several issues about the candidate weights matrix and candidate model. First, the MSAR model is considered as the basis model in our candidate model set and the high-order MSAR is considered as a competitor. It seems possible to consider the high-order MSAR model itself as the basis model. Then $K$ candidate spatial weights matrices will lead to $2^K$ possible high-order MSAR candidate models. However, we believe this approach is infeasible since high-order SAR model for the multivariate response performs unstable, as we have pointed out in Section \ref{sec5.4} and Section \ref{sec6}. Second, an interesting problem is that what if different response variables exhibit distinct spatial dependencies in the MSAR model. We may modify (\ref{MSARmodel}) to $\bm{Y}=(\bm{W}_1\bm{Y}\bm{D}_1,\cdots,\bm{W}_q\bm{Y}\bm{D}_q)+\bm{X}\bm{B}+\bm{E},$ where $\bm{D}_j$ is the $j$-th column of $\bm{D}$, and $\bm{W}_j$ is the spatial weights matrix for the $j$-th response variable, $j=1,\cdots,q$. But further explorations are needed. Third, our approach is to conduct model averaging with a set of given candidate
weights matrices $\mathcal{W} = \big \{ \bm{W}_1, \cdots, \bm{W}_K \big \}$, which can reduce the mis-specification risk of a single weights matrix. It is possible to estimate $\bm{W}$ assuming that certain elements are zero. A recent work \citep{de2024identifying} provides a comprehensive discussion about this problem when a panel data is available.
 Therefore, if another dataset of the same spatial units is collected with repeated observations on each unit, it is possible to identify and estimate the unknown spatial weights matrix (or a set of unknown spatial weights matrices). Subsequently, our model averaging method can still be implemented based on this candidate set. }

The current work can also be extended in several other ways. First, although the number of candidates can be diverging theoretically, empirically we only consider fixed number of candidates. If there are many candidate spatial weights matrices, we may need to introduce a model screening method to remove the poorest model before averaging \citep{yuan2005combining}, which will contribute to greater estimation and predictive efficiency. {\color{black}Second, the MSAR model may have a computational difficulty when $q$ is large. To address the challenge, we may simplify $\bm{D}$ to a diagonal matrix or assume the regression parameter matrix $\bm{B}$ is sparse and consider a penalized least square estimation based on \cite{zhu2020multivariate}.} Third, so far we only consider a cross-sectional MSAR model. In practice, since many responses can be observed in time series, MS and MA methods should consider the time dynamics. Fourth, so far we only consider continuous responses. In real data analysis, the discrete responses can be frequently encountered, which needs to be further studied.

\section*{Acknowledgements}
We would like to express our appreciation to Editor Yuya Sasaki, the Associate Editor, and two anonymous referees for their valuable comments that greatly improved the article. The code and data that support the findings of this study are openly available in a online supplementary material.

\section*{Funding}

Fang Fang gratefully acknowledges research support from the National Key R\&D Program of China (2021YFA1000100, 2021YFA1000101), National Natural Science Foundation of China (72331005, 12071143), and the Basic Research Project of Shanghai Science and Technology Commission (22JC1400800, TQ20220105).

\bibliographystyle{elsarticle-harv}
\bibliography{MSAR}

\section*{Appendices}


\appendix

\section{\texorpdfstring{Derivation of $\frac{\partial \text{vec} (\widehat{\bm{D}}_k)}{\partial \bm{y}^T}$}{Derivation of partial vec(D_k) / partial y^T}}

By minimizing (\ref{loss}) with $\bm{W}$ being replaced by $\bm{W}_k$, $\bm{\theta}_k=(\bm{D}_k, \bm{\beta}_k, \bm{\Sigma_e})$ can be estimated as $\widehat{\bm{\theta}}_k=(\widehat{\bm{D}}_k, \widehat{\bm{\beta}}_k, \widehat{\bm{\Sigma}}_k)$ in the $k$-th model. From the first-order condition of (\ref{loss}), we have \begin{equation}\label{temmppp}
 \frac{\partial Q(\bm{\theta}_k)}{\partial \text{vec} (\bm{D}_k)} \Big|_{\bm{\theta}_k = \widehat{\bm{\theta}}_k} = 0.
 \end{equation}
 It's very difficult to derive the exact form of $\frac{\partial \text{vec} (\widehat{\bm{D}}_k)}{\partial \bm{y}^T}$ since $\widehat{\bm{\beta}}_k$ and $\widehat{\bm{\Sigma}}_k$ both are related to $\widehat{\bm{D}}_k$. We propose to ignore the randomness of $\widehat{\bm{\Sigma}}_k$ and derive an approximated form of $\frac{\partial \text{vec} (\widehat{\bm{D}}_k)}{\partial \bm{y}^T}$. In such a case, $\widehat{\bm{\beta}}_k$ is a function of $\widehat{\bm{D}}_k$ and $\bm{y}$ by (\ref{temmppp}). Express $\frac{\partial Q(\bm{\theta}_k)}{\partial \text{vec} (\bm{D}_k)} \Big|_{\bm{\theta_k} = \widehat{\bm{\theta}}_k}$ as a function of $\widehat{\bm{D}}_k$ and $\bm{y}$, that is $\widetilde{Q} (\widehat{\bm{d}}_k, \bm{y}) = \frac{\partial Q(\bm{\theta}_k)}{\partial \text{vec} (\bm{D}_k)} \Big|_{\bm{\theta}_k = \widehat{\bm{\theta}}_k}=0$, where $\widehat{\bm{d}}_k = \text{vec} (\widehat{\bm{D}}_k)$.  Taking the derivative with respect to $\bm{y}^T$, we have
$$
\frac{\partial \widetilde{Q} (\widehat{\bm{d}}_k, \bm{y})}{\partial \bm{y}^T} + \frac{\partial \widetilde{Q} (\widehat{\bm{d}}_k, \bm{y})}{\partial \widehat{\bm{d}}_k^T} \frac{\partial \widehat{\bm{d}}_k}{\partial \bm{y}^T} = 0.
$$
So
\begin{equation*}
\frac{\partial \text{vec} (\widehat{\bm{D}}_k)}{\partial \bm{y}^T} = - \left \{ \frac{\partial \widetilde{Q} (\widehat{\bm{d}}_k, \bm{y})}{\partial \widehat{\bm{d}}_k^T} \right \}^{-1} \frac{\partial \widetilde{Q} (\widehat{\bm{d}}_k, \bm{y})}{\partial \bm{y}^T}.
\end{equation*}

First we derive $\widetilde{Q} (\widehat{\bm{d}}_k, \bm{y})$. Let $\bm{M}_k\!=\! \{ \text{diag} (\bm{\widehat{\Sigma}}_k^{-1}) \otimes \bm{I}_n + \text{diag} (\widehat{\bm{D}}_k \bm{\widehat{\Sigma}}_k^{-1} \widehat{\bm{D}}_k^T ) \otimes \text{diag} (\bm{W}_k^T \bm{W}_k) \}^{-1}$, $\bm{F} = \bm{M} \bm{S}^T (\bm{\Sigma_e}^{-1} \otimes \bm{I}_n) (\bm{S} \bm{y} - \bm{\widetilde{X}} \bm{\beta}_k)$, $\bm{F}_k = \bm{M}_k \bm{S}_k^T (\bm{\widehat{\Sigma}}_k^{-1} \otimes \bm{I}_n) (\bm{S}_k \bm{y} - \bm{\widetilde{X}} \widehat{\bm{\beta}}_k)$, $\bm{V}_k = (\widehat{\bm{D}}_k \bm{\widehat{\Sigma}}_k^{-1} \widehat{\bm{D}}_k^T) \otimes (\bm{W}_k^T \bm{W}_k)$. Let $d_{ij}^k$  be the $(i, j)$-th element of ${\bm{D}}_k$. Since $\frac{\partial \bm{V}_k}{\partial d_{ij}^k} = \left( \bm{I}_{ij} \bm{\widehat{\Sigma}}_k^{-1} \widehat{\bm{D}}_k^T + \widehat{\bm{D}}_k \bm{\widehat{\Sigma}}_k^{-1} \bm{I}_{ji} \right) \otimes (\bm{W}_k^T \bm{W}_k)$ and $\frac{\partial \bm{M}_k}{\partial d_{ij}^k} = - \bm{M}_k \text{diag} \left( \frac{\partial \bm{V}_k}{\partial d_{ij}^k} \right)\bm{M}_k$, we can get
\begin{eqnarray*}
\bm{F}_{ij}^k & =& \frac{\partial \bm{F}}{\partial d_{ij}^k} \Big|_{\bm{\theta}_k = \widehat{\bm{\theta}}_k}  =  \frac{\partial \bm{M}_k}{\partial d_{ij}^k} \bm{S}_k^T (\bm{\widehat{\Sigma}}_k^{-1} \otimes \bm{I}_n)(\bm{S}_k\bm{y}- \widetilde{\bm{X}} \widehat{\bm{\beta}}_k)
\nonumber\\
&&  + \bm{M}_k \frac{\partial \bm{S}_k^T}{\partial d_{ij}^k} (\bm{\widehat{\Sigma}}_k^{-1} \otimes \bm{I}_n)(\bm{S}_k\bm{y} - \widetilde{\bm{X}} \widehat{\bm{\beta}}_k) +  \bm{M}_k \bm{S}_k^T (\bm{\widehat{\Sigma}}_k^{-1} \otimes \bm{I}_n) \frac{\partial \bm{S}_k}{\partial d_{ij}^k} \bm{y},
\end{eqnarray*}
where $\frac{\partial \bm{S}_k}{\partial d_{ij}^k} = - \bm{I}_{ji} \otimes \bm{W}_k$ has been calculated in Section 3.
Then $\widetilde{Q} (\widehat{\bm{d}}_k, \bm{y})$ can be expressed as the combination of $2 \bm{F}_k^T \bm{F}_{ij}^k$, $i=1,\cdots,q$, $j=1,\cdots,q$.

Then we calculate $\frac{\partial \widetilde{Q} (\widehat{\bm{d}}_k, \bm{y})}{\partial \bm{y}^T}$. Let $y_t$ be the $t$-th element of $\bm{y}$, $\bm{J}_t$ be a $np \times 1$ vector with all zero elements except for the $t$-th entry which is 1. By $\frac{\partial \bm{F}_k}{\partial y_t} = \bm{M}_k \bm{S}_k^T (\bm{\widehat{\Sigma}_k}^{-1} \otimes \bm{I}_n)\bm{S}_k\bm{J}_t$ and
\begin{eqnarray*}
\frac{\partial \bm{F}_{ij}^k}{\partial y_t} & = & \frac{\partial \bm{M}_k}{\partial d_{ij}^k} \bm{S}_k^T (\bm{\widehat{\Sigma}}_k^{-1} \otimes \bm{I}_n) \bm{S}_k \bm{J}_t + \bm{M}_k \frac{\partial \bm{S}_k^T}{\partial d_{ij}^k} (\bm{\widehat{\Sigma}}_k^{-1} \otimes \bm{I}_n) \bm{S}_k \bm{J}_t + \bm{M}_k \bm{S}_k^T (\bm{\widehat{\Sigma}}_k^{-1} \otimes \bm{I}_n) \frac{\partial \bm{S}_k}{\partial d_{ij}^k} \bm{J}_t,
\end{eqnarray*}
we can get $\frac{\partial (2 \bm{F}_k^T \bm{F}_{ij}^k)}{ \partial y_t} = 2 \frac{\partial \bm{F}_k^T}{\partial y_t} \bm{F}_{ij}^k +2 \bm{F}_k^T \frac{\partial \bm{F}_{ij}^k}{\partial y_t}$. Then $\frac{\partial \widetilde{Q} (\widehat{\bm{d}}_k, \bm{y})}{\partial \bm{y}^T}$ can be easily computed.

Finally, we calculate $\frac{\partial \widetilde{Q} (\widehat{\bm{d}}_k, \bm{y})}{\partial \widehat{\bm{d}}_k}$. Let $\bm{A}_k = \widetilde{\bm{X}}^T \bm{M}_k \bm{S}_k^T (\bm{\widehat{\Sigma}}_k^{-1} \otimes \bm{I}_n)$. By
\begin{eqnarray*}
 \frac{\partial^2 \bm{V}_k}{\partial d_{ij}^k \partial d_{st}^k} &=& \left( \bm{I}_{ij} \bm{\widehat{\Sigma}}_k^{-1} \bm{I}_{ts} + \bm{I}_{st} \bm{\widehat{\Sigma}}_k^{-1} \bm{I}_{ji}  \right) \otimes (\bm{W}_k^T \bm{W}_k),
\\
 \frac{\partial^2 \bm{M}_k}{\partial d_{ij}^k \partial d_{st}^k} &=& -2 \bm{M}_k \frac{\partial \bm{M}_k}{\partial d_{st}^k} \text{diag} \left( \frac{\partial \bm{V}_k}{\partial d_{ij}^k} \right) - \bm{M}_k^2 \text{diag} \left( \frac{\partial^2 \bm{V}_k}{\partial d_{ij}^k \partial d_{st}^k} \right),
\\
 \widehat{\bm{\beta}}_k &=& (\bm{A}_k \widetilde{\bm{X}})^{-1} \bm{A}_k \bm{S}_k \bm{y} ,
\\
 \frac{\partial \bm{A}_k}{\partial d_{st}^k} &=& \Big \{ (\widehat{\bm{\Sigma}}_k^{-1} \otimes \bm{I}_n) \otimes (\widetilde{\bm{X}}^T \bm{M}_k) \Big \} \frac{\partial \bm{S}_k^T}{\partial d_{st}^k} + \Big \{ ((\widehat{\bm{\Sigma}}_k^{-1} \otimes \bm{I}_n) \bm{S}_k ) \otimes \widetilde{\bm{X}}^T \Big \} \frac{\partial\bm{M}_k}{\partial d_{st}^k},
\\
 \frac{\partial (\bm{A}_k \widetilde{\bm{X}})^{-1}}{\partial d_{st}^k} &=& -  (\bm{A}_k \widetilde{\bm{X}})^{-1} \frac{\partial \bm{A}_k}{\partial d_{st}^k}\widetilde{\bm{X}} (\bm{A}_k \widetilde{\bm{X}})^{-1},
\\
 \frac{\partial \widehat{\bm{\beta}}_k}{\partial d_{st}^k}&=& \left( \bm{y}^T \otimes \left( (\bm{A}_k \widetilde{\bm{X}})^{-1} \bm{A}_k \right) \right) \frac{\partial \text{vec} (\bm{S}_k)}{\partial d_{st}^k} + \left( (\bm{S}_k \bm{y})^T \otimes (\bm{A}_k \widetilde{\bm{X}})^{-1} \right) \frac{\partial \text{vec} (\bm{A}_k)}{\partial d_{st}^k}
\\
&&+ \left( \left( \bm{A}_k \bm{S}_k \bm{y} \right)^T \otimes \bm{I}_{pq} \right) \frac{\partial (\bm{A}_k \widetilde{\bm{X}})^{-1}}{\partial d_{st}^k},
\\
\frac{\partial \bm{F}_{ij}^k}{\partial d_{st}^k}&=& \frac{\partial^2 \bm{M}_k}{\partial d_{ij}^k \partial d_{st}^k} \bm{S}_k^T (\bm{\widehat{\Sigma}}_k^{-1} \otimes \bm{I}_n)(\bm{S}_k\bm{y}- \widetilde{\bm{X}} \widehat{\bm{\beta}}_k) + \frac{\partial \bm{M}_k}{\partial d_{ij}^k} \frac{\partial \bm{S}_k^T}{\partial d_{st}^k} (\bm{\widehat{\Sigma}}_k^{-1} \otimes \bm{I}_n)(\bm{S}_k\bm{y}- \widetilde{\bm{X}} \widehat{\bm{\beta}}_k)
\\
&& + \frac{\partial \bm{M}_k}{\partial d_{ij}^k} \bm{S}_k^T (\bm{\widehat{\Sigma}}_k^{-1} \otimes \bm{I}_n) (\frac{\partial \bm{S}_k}{\partial d_{st}^k} \bm{y} - \widetilde{\bm{X}} \frac{\partial \widehat{\bm{\beta}}_k}{\partial d_{st}^k})
+ \frac{\partial \bm{M}_k}{\partial d_{st}^k} \frac{\partial \bm{S}_k^T}{\partial d_{ij}^k} (\bm{\widehat{\Sigma}}_k^{-1} \otimes \bm{I}_n)(\bm{S}_k\bm{y}  - \widetilde{\bm{X}} \widehat{\bm{\beta}}_k)
\\
&& + \bm{M}_k \frac{\partial \bm{S}_k^T}{\partial d_{ij}^k} (\bm{\widehat{\Sigma}}_k^{-1} \otimes \bm{I}_n) (\frac{\partial \bm{S}_k}{\partial d_{st}^k} \bm{y} - \widetilde{\bm{X}} \frac{\partial \widehat{\bm{\beta}}_k}{\partial d_{st}^k}) + \frac{\partial \bm{M}_k}{\partial d_{st}^k} \bm{S}_k^T (\bm{\widehat{\Sigma}}_k^{-1} \otimes \bm{I}_n) \frac{\partial \bm{S}_k}{\partial d_{ij}^k} \bm{y}
\\
&& + \bm{M}_k \frac{\partial \bm{S}_k^T}{\partial d_{st}^k} (\bm{\widehat{\Sigma}}_k^{-1} \otimes \bm{I}_n) \frac{\partial \bm{S}_k}{\partial d_{ij}^k} \bm{y},
\end{eqnarray*}
we can get $\frac{\partial  (2 \bm{F}_k^T \bm{F}_{ij}^k)}{\partial d_{st}^k} = 2 \frac{\partial \bm{F}_k^T}{\partial d_{st}^k} \bm{F}_{ij}^k + 2 \bm{F}_k^T \frac{\partial\bm{F}_{ij}^k}{\partial d_{st}^k}=2 (\bm{F}_{st}^k)^T \bm{F}_{ij}^k + 2 \bm{F}_k^T \frac{\partial\bm{F}_{ij}^k}{\partial d_{st}^k}$. Then $\frac{\partial \widetilde{Q} (\widehat{\bm{d}}_k, \bm{y})}{\partial \widehat{\bm{d}}_k}$ can be easily computed.

\section{Proof of Theorem \ref{Th1}}

Define $\tilde{\bm{\epsilon}} = (\bm{I}_{nq} - \bm{D}^T \otimes \bm{W})^{-1} \bm{\varepsilon}$ such that $\bm{y} = \bm{\mu} + \tilde{\bm{\epsilon}}$. Note that
\begin{align*}
R_k^* = E \Vert \widetilde{\bm{\mu}}_k |_{\widehat{\bm{D}}_k = \bm{D}_k^*} - \bm{\mu} \Vert^2 = E \Vert \overline{\bm{P}}_k \bm{y} - \bm{\mu} \Vert^2 = \Vert \overline{\bm{P}}_k \bm{\mu} - \bm{\mu} \Vert^2 + \text{tr} \left( \overline{\bm{P}}_k \bm{\Omega} \overline{\bm{P}}_k^T \right).
\end{align*}
So
\begin{equation}\label{lbound}
R_k^*\geq \Vert \overline{\bm{P}}_k \bm{\mu} - \bm{\mu} \Vert^2 \mbox{ and } R_k^*\geq  \text{tr} \left( \overline{\bm{P}}_k \bm{\Omega} \overline{\bm{P}}_k^T \right).
\end{equation}

Therefore, we have
\begin{align*}
|L_k - R_k^*|  = & \left| \Vert \widetilde{\bm{P}}_k \bm{y} - \bm{\mu} \Vert^2 - \Vert \overline{\bm{P}}_k \bm{\mu} - \bm{\mu} \Vert^2 -  \text{tr} \left( \overline{\bm{P}}_k \bm{\Omega} \overline{\bm{P}}_k^T \right)\right|
\\
= & \Big| \left \Vert ( \widetilde{\bm{P}}_k - \overline{\bm{P}} _k) \bm{\mu} + ( \widetilde{\bm{P}}_k - \overline{\bm{P}}_k ) \tilde{\bm{\epsilon}} + \overline{\bm{P}}_k \tilde{\bm{\epsilon}} +  \overline{\bm{P}}_k \bm{\mu} - \bm{\mu} \right \Vert^2
\\
&  - \Vert \overline{\bm{P}}_k \bm{\mu} - \bm{\mu} \Vert^2 -  \text{tr} \left( \overline{\bm{P}}_k \bm{\Omega} \overline{\bm{P}}_k^T \right)  \Big|
\\
= & \Big|  \Vert ( \widetilde{\bm{P}}_k - \overline{\bm{P}}_k ) \bm{\mu} + ( \widetilde{\bm{P}}_k - \overline{\bm{P}}_k ) \tilde{\bm{\epsilon}} + \overline{\bm{P}}_k \tilde{\bm{\epsilon}} \Vert^2  - \text{tr} \left( \overline{\bm{P}}_k \bm{\Omega} \overline{\bm{P}}_k^T \right)
\\
& + 2 \left \{  ( \widetilde{\bm{P}}_k - \overline{\bm{P}}_k ) \bm{\mu} + ( \widetilde{\bm{P}}_k - \overline{\bm{P}}_k ) \tilde{\bm{\epsilon}} + \overline{\bm{P}}_k \tilde{\bm{\epsilon}}  \right \}^T (\overline{\bm{P}}_k \bm{\mu} - \bm{\mu}) \Big|
\\
\leq & \Big| \Vert \overline{\bm{P}}_k  \tilde{\bm{\epsilon}} \Vert^2 -  \text{tr} \left( \overline{\bm{P}}_k \bm{\Omega} \overline{\bm{P}}_k^T \right)  \Big| + 2 \Big| (\overline{\bm{P}}_k \bm{\mu} - \bm{\mu})^T  \overline{\bm{P}} _k \tilde{\bm{\epsilon}}  \Big| + \Vert ( \widetilde{\bm{P}}_k - \overline{\bm{P}}_k ) \bm{\mu} \Vert^2
\\
+ & \Vert( \widetilde{\bm{P}}_k - \overline{\bm{P}}_k ) \tilde{\bm{\epsilon}}  \Vert^2
+ 2 \Big| (\overline{\bm{P}}_k \bm{\mu} - \bm{\mu})^T ( \widetilde{\bm{P}}_k - \overline{\bm{P}}_k ) \bm{\mu}  \Big|
+ 2 \Big| (\overline{\bm{P}}_k \bm{\mu} - \bm{\mu})^T ( \widetilde{\bm{P}}_k - \overline{\bm{P}}_k ) \tilde{\bm{\epsilon}} \Big|
\\
+ & 2 \Big| \bm{\mu}^T ( \widetilde{\bm{P}}_k - \overline{\bm{P}}_k)^T \overline{\bm{P}}_k \tilde{\bm{\epsilon}} \Big| + 2 \Big| \tilde{\bm{\epsilon}}^T ( \widetilde{\bm{P}}_k - \overline{\bm{P}}_k)^T \overline{\bm{P}}_k \tilde{\bm{\epsilon}} \Big| + 2 \Big| \bm{\mu}^T ( \widetilde{\bm{P}}_k - \overline{\bm{P}}_k)^T ( \widetilde{\bm{P}}_k - \overline{\bm{P}}_k) \tilde{\bm{\epsilon}} \Big|
\\
\leq & \Big| \Vert \overline{\bm{P}}_k  \tilde{\bm{\epsilon}} \Vert^2 -  \text{tr} \left( \overline{\bm{P}}_k \bm{\Omega} \overline{\bm{P}}_k^T \right)  \Big| + 2 \Big| (\overline{\bm{P}}_k \bm{\mu} - \bm{\mu})^T  \overline{\bm{P}} _k \tilde{\bm{\epsilon}}  \Big| + \Vert ( \widetilde{\bm{P}}_k - \overline{\bm{P}}_k ) \bm{\mu} \Vert^2
\\
+ & \Vert( \widetilde{\bm{P}}_k - \overline{\bm{P}}_k ) \tilde{\bm{\epsilon}}  \Vert^2 + 2 R_k^{*1/2} \Vert ( \widetilde{\bm{P}}_k - \overline{\bm{P}}_k ) \bm{\mu} \Vert + 2 R_k^{*1/2} \Vert ( \widetilde{\bm{P}}_k - \overline{\bm{P}}_k ) \tilde{\bm{\epsilon}} \Vert
\\
+ & 2 \Big| \bm{\mu}^T ( \widetilde{\bm{P}}_k - \overline{\bm{P}}_k)^T \overline{\bm{P}}_k \tilde{\bm{\epsilon}} \Big| + 2 \Big| \tilde{\bm{\epsilon}}^T ( \widetilde{\bm{P}}_k - \overline{\bm{P}}_k)^T \overline{\bm{P}}_k \tilde{\bm{\epsilon}} \Big| + 2 \Big| \bm{\mu}^T ( \widetilde{\bm{P}}_k - \overline{\bm{P}}_k)^T ( \widetilde{\bm{P}}_k - \overline{\bm{P}}_k) \tilde{\bm{\epsilon}} \Big|.
\end{align*}

Let $\widehat{C}_k^* = \widehat{C}_k - \Vert  \tilde{\bm{\epsilon}} \Vert^2$. We have
\begin{align*}
|\widehat{C}_k^* - L_k| = & \Big| \Vert \widetilde{\bm{P}}_k \bm{y} -  \bm{y} \Vert^2 + 2 \text{tr}(\widetilde{\bm{P}}_k \widehat{\bm{\Omega}}) - \Vert \widetilde{\bm{P}}_k \bm{y} - \bm{\mu}  \Vert^2
\\
& + 2 \text{tr} \left( \frac{\partial \text{vec}(\widehat{\bm{D}}_k)}{\partial \bm{y}^T} (\bm{y}^T \otimes \widehat{\bm{\Omega}}) \frac{\partial \text{vec}(\widetilde{\bm{P}}_k)}{\partial \text{vec}(\widehat{\bm{D}}_k)^T} \right) - \Vert  \tilde{\bm{\epsilon}} \Vert^2 \Big|
\\
= & \Big| \Vert \widetilde{\bm{P}}_k \bm{y} -  \bm{\mu} - \tilde{\bm{\epsilon}} \Vert^2 + 2 \text{tr}(\widetilde{\bm{P}}_k \widehat{\bm{\Omega}}) - \Vert \widetilde{\bm{P}}_k \bm{y} - \bm{\mu}  \Vert^2
\\
& + 2 \text{tr} \left( \frac{\partial \text{vec}(\widehat{\bm{D}}_k)}{\partial \bm{y}^T} (\bm{y}^T \otimes \widehat{\bm{\Omega}}) \frac{\partial \text{vec}(\widetilde{\bm{P}}_k)}{\partial \text{vec}(\widehat{\bm{D}}_k)^T} \right) - \Vert  \tilde{\bm{\epsilon}} \Vert^2 \Big|
\\
= & \Big| -2(\widetilde{\bm{P}}_k \bm{\mu} - \bm{\mu})^T \tilde{\bm{\epsilon}} - 2 \tilde{\bm{\epsilon}}^T \widetilde{\bm{P}}_k^T \tilde{\bm{\epsilon}} + 2 \text{tr}(\widetilde{\bm{P}}_k \widehat{\bm{\Omega}})
\\
& + 2 \text{tr} \left( \frac{\partial \text{vec}(\widehat{\bm{D}}_k)}{\partial \bm{y}^T} (\bm{y}^T \otimes \widehat{\bm{\Omega}}) \frac{\partial \text{vec}(\widetilde{\bm{P}}_k)}{\partial \text{vec}(\widehat{\bm{D}}_k)^T} \right)  \Big|
\\
\leq &  2 \Big| \text{tr} \left( \frac{\partial \text{vec}(\widehat{\bm{D}}_k)}{\partial \bm{y}^T} (\bm{y}^T \otimes \widehat{\bm{\Omega}}) \frac{\partial \text{vec}(\widetilde{\bm{P}}_k)}{\partial \text{vec}(\widehat{\bm{D}}_k)^T} \right)  \Big| + 2 |(\overline{\bm{P}}_k \bm{\mu} - \bm{\mu})^T \tilde{\bm{\epsilon}}|
\\
& + 2 \Big| \bm{\mu}^T ( \widetilde{\bm{P}}_k - \overline{\bm{P}}_k )^T \tilde{\bm{\epsilon}} \Big| + 2 \Big| \tilde{\bm{\epsilon}}^T \overline{\bm{P}}_k^T \tilde{\bm{\epsilon}} - \text{tr} \left( \overline{\bm{P}}_k \bm{\Omega} \right)\Big| + 2 \Big| \text{tr} \left( ( \widetilde{\bm{P}}_k - \overline{\bm{P}}_k)  \widehat{\bm{\Omega}} \right) \Big|
\\
& + 2 \Big| \text{tr} \left( \overline{\bm{P}}_k (\bm{\Omega} - \widehat{\bm{\Omega}}) \right)  \Big|  + 2 \Big| \tilde{\bm{\epsilon}}^T ( \widetilde{\bm{P}}_k - \overline{\bm{P}}_k)^T \tilde{\bm{\epsilon}}  \Big|
\\
\leq &  2 \Big| \text{tr} \left( \frac{\partial \text{vec}(\widehat{\bm{D}}_k)}{\partial \bm{y}^T} (\bm{y}^T \otimes \widehat{\bm{\Omega}}) \frac{\partial \text{vec}(\widetilde{\bm{P}}_k)}{\partial \text{vec}(\widehat{\bm{D}}_k)^T} \right)  \Big| + 2 |(\overline{\bm{P}}_k \bm{\mu} - \bm{\mu})^T \tilde{\bm{\epsilon}}|
\\
& + 2 \Vert \bm{\mu} \Vert \Vert ( \widetilde{\bm{P}}_k - \overline{\bm{P}}_k )^T \tilde{\bm{\epsilon}} \Vert + 2 \Big| \tilde{\bm{\epsilon}}^T \overline{\bm{P}}_k^T \tilde{\bm{\epsilon}} - \text{tr} \left( \overline{\bm{P}}_k \bm{\Omega} \right)\Big| + 2 \Big| \text{tr} \left( ( \widetilde{\bm{P}}_k - \overline{\bm{P}}_k)  \widehat{\bm{\Omega}} \right) \Big|
\\
&  + 2 \Big| \text{tr} \left( \overline{\bm{P}}_k (\bm{\Omega} - \widehat{\bm{\Omega}}) \right)  \Big|   + 2 \Vert \tilde{\bm{\epsilon}}\Vert \Vert ( \widetilde{\bm{P}}_k - \overline{\bm{P}}_k )^T \tilde{\bm{\epsilon}} \Vert.
\end{align*}

Similar to Theorem 2.1 in \cite{li1987} and Theorem 1 in \cite{zhang2018spatial}, we just need to verify that
\begin{subequations}
\begin{eqnarray}
\label{Th1-verify1} && \underset{k}{\sup} R_k^{* -1} \Big|  \text{tr} \left( \frac{\partial \text{vec}(\widehat{\bm{D}}_k)}{\partial \bm{y}^T} (\bm{y}^T \otimes \widehat{\bm{\Omega}}) \frac{\partial \text{vec}(\widetilde{\bm{P}}_k)}{\partial \text{vec}(\widehat{\bm{D}}_k)^T} \right) \Big| = o_p (1),
\\
\label{Th1-verify2} && \underset{k}{\sup} R_k^{* -1} \Big| \Vert \overline{\bm{P}}_k  \tilde{\bm{\epsilon}} \Vert^2 -  \text{tr} \left( \overline{\bm{P}}_k \bm{\Omega} \overline{\bm{P}}_k^T \right)  \Big| = o_p (1),
\\
\label{Th1-verify3} && \underset{k}{\sup} R_k^{* -1} \Big| (\overline{\bm{P}}_k \bm{\mu} - \bm{\mu})^T  \overline{\bm{P}}_k \tilde{\bm{\epsilon}}  \Big| = o_p (1),
\\
\label{Th1-verify4} && \underset{k}{\sup} R_k^{* -1} \Big| \tilde{\bm{\epsilon}}^T \overline{\bm{P}}_k^T \tilde{\bm{\epsilon}} - \text{tr} \left( \overline{\bm{P}}_k \bm{\Omega} \right)\Big| = o_p (1),
\\
\label{Th1-verify5} && \underset{k}{\sup} R_k^{* -1} \Big|(\overline{\bm{P}}_k \bm{\mu} - \bm{\mu})^T \tilde{\bm{\epsilon}} \Big| = o_p (1),
\\
\label{Th1-verify6} && \underset{k}{\sup} R_k^{* -1} \Vert ( \widetilde{\bm{P}}_k - \overline{\bm{P}}_k ) \bm{\mu} \Vert^2 = o_p (1),
\\
\label{Th1-verify7} && \underset{k}{\sup} R_k^{* -1} \Vert ( \widetilde{\bm{P}}_k - \overline{\bm{P}}_k ) \tilde{\bm{\epsilon}} \Vert^2 = o_p (1),
\\
\label{Th1-verify8} && \underset{k}{\sup} R_k^{* -1} \Big| \text{tr} \left( ( \widetilde{\bm{P}}_k - \overline{\bm{P}}_k )  \widehat{\bm{\Omega}} \right) \Big| = o_p (1),
\\
\label{Th1-verify9} && \underset{k}{\sup} R_k^{* -1} \Big| \text{tr} \left( \overline{\bm{P}}_k (\bm{\Omega} - \widehat{\bm{\Omega}}) \right)  \Big| = o_p (1),
\\
\label{Th1-verify10} && \underset{k}{\sup} R_k^{* -1} \Vert \bm{\mu} \Vert \Vert ( \widetilde{\bm{P}}_k - \overline{\bm{P}}_k )^T \tilde{\bm{\epsilon}} \Vert = o_p (1),
\\
\label{Th1-verify11} && \underset{k}{\sup} R_k^{* -1} \Vert \tilde{\bm{\epsilon}}\Vert \Vert ( \widetilde{\bm{P}}_k - \overline{\bm{P}}_k  )^T \tilde{\bm{\epsilon}} \Vert = o_p (1),
\\
\label{Th1-verify12} && \underset{k}{\sup} R_k^{* -1} \Big| \bm{\mu}^T ( \widetilde{\bm{P}}_k - \overline{\bm{P}}_k)^T \overline{\bm{P}}_k \tilde{\bm{\epsilon}} \Big| = o_p(1),
\\
\label{Th1-verify13} && \underset{k}{\sup} R_k^{* -1}  \Big| \tilde{\bm{\epsilon}}^T ( \widetilde{\bm{P}}_k - \overline{\bm{P}}_k)^T \overline{\bm{P}}_k \tilde{\bm{\epsilon}} \Big| = o_p(1),
\\
\label{Th1-verify14} && \underset{k}{\sup} R_k^{* -1} \Big| \bm{\mu}^T ( \widetilde{\bm{P}}_k - \overline{\bm{P}}_k)^T ( \widetilde{\bm{P}}_k - \overline{\bm{P}}_k) \tilde{\bm{\epsilon}} \Big| = o_p(1).
\end{eqnarray}
\end{subequations}

By condition (\ref{C5}), we can directly obtain (\ref{Th1-verify1}). From condition (\ref{C4}) and $\lambda_{\max} (\bm{P}) = 1$, we have
\begin{align}\label{barP}
\underset{k}{\sup} \left \{  \lambda_{\max} (\overline{\bm{P}}_k) \right \} & = \underset{k}{\sup} \left[  \lambda_{\max} \left \{  (\bm{I}_{nq} - \bm{D}_k^{*T} \otimes \bm{W}_k)^{-1} \bm{P} (\bm{I}_{nq} - \bm{D}_k^{*T} \otimes \bm{W}_k)  \right \} \right]
\nonumber\\
& \leq \underset{k}{\sup} \left[ \lambda_{\max} \left \{  (\bm{I}_{nq} - \bm{D}_k^{*T} \otimes \bm{W}_k)^{-1}  \right \} \right] [ 1 + \underset{k}{\sup} \lambda_{\max} (\bm{D}_k^{*T} \otimes \bm{W}_k) ] = O(1).
\end{align}

By Chebyshev's inequality, Theorem 2 of \cite{whittle1960bounds} and (\ref{lbound}), for any $\delta > 0$, we have
\begin{align*}
P & \left \{ \underset{k}{\sup} R_k^{* -1} \Big| \Vert \overline{\bm{P}}_k  \tilde{\bm{\epsilon}} \Vert^2 -  \text{tr} \left( \overline{\bm{P}}_k \bm{\Omega} \overline{\bm{P}}_k^T \right)  \Big| > \delta \right \}
\\
& \leq \sum_{k = 1}^K P \left \{  \Big| \tilde{\bm{\epsilon}}^T \overline{\bm{P}}_k^T \overline{\bm{P}}_k \tilde{\bm{\epsilon}} - \text{tr} \left(  \overline{\bm{P}}_k \bm{\Omega} \overline{\bm{P}}_k^T \right) \Big| > \delta R_k^* \right \}
\\
& \leq \sum_{k = 1}^K \delta^{-2G} R_k^{* -2G} E \left[ \tilde{\bm{\epsilon}}^T \overline{\bm{P}}_k^T \overline{\bm{P}}_k \tilde{\bm{\epsilon}} - \text{tr} \left(  \overline{\bm{P}}_k \bm{\Omega} \overline{\bm{P}}_k^T  \right) \right]^{2G}
\\
& = \sum_{k = 1}^K \delta^{-2G} R_k^{* -2G} E \left[ (\bm{\Sigma}^{-\frac{1}{2}} \bm{\varepsilon})^T \bm{\Sigma}^{\frac{1}{2}} (\bm{S}^{-1})^T \overline{\bm{P}}_k^T \overline{\bm{P}}_k \bm{S}^{-1} \bm{\Sigma}^{\frac{1}{2}} (\bm{\Sigma}^{-\frac{1}{2}} \bm{\varepsilon}) - \text{tr} \left(  \overline{\bm{P}}_k \bm{\Omega} \overline{\bm{P}}_k^T  \right) \right]^{2G}
\\
& \leq \sum_{k = 1}^K \delta^{-2G} R_k^{* -2G} C_1 \Big \{ \text{tr} \left( \bm{\Sigma}^{\frac{1}{2}} (\bm{S}^{-1})^T \overline{\bm{P}}_k^T \overline{\bm{P}}_k \bm{S}^{-1} \bm{\Sigma}^{\frac{1}{2}} \bm{\Sigma}^{\frac{1}{2}}    (\bm{S}^{-1})^T \overline{\bm{P}}_k^T \overline{\bm{P}}_k \bm{S}^{-1} \bm{\Sigma}^{\frac{1}{2}}  \right) \Big \}^G
\\
& \leq \sum_{k = 1}^K \delta^{-2G} R_k^{* -2G} C_1 \lambda_{\max}^G (\bm{\Omega}) \lambda_{\max}^{2G} (\overline{\bm{P}}_k) \text{tr}^G \left( \overline{\bm{P}}_k \bm{\Omega} \overline{\bm{P}}_k^T \right)
\\
& \leq  \lambda_{\max}^G (\bm{\Omega})  C_1 \delta^{-2G}  \sum_{k=1}^K R_k^{* -2G} \lambda_{\max}^{2G} (\overline{\bm{P}}_k) (R_k^*)^G
\\
& \leq  \lambda_{\max}^G (\bm{\Omega})  C_1 \delta^{-2G}  \sum_{k=1}^K \lambda_{\max}^{2G} (\overline{\bm{P}}_k) (R_k^*)^{-G},
\end{align*}
where $C_1$ is a constant. Then from (\ref{barP}) and condition (\ref{C1}), (\ref{C2}), we obtain (\ref{Th1-verify2}). The equations (\ref{Th1-verify3})-(\ref{Th1-verify5}) can be proved similarly.

By condition (\ref{C2}) and (\ref{C6}), we have
\begin{align*}
\underset{k}{\sup} R_k^{* -1} \Vert ( \widetilde{\bm{P}}_k - \overline{\bm{P}}_k ) \bm{\mu} \Vert^2 & \leq \xi_n^{-1}  \Vert \bm{\mu} \Vert^2 \underset{k}{\sup} \lambda_{\max}^2 \left(  \widetilde{\bm{P}}_k - \overline{\bm{P}}_k \right) = o_p (1),
\end{align*}
which is (\ref{Th1-verify6}). Similarly, we can obtain (\ref{Th1-verify7}).

By condition (\ref{C2}) and (\ref{C6}), we have
\begin{align*}
\underset{k}{\sup} R_k^{* -1} \Big| \text{tr} \left( ( \widetilde{\bm{P}}_k - \overline{\bm{P}}_k)  \widehat{\bm{\Omega}} \right) \Big| & \leq \xi_n^{-1} \underset{k}{\sup} \left \{  \lambda_{\max} (\widetilde{\bm{P}}_k - \overline{\bm{P}}_k) \lambda_{\max} (\widehat{\bm{\Omega}}) \text{rank} (\widetilde{\bm{P}}_k - \overline{\bm{P}}_k)  \right \}
\\
& \leq 2pq \xi_n^{-1} \underset{k}{\sup} \left \{  \lambda_{\max} (\widetilde{\bm{P}}_k - \overline{\bm{P}}_k) \right \} \lambda_{\max} (\widehat{\bm{\Omega}}) = o_p (1),
\end{align*}
which is (\ref{Th1-verify8}).

For (\ref{Th1-verify9}), using (\ref{barP}), (\ref{C2}) and (\ref{C6}),  we have
\begin{align*}
\underset{k}{\sup} R_k^{* -1} \Big| \text{tr} \left( \overline{\bm{P}}_k (\bm{\Omega} - \widehat{\bm{\Omega}}) \right)  \Big| & \leq \xi_n^{-1} \underset{k}{\sup}  \left| \text{tr} \left( \overline{\bm{P}}_k (\bm{\Omega} - \widehat{\bm{\Omega}}) \right) \right|
\\
& \leq pq \xi_n^{-1} (\lambda_{\max} (\bm{\Omega}) + \lambda_{\max} (\widehat{\bm{\Omega}})) \underset{k}{\sup} \lambda_{\max} (\overline{\bm{P}}_k) = o_p (1).
\end{align*}

Similarly, we can get
\begin{subequations}
\begin{eqnarray*}
&& \underset{k}{\sup} R_k^{* -1} \Vert \bm{\mu} \Vert \Vert ( \widetilde{\bm{P}}_k - \overline{\bm{P}}_k )^T \tilde{\bm{\epsilon}} \Vert \leq \xi_n^{-1} \Vert \bm{\mu} \Vert \underset{k}{\sup} \lambda_{\max} (\widetilde{\bm{P}}_k - \overline{\bm{P}}_k) \Vert \tilde{\bm{\epsilon}} \Vert = o_p (1),
\nonumber\\
&& \underset{k}{\sup} R_k^{* -1}  \Vert \tilde{\bm{\epsilon}}\Vert \Vert ( \widetilde{\bm{P}}_k - \overline{\bm{P}}_k  )^T \tilde{\bm{\epsilon}} \Vert \leq \xi_n^{-1} \Vert \tilde{\bm{\epsilon}} \Vert^2 \underset{k}{\sup} \lambda_{\max} (\widetilde{\bm{P}}_k - \overline{\bm{P}}_k) = o_p (1),
\nonumber\\
&& \underset{k}{\sup} R_k^{* -1} \Big| \bm{\mu}^T ( \widetilde{\bm{P}}_k - \overline{\bm{P}}_k)^T \overline{\bm{P}}_k \tilde{\bm{\epsilon}} \Big| \leq \xi_n^{-1} \Vert \bm{\mu} \Vert \Vert \tilde{\bm{\epsilon}} \Vert \underset{k}{\sup} \lambda_{\max} (\widetilde{\bm{P}}_k - \overline{\bm{P}}_k) \underset{k}{\sup} \lambda_{\max} (\overline{\bm{P}}_k) = o_p(1),
\nonumber\\
&& \underset{k}{\sup} R_k^{* -1}  \Big| \tilde{\bm{\epsilon}}^T ( \widetilde{\bm{P}}_k - \overline{\bm{P}}_k)^T \overline{\bm{P}}_k \tilde{\bm{\epsilon}} \Big| \leq \xi_n^{-1} \Vert \tilde{\bm{\epsilon}} \Vert^2 \underset{k}{\sup} \lambda_{\max} (\widetilde{\bm{P}}_k - \overline{\bm{P}}_k) \underset{k}{\sup} \lambda_{\max} (\overline{\bm{P}}_k) = o_p(1),
\nonumber\\
&&  \underset{k}{\sup} R_k^{* -1} \Big| \bm{\mu}^T ( \widetilde{\bm{P}}_k - \overline{\bm{P}}_k)^T ( \widetilde{\bm{P}}_k - \overline{\bm{P}}_k) \tilde{\bm{\epsilon}} \Big| \leq \xi_n^{-1} \Vert \bm{\mu} \Vert \Vert \tilde{\bm{\epsilon}} \Vert \underset{k}{\sup} \lambda_{\max}^2 (\widetilde{\bm{P}}_k - \overline{\bm{P}}_k) = o_p(1),
\end{eqnarray*}
\end{subequations}
which is (\ref{Th1-verify10}) - (\ref{Th1-verify14}). This completes the proof.

\section{Proof of Theorem \ref{Th2}}

Rewrite $\widehat{C}_k^*$ as
\begin{align*}
& \widehat{C}_k^* = \left \Vert \widetilde{\bm{P}}_k \bm{y} - \bm{y} \right \Vert^2 + 2 \left \{ \text{tr}(\widetilde{\bm{P}}_k \widehat{\bm{\Omega}}) + \text{tr} \left( \frac{\partial \text{vec}(\widehat{\bm{D}}_k)}{\partial \bm{y}^T} (\bm{y}^T \otimes \widehat{\bm{\Omega}}) \frac{\partial \text{vec}(\widetilde{\bm{P}}_k)}{\partial \text{vec}(\widehat{\bm{D}}_k)^T} \right) \right \} - \Vert \tilde{\bm{\epsilon}} \Vert^2
\\
& = L_k \!+\! \left \Vert \widetilde{\bm{P}}_k \bm{y} - \bm{y} \right \Vert^2\! -\! \Vert \widetilde{\bm{P}}_k \bm{y} - \bm{\mu}  \Vert^2\! + \!2 \text{tr}(\widetilde{\bm{P}}_k \widehat{\bm{\Omega}}) + 2 \text{tr} \left( \frac{\partial \text{vec}(\widehat{\bm{D}}_k)}{\partial \bm{y}^T} (\bm{y}^T \otimes \widehat{\bm{\Omega}}) \frac{\partial \text{vec}(\widetilde{\bm{P}}_k)}{\partial \text{vec}(\widehat{\bm{D}}_k)^T} \right) - \Vert \tilde{\bm{\epsilon}} \Vert^2
\\
& = L_k \!-\! 2 (\widetilde{\bm{P}}_k \bm{\mu} - \bm{\mu})^T \tilde{\bm{\epsilon}} - 2 \tilde{\bm{\epsilon}}^T \widetilde{\bm{P}}_k^T \tilde{\bm{\epsilon}} + 2 \text{tr}(\widetilde{\bm{P}}_k \widehat{\bm{\Omega}}) + 2 \text{tr} \left( \frac{\partial \text{vec}(\widehat{\bm{D}}_k)}{\partial \bm{y}^T} (\bm{y}^T \otimes \widehat{\bm{\Omega}}) \frac{\partial \text{vec}(\widetilde{\bm{P}}_k)}{\partial \text{vec}(\widehat{\bm{D}}_k)^T} \right)
\\
& = R_k^* + b_k,
\end{align*}
where $b_k = L_k - R_k^* - 2 (\widetilde{\bm{P}}_k \bm{\mu} - \bm{\mu})^T \tilde{\bm{\epsilon}} - 2 \tilde{\bm{\epsilon}}^T \widetilde{\bm{P}}_k^T \tilde{\bm{\epsilon}} + 2 \text{tr}(\widetilde{\bm{P}}_k \widehat{\bm{\Omega}}) + 2 \text{tr} \left( \frac{\partial \text{vec}(\widehat{\bm{D}}_k)}{\partial \bm{y}^T} (\bm{y}^T \otimes \widehat{\bm{\Omega}}) \frac{\partial \text{vec}(\widetilde{\bm{P}}_k)}{\partial \text{vec}(\widehat{\bm{D}}_k)^T} \right)$. Without loss of generality, we assume $\bm{W}_1 = \bm{W}^*$, i.e. the first model is true. Then we will show $\underset{k \in \mathcal{S}}{\sup} R_k^{* -1} |b_k| = o_p(1)$ and $|b_1| / \xi_n^* = o_p (1)$.

 It directly follows  (\ref{C8}) that $ \underset{k \in \mathcal{S}}{\sup} R_k^{* -1} \left| \text{tr} \left( \frac{\partial \text{vec} (\widehat{\bm{D}}_k)}{\partial \bm{y}^T} (\bm{y}^T \otimes \widehat{\bm{\Omega}}) \frac{\partial \text{vec} \widetilde{\bm{P}}_k}{\partial \widehat{\bm{D}}_k} \right)  \right| = o_p (1)$. Next we will show that $\underset{k \in \mathcal{S}}{\sup} R_k^{* -1} \left|(\widetilde{\bm{P}}_k \bm{\mu} - \bm{\mu})^T \tilde{\bm{\epsilon}} \right| = o_p(1)$. Note that $\left|(\widetilde{\bm{P}}_k \bm{\mu} - \bm{\mu})^T \tilde{\bm{\epsilon}} \right| \leq \left|(\widetilde{\bm{P}}_k \bm{\mu} - \overline{\bm{P}}_k \bm{\mu})^T \tilde{\bm{\epsilon}} \right| + \left|(\overline{\bm{P}}_k \bm{\mu} - \bm{\mu})^T \tilde{\bm{\epsilon}} \right| $. It follows Chebyshev's inequality, Theorem 2 of \cite{whittle1960bounds}, (\ref{lbound}), (\ref{C2}) and (\ref{C7}) that
\begin{align*}
P \left \{ \underset{k \in \mathcal{S}}{\sup} R_k^{* -1} \left|(\overline{\bm{P}}_k \bm{\mu} - \bm{\mu})^T \tilde{\bm{\epsilon}} \right| > \delta \right \} & \leq \sum_{k \in \mathcal{S}} P \Big \{ \left|(\overline{\bm{P}}_k \bm{\mu} - \bm{\mu})^T \tilde{\bm{\epsilon}} \right| > \delta R_k^* \Big \}
\nonumber\\
& \leq \delta^{-2G} \sum_{k \in \mathcal{S}} R_k^{* -2G}  E \left|(\overline{\bm{P}}_k \bm{\mu} - \bm{\mu})^T \tilde{\bm{\epsilon}} \right|^{2G}
\nonumber\\
& = \delta^{-2G} \sum_{k \in \mathcal{S}} R_k^{* -2G}  E \left|(\overline{\bm{P}}_k \bm{\mu} - \bm{\mu})^T \bm{S}^{-1} \bm{\Sigma}^{\frac{1}{2}} (\bm{\Sigma}^{- \frac{1}{2}} \bm{\varepsilon}) \right|^{2G}
\nonumber\\
& \leq C_2 \delta^{-2G} \sum_{k \in \mathcal{S}} R_k^{* -2G} \Vert \bm{\Sigma}^{\frac{1}{2}} (\bm{S}^{-1})^T  (\overline{\bm{P}}_k \bm{\mu} - \bm{\mu}) \Vert^{2G}
\nonumber\\
& \leq C_2 \delta^{-2G}  \lambda_{\max}^G (\bm{\Omega}) \sum_{k \in \mathcal{S}} R_k^{* -2G} (R_k^*)^G
\nonumber\\
& =  C_2 \delta^{-2G}  \lambda_{\max}^G (\bm{\Omega}) \sum_{k \in \mathcal{S}} (R_k^*)^{-G}  = o(1),
\end{align*}
where $C_2$ is a constant. Also, we have
\begin{align*}
 \underset{k \in \mathcal{S}}{\sup} R_k^{* -1 } \left|(\widetilde{\bm{P}}_k \bm{\mu} - \overline{\bm{P}}_k \bm{\mu})^T \tilde{\bm{\epsilon}} \right|  \leq \xi_n^{* -1} \Vert \bm{\mu} \Vert  \Vert  \tilde{\bm{\epsilon}} \Vert \underset{k \in \mathcal{S}}{\sup} \lambda_{\max} (\widetilde{\bm{P}}_k - \overline{\bm{P}}_k) = o_p (1).
\end{align*}

Next, we consider $ \underset{k \in \mathcal{S}}{\sup} R_k^{* -1} \left| \text{tr}(\widetilde{\bm{P}}_k \widehat{\bm{\Omega}}) - \tilde{\bm{\epsilon}}^T \widetilde{\bm{P}}_k^T \tilde{\bm{\epsilon}} \right| $. Note that $\left| \text{tr}(\widetilde{\bm{P}}_k \widehat{\bm{\Omega}}) - \tilde{\bm{\epsilon}}^T \widetilde{\bm{P}}_k^T \tilde{\bm{\epsilon}} \right| \leq \left| \text{tr}(\overline{\bm{P}}_k \bm{\Omega})\right.$ $ \left.- \tilde{\bm{\epsilon}}^T \overline{\bm{P}}_k^T \tilde{\bm{\epsilon}} \right| + \left| \text{tr} (\overline{\bm{P}}_k (\widehat{\bm{\Omega}} - \bm{\Omega})) \right| + \left| \text{tr} ((\widetilde{\bm{P}}_k - \overline{\bm{P}}_k) \widehat{\bm{\Omega}}) \right| + \left| \tilde{\bm{\epsilon}}^T (\widetilde{\bm{P}}_k - \overline{\bm{P}}_k)^T \tilde{\bm{\epsilon}} \right|$. For the first term, by Chebyshev's inequality, Theorem 2 of \cite{whittle1960bounds}, (\ref{C2}) and (\ref{C7}), we have
\begin{align*}
& P \left \{ \underset{k \in \mathcal{S}}{\sup} R_k^{* -1} \left| \text{tr}(\overline{\bm{P}}_k \bm{\Omega}) - \tilde{\bm{\epsilon}}^T \overline{\bm{P}}_k^T \tilde{\bm{\epsilon}} \right|  > \delta \right \}
\nonumber\\
& \leq \sum_{k \in \mathcal{S}} \delta^{-2G} R_k^{* -2G}  E \left| \tilde{\bm{\epsilon}}^T \overline{\bm{P}}_k \tilde{\bm{\epsilon}} - \text{tr}(\overline{\bm{P}}_k \bm{\Omega}) \right|^{2G}
\nonumber\\
& = \sum_{k \in \mathcal{S}} \delta^{-2G} R_k^{* -2G}  E \left| (\bm{\Sigma}^{- \frac{1}{2}} \bm{\varepsilon})^T \bm{\Sigma}^{\frac{1}{2}} (\bm{S}^{-1})^T \overline{\bm{P}}_k \bm{S}^{-1} \bm{\Sigma}^{\frac{1}{2}} (\bm{\Sigma}^{- \frac{1}{2}} \bm{\varepsilon})  - \text{tr}(\overline{\bm{P}}_k \bm{\Omega}) \right|^{2G}
\nonumber\\
& \leq C_3 \delta^{-2G} \sum_{k \in \mathcal{S}} R_k^{* -2G} \Big \{ \text{tr} \left( \bm{\Sigma}^{\frac{1}{2}} (\bm{S}^{-1})^T \overline{\bm{P}}_k \bm{S}^{-1} \bm{\Sigma}^{\frac{1}{2}} \bm{\Sigma}^{\frac{1}{2}} (\bm{S}^{-1})^T \overline{\bm{P}}_k^T \bm{S}^{-1}  \bm{\Sigma}^{\frac{1}{2}} \right) \Big \}^G
\nonumber\\
& \leq C_3 \delta^{-2G} \lambda_{\max}^G (\bm{\Omega}) \sum_{k \in \mathcal{S}} R_k^{* -2G} \Big \{  \text{tr} (\overline{\bm{P}}_k \bm{\Omega} \overline{\bm{P}}_k^T) \Big \}^G
\nonumber\\
& \leq C_3 \delta^{-2G} \lambda_{\max}^G (\bm{\Omega}) \sum_{k \in \mathcal{S}} (R_k^*)^{-G} = o(1),
\end{align*}
where $C_3$ is a constant. For the second term, we have
\begin{align*}
\underset{k \in \mathcal{S}}{\sup} R_k^{* -1} \left| \text{tr} (\overline{\bm{P}}_k (\widehat{\bm{\Omega}} - \bm{\Omega})) \right| & \leq \xi_n^{* -1} \lambda_{\max} (\widehat{\bm{\Omega}} + \bm{\Omega}) \underset{k \in \mathcal{S}}{\sup} \lambda_{\max} (\overline{\bm{P}}_k) \text{rank} (\overline{\bm{P}}_k)
\nonumber\\
& \leq 2 pq \xi_n^{* -1} \lambda_{\max} (\widehat{\bm{\Omega}} + \bm{\Omega}) \underset{k \in \mathcal{S}}{\sup} \lambda_{\max} (\overline{\bm{P}}_k) = o (1).
\end{align*}

\noindent Similarly, we have
\begin{align*} \label{Th2-neq5}
\underset{k \in \mathcal{S}}{\sup} R_k^{* -1} \left| \text{tr} ((\widetilde{\bm{P}}_k - \overline{\bm{P}}_k) \widehat{\bm{\Omega}}) \right| & \leq \xi_n^{* -1} \underset{k \in \mathcal{S}}{\sup} \left( \lambda_{\max} (\widetilde{\bm{P}}_k - \overline{\bm{P}}_k) \lambda_{\max} (\widehat{\bm{\Omega}}) \text{rank} (\widetilde{\bm{P}}_k - \overline{\bm{P}}_k) \right)
\nonumber\\
& \leq 2 pq \xi_n^{* -1} \lambda_{\max} (\widehat{\bm{\Omega}})  \underset{k \in \mathcal{S}}{\sup} \lambda_{\max} (\widetilde{\bm{P}}_k - \overline{\bm{P}}_k) = o_p (1).
\end{align*}
For the last term, we have $\underset{k \in \mathcal{S}}{\sup} R_k^{* -1} \left| \tilde{\bm{\epsilon}}^T (\widetilde{\bm{P}}_k - \overline{\bm{P}}_k) \tilde{\bm{\epsilon}} \right| \leq \xi_n^{* -1} \Vert \tilde{\bm{\epsilon}} \Vert^2 \underset{k \in \mathcal{S}}{\sup} \lambda_{\max} (\widetilde{\bm{P}}_k - \overline{\bm{P}}_k) = o_p (1) $. Thus, $\underset{k \in \mathcal{S}}{\sup} R_k^{* -1} \left| \text{tr}(\widetilde{\bm{P}}_k \widehat{\bm{\Omega}}) - \tilde{\bm{\epsilon}}^T \widetilde{\bm{P}}_k^T \tilde{\bm{\epsilon}} \right| = o_p (1)$.

Finally, $\underset{k \in \mathcal{S}}{\sup} R_k^{* -1} \left| L_k - R_k^* \right|=o_p(1)$ can be shown with similar arguments in the proof of Theorem \ref{Th1}. Thus, we have shown that $\widehat{C}_k^* = R_k^* + b_k$, where $\underset{k \in \mathcal{S}}{\sup} R_k^{* -1} |b_k| = o_p (1)$.  Similarly, we can show $|b_1| / \xi_n^*=o_p(1)$.

For $k \in \mathcal{S}$, we have $\widehat{C}_k^* / \xi_n^*=R_k^* / \xi_n^* + b_k / \xi_n^* = R_k^* / \xi_n^*(1+b_k/R_k^*) = R_k^* / \xi_n^*(1+o_p(1))$ uniformly. From the discussion of condition (\ref{C1}), we know $R_1^* = O(1)$.  Then $\widehat{C}_1^* / \xi_n^* =R_1^*/\xi_n^*+b_1/\xi_n^* \rightarrow_p 0$. Note that $R_k^* / \xi_n^*\geq 1$ for $k\in \mathcal{S}$. Then we have $P \big(\inf_{k\in\mathcal{S}}\widehat{C}_k^*>\widehat{C}_1^*\big)  \rightarrow 1$, which finishes the proof.

\section{Proof of Theorem \ref{Th3}}

Define $\overline{\bm{P}} (\w) = \sum_{k=1}^K w_k \overline{\bm{P}}_k$. We have
\begin{eqnarray*}
R^* (\w) &=& E \Vert \overline{\bm{P}} (\w) \bm{y} - \bm{\mu}  \Vert^2 = \Vert \overline{\bm{P}} (\w) \bm{\mu} - \bm{\mu}  \Vert^2 + \text{tr} \left( \overline{\bm{P}} (\w) \bm{\Omega} \overline{\bm{P}} (\w)^T \right)\nonumber\\
&\geq& \Vert \overline{\bm{P}} (\w) \bm{\mu} - \bm{\mu}  \Vert^2 \mbox{ and } \text{tr} \left( \overline{\bm{P}} (\w) \bm{\Omega} \overline{\bm{P}} (\w)^T \right).
\end{eqnarray*}
Therefore, we have
\begin{align*}
|L (\w) - R^* (\w)|  = & \left| \Vert \widetilde{\bm{P}} (\w) \bm{y} - \bm{\mu} \Vert^2 - \Vert \overline{\bm{P}} (\w) \bm{\mu} - \bm{\mu} \Vert^2 -  \text{tr} \left( \overline{\bm{P}} (\w) \bm{\Omega} \overline{\bm{P}} (\w)^T \right)\right|
\\
= & \Big| \left \Vert ( \widetilde{\bm{P}} (\w) - \overline{\bm{P}} (\w) ) \bm{\mu} + ( \widetilde{\bm{P}} (\w) - \overline{\bm{P}} (\w) ) \tilde{\bm{\epsilon}} + \overline{\bm{P}} (\w) \tilde{\bm{\epsilon}} +  \overline{\bm{P}} (\w) \bm{\mu} - \bm{\mu} \right \Vert^2
\\
&  - \Vert \overline{\bm{P}} (\w) \bm{\mu} - \bm{\mu} \Vert^2 -  \text{tr} \left( \overline{\bm{P}} (\w) \bm{\Omega} \overline{\bm{P}} (\w)^T \right)  \Big|
\\
= & \Big|  \Vert ( \widetilde{\bm{P}} (\w) - \overline{\bm{P}} (\w) ) \bm{\mu} + ( \widetilde{\bm{P}} (\w) - \overline{\bm{P}} (\w) ) \tilde{\bm{\epsilon}} + \overline{\bm{P}} (\w) \tilde{\bm{\epsilon}} \Vert^2  - \text{tr} \left( \overline{\bm{P}} (\w) \bm{\Omega} \overline{\bm{P}} (\w)^T \right)
\\
& + 2 \left \{  ( \widetilde{\bm{P}} (\w) - \overline{\bm{P}} (\w) ) \bm{\mu} + ( \widetilde{\bm{P}} (\w) - \overline{\bm{P}} (\w) ) \tilde{\bm{\epsilon}} + \overline{\bm{P}} (\w) \tilde{\bm{\epsilon}}  \right \}^T (\overline{\bm{P}} (\w) \bm{\mu} - \bm{\mu}) \Big|
\\
\leq & \Big| \Vert \overline{\bm{P}} (\w)  \tilde{\bm{\epsilon}} \Vert^2 -  \text{tr} \left( \overline{\bm{P}} (\w) \bm{\Omega} \overline{\bm{P}} (\w)^T \right)  \Big| + 2 \Big| (\overline{\bm{P}} (\w) \bm{\mu} - \bm{\mu})^T  \overline{\bm{P}} (\w) \tilde{\bm{\epsilon}}  \Big|
\\
+ & \Vert ( \widetilde{\bm{P}} (\w) - \overline{\bm{P}} (\w) ) \bm{\mu} \Vert^2 + \Vert( \widetilde{\bm{P}} (\w) - \overline{\bm{P}} (\w) ) \tilde{\bm{\epsilon}}  \Vert^2
\\
+ & 2 \Big| (\overline{\bm{P}} (\w) \bm{\mu} - \bm{\mu})^T ( \widetilde{\bm{P}} (\w) - \overline{\bm{P}} (\w) ) \bm{\mu}  \Big|
\\
+ & 2 \Big| (\overline{\bm{P}} (\w) \bm{\mu} - \bm{\mu})^T ( \widetilde{\bm{P}} (\w) - \overline{\bm{P}} (\w) ) \tilde{\bm{\epsilon}} \Big| + 2 \Big| \tilde{\bm{\epsilon}}^T ( \widetilde{\bm{P}} (\w) - \overline{\bm{P}} (\w) )^T \overline{\bm{P}} (\w) \tilde{\bm{\epsilon}} \Big|
\\
+ & 2 \Big| \bm{\mu}^T ( \widetilde{\bm{P}} (\w) - \overline{\bm{P}} (\w) )^T ( \widetilde{\bm{P}} (\w) - \overline{\bm{P}} (\w) ) \tilde{\bm{\epsilon}} \Big| + 2 \Big| \bm{\mu}^T ( \widetilde{\bm{P}} (\w) - \overline{\bm{P}} (\w) )^T \overline{\bm{P}} (\w) \tilde{\bm{\epsilon}} \Big|
\\
\leq & \Big| \Vert \overline{\bm{P}} (\w)  \tilde{\bm{\epsilon}} \Vert^2 -  \text{tr} \left( \overline{\bm{P}} (\w) \bm{\Omega} \overline{\bm{P}} (\w)^T \right)  \Big| + 2 \Big| (\overline{\bm{P}} (\w) \bm{\mu} - \bm{\mu})^T  \overline{\bm{P}} (\w) \tilde{\bm{\epsilon}}  \Big|
\\
+ & \Vert ( \widetilde{\bm{P}} (\w) - \overline{\bm{P}} (\w) ) \bm{\mu} \Vert^2 + \Vert( \widetilde{\bm{P}} (\w) - \overline{\bm{P}} (\w) ) \tilde{\bm{\epsilon}}  \Vert^2
\\
+ & 2 R^{*1/2}(\w) \Vert ( \widetilde{\bm{P}} (\w) - \overline{\bm{P}} (\w) ) \bm{\mu} \Vert + 2 R^{*1/2}(\w) \Vert ( \widetilde{\bm{P}} (\w) - \overline{\bm{P}} (\w) ) \tilde{\bm{\epsilon}} \Vert
\\
+ & 2 \Big| \tilde{\bm{\epsilon}}^T ( \widetilde{\bm{P}} (\w) - \overline{\bm{P}} (\w) )^T \overline{\bm{P}} (\w) \tilde{\bm{\epsilon}} \Big| + 2  \Big| \bm{\mu}^T ( \widetilde{\bm{P}} (\w) - \overline{\bm{P}} (\w) )^T ( \widetilde{\bm{P}} (\w) - \overline{\bm{P}} (\w) ) \tilde{\bm{\epsilon}} \Big|
\\
+ & 2 \Big| \bm{\mu}^T ( \widetilde{\bm{P}} (\w) - \overline{\bm{P}} (\w) )^T \overline{\bm{P}} (\w) \tilde{\bm{\epsilon}} \Big|.
\end{align*}

Let $\widehat{C}^* (\w) = \widehat{C} (\w) - \Vert  \tilde{\bm{\epsilon}} \Vert^2$. We have
\begin{align*}
|\widehat{C}^* (\w) - L(\w)| = & \Big| \Vert \widetilde{\bm{P}} (\w) \bm{y} -  \bm{y} \Vert^2 + 2 \text{tr}(\widetilde{\bm{P}}(\w) \widehat{\bm{\Omega}}) - \Vert \widetilde{\bm{P}} (\w) \bm{y} - \bm{\mu}  \Vert^2
\\
& + 2 \sum_{k=1}^K w_k \text{tr} \left( \frac{\partial \text{vec}(\widehat{\bm{D}}_k)}{\partial \bm{y}^T} (\bm{y}^T \otimes \widehat{\bm{\Omega}}) \frac{\partial \text{vec}(\widetilde{\bm{P}}_k)}{\partial \text{vec}(\widehat{\bm{D}}_k)^T} \right) - \Vert  \tilde{\bm{\epsilon}} \Vert^2 \Big|
\\
= & \Big| \Vert \widetilde{\bm{P}} (\w) \bm{y} -  \bm{\mu} - \tilde{\bm{\epsilon}} \Vert^2 + 2 \text{tr}(\widetilde{\bm{P}}(\w) \widehat{\bm{\Omega}}) - \Vert \widetilde{\bm{P}} (\w) \bm{y} - \bm{\mu}  \Vert^2
\\
& + 2 \sum_{k=1}^K w_k \text{tr} \left( \frac{\partial \text{vec}(\widehat{\bm{D}}_k)}{\partial \bm{y}^T} (\bm{y}^T \otimes \widehat{\bm{\Omega}}) \frac{\partial \text{vec}(\widetilde{\bm{P}}_k)}{\partial \text{vec}(\widehat{\bm{D}}_k)^T} \right) - \Vert  \tilde{\bm{\epsilon}} \Vert^2 \Big|
\\
= & \Big| -2(\widetilde{\bm{P}} (\w) \bm{\mu} - \bm{\mu})^T \tilde{\bm{\epsilon}} - 2 \tilde{\bm{\epsilon}}^T \widetilde{\bm{P}} (\w) \tilde{\bm{\epsilon}} + 2 \text{tr}(\widetilde{\bm{P}}(\w) \widehat{\bm{\Omega}})
\\
& + 2 \sum_{k=1}^K w_k \text{tr} \left( \frac{\partial \text{vec}(\widehat{\bm{D}}_k)}{\partial \bm{y}^T} (\bm{y}^T \otimes \widehat{\bm{\Omega}}) \frac{\partial \text{vec}(\widetilde{\bm{P}}_k)}{\partial \text{vec}(\widehat{\bm{D}}_k)^T} \right)  \Big|
\\
\leq &  2 \Big| \sum_{k=1}^K w_k \text{tr} \left( \frac{\partial \text{vec}(\widehat{\bm{D}}_k)}{\partial \bm{y}^T} (\bm{y}^T \otimes \widehat{\bm{\Omega}}) \frac{\partial \text{vec}(\widetilde{\bm{P}}_k)}{\partial \text{vec}(\widehat{\bm{D}}_k)^T} \right)  \Big| + 2 |(\overline{\bm{P}} (\w) \bm{\mu} - \bm{\mu})^T \tilde{\bm{\epsilon}}|
\\
& + 2 \Big| \bm{\mu}^T ( \widetilde{\bm{P}} (\w) - \overline{\bm{P}} (\w) )^T \tilde{\bm{\epsilon}} \Big| + 2 \Big| \tilde{\bm{\epsilon}}^T \overline{\bm{P}} (\w)^T \tilde{\bm{\epsilon}} - \text{tr} \left( \overline{\bm{P}} (\w) \bm{\Omega} \right)\Big|
\\
& + 2 \Big| \text{tr} \left( ( \widetilde{\bm{P}} (\w) - \overline{\bm{P}} (\w))  \widehat{\bm{\Omega}} \right) \Big| + 2 \Big| \text{tr} \left( \overline{\bm{P}} (\w) (\bm{\Omega} - \widehat{\bm{\Omega}}) \right)  \Big|
\\
& + 2 \Big| \tilde{\bm{\epsilon}}^T ( \widetilde{\bm{P}} (\w) - \overline{\bm{P}} (\w)  )^T \tilde{\bm{\epsilon}}  \Big|
\\
\leq &  2 \Big| \sum_{k=1}^K w_k \text{tr} \left( \frac{\partial \text{vec}(\widehat{\bm{D}}_k)}{\partial \bm{y}^T} (\bm{y}^T \otimes \widehat{\bm{\Omega}}) \frac{\partial \text{vec}(\widetilde{\bm{P}}_k)}{\partial \text{vec}(\widehat{\bm{D}}_k)^T} \right)  \Big| + 2 |(\overline{\bm{P}} (\w) \bm{\mu} - \bm{\mu})^T \tilde{\bm{\epsilon}}|
\\
& + 2 \Vert \bm{\mu} \Vert \Vert ( \widetilde{\bm{P}} (\w) - \overline{\bm{P}} (\w) )^T \tilde{\bm{\epsilon}} \Vert + 2 \Big| \tilde{\bm{\epsilon}}^T \overline{\bm{P}} (\w)^T \tilde{\bm{\epsilon}} - \text{tr} \left( \overline{\bm{P}} (\w) \bm{\Omega} \right)\Big|
\\
& + 2 \Big| \text{tr} \left( ( \widetilde{\bm{P}} (\w) - \overline{\bm{P}} (\w))  \widehat{\bm{\Omega}} \right) \Big| + 2 \Big| \text{tr} \left( \overline{\bm{P}} (\w) (\bm{\Omega} - \widehat{\bm{\Omega}}) \right)  \Big|
\\
& + 2 \Vert \tilde{\bm{\epsilon}}\Vert \Vert ( \widetilde{\bm{P}} (\w) - \overline{\bm{P}} (\w)  )^T \tilde{\bm{\epsilon}} \Vert
\end{align*}

Let $\sup_{\w}$ indicate supremum over $\w \in \mathcal{H}$. Similar to Theorem 2.1 in \cite{li1987} and Theorem 1 in \cite{zhang2018spatial}, we just need to verify that
\begin{subequations}
\begin{eqnarray}
\label{verify1} && \underset{\w}{\sup} R^* (\w)^{-1} \Big| \sum_{k=1}^K w_k \text{tr} \left( \frac{\partial \text{vec}(\widehat{\bm{D}}_k)}{\partial \bm{y}^T} (\bm{y}^T \otimes \bm{\Omega}) \frac{\partial \text{vec}(\widetilde{\bm{P}}_k)}{\partial \text{vec}(\widehat{\bm{D}}_k)^T} \right) \Big| = o_p (1),
\\
\label{verify2} && \underset{\w}{\sup} R^* (\w)^{-1} \Big| \Vert \overline{\bm{P}} (\w)  \tilde{\bm{\epsilon}} \Vert^2 -  \text{tr} \left( \overline{\bm{P}} (\w) \bm{\Omega} \overline{\bm{P}} (\w)^T \right)  \Big| = o_p (1),
\\
\label{verify3} && \underset{\w}{\sup} R^* (\w)^{-1} \Big| (\overline{\bm{P}} (\w) \bm{\mu} - \bm{\mu})^T  \overline{\bm{P}} (\w) \tilde{\bm{\epsilon}}  \Big| = o_p (1),
\\
\label{verify4} && \underset{\w}{\sup} R^* (\w)^{-1} \Big| \tilde{\bm{\epsilon}}^T \overline{\bm{P}} (\w)^T \tilde{\bm{\epsilon}} - \text{tr} \left( \overline{\bm{P}} (\w) \bm{\Omega} \right)\Big| = o_p (1),
\\
\label{verify5} && \underset{\w}{\sup} R^* (\w)^{-1} \Big|(\overline{\bm{P}} (\w) \bm{\mu} - \bm{\mu})^T \tilde{\bm{\epsilon}} \Big| = o_p (1),
\\
\label{verify6} && \underset{\w}{\sup} R^* (\w)^{-1} \Vert ( \widetilde{\bm{P}} (\w) - \overline{\bm{P}} (\w) ) \bm{\mu} \Vert^2 = o_p (1),
\\
\label{verify7} && \underset{\w}{\sup} R^* (\w)^{-1} \Vert ( \widetilde{\bm{P}} (\w) - \overline{\bm{P}} (\w) ) \tilde{\bm{\epsilon}} \Vert^2 = o_p (1),
\\
\label{verify8} && \underset{\w}{\sup} R^* (\w)^{-1} \Big| \text{tr} \left( ( \widetilde{\bm{P}} (\w) - \overline{\bm{P}} (\w))  \widehat{\bm{\Omega}} \right) \Big| = o_p (1),
\\
\label{verify9} && \underset{\w}{\sup} R^* (\w)^{-1} \Big| \text{tr} \left( \overline{\bm{P}} (\w) (\bm{\Omega} - \widehat{\bm{\Omega}}) \right)  \Big| = o_p (1),
\\
\label{verify10} && \underset{\w}{\sup} R^* (\w)^{-1} \Vert \bm{\mu} \Vert \Vert ( \widetilde{\bm{P}} (\w) - \overline{\bm{P}} (\w) )^T \tilde{\bm{\epsilon}} \Vert = o_p (1),
\\
\label{verify11} && \underset{\w}{\sup} R^* (\w)^{-1} \Vert \tilde{\bm{\epsilon}}\Vert \Vert ( \widetilde{\bm{P}} (\w) - \overline{\bm{P}} (\w)  )^T \tilde{\bm{\epsilon}} \Vert = o_p (1),
\\
\label{verify12} && \underset{\w}{\sup} R^* (\w)^{-1} \Big| \tilde{\bm{\epsilon}}^T ( \widetilde{\bm{P}} (\w) - \overline{\bm{P}} (\w) )^T \overline{\bm{P}} (\w) \tilde{\bm{\epsilon}} \Big| = o_p (1),
\\
\label{verify13} && \underset{\w}{\sup} R^* (\w)^{-1} \Big| \bm{\mu}^T ( \widetilde{\bm{P}} (\w) - \overline{\bm{P}} (\w) )^T ( \widetilde{\bm{P}} (\w) - \overline{\bm{P}} (\w) ) \tilde{\bm{\epsilon}} \Big| = o_p (1),
\\
\label{verify14} && \underset{\w}{\sup} R^* (\w)^{-1} \Big| \bm{\mu}^T ( \widetilde{\bm{P}} (\w) - \overline{\bm{P}} (\w) )^T \overline{\bm{P}} (\w) \tilde{\bm{\epsilon}} \Big| = o_p (1).
\end{eqnarray}
\end{subequations}

By condition (\ref{C11}), we can directly obtain (\ref{verify1}). Using the triangle inequality,  Chebyshev's inequality, Theorem 2 of \cite{whittle1960bounds}, (\ref{barP}) and condition (\ref{C2}), (\ref{C10}), we observe, for any $\delta > 0$,
\begin{align*}
P & \left \{ \underset{\w}{\sup} R^* (\w)^{-1} \Big| \Vert \overline{\bm{P}} (\w)  \tilde{\bm{\epsilon}} \Vert^2 -  \text{tr} \left( \overline{\bm{P}} (\w) \bm{\Omega} \overline{\bm{P}} (\w)^T \right)  \Big| > \delta \right \}
\\
& \leq P \left \{ \underset{\w}{\sup} \Big| \Vert \overline{\bm{P}} (\w)  \tilde{\bm{\epsilon}} \Vert^2 -  \text{tr} \left( \overline{\bm{P}} (\w) \bm{\Omega} \overline{\bm{P}} (\w)^T \right)  \Big| > \delta \widetilde{\xi}_n \right \}
\\
& \leq P \left \{ \underset{\w}{\sup} \sum_{k_1 = 1}^K \sum_{k_2 = 1}^K w_{k_1} w_{k_2} \Big| \tilde{\bm{\epsilon}}^T \overline{\bm{P}}_{k_1}^T \overline{\bm{P}}_{k_2} \tilde{\bm{\epsilon}} - \text{tr} \left(  \overline{\bm{P}}_{k_1} \bm{\Omega} \overline{\bm{P}}_{k_2}^T \right) \Big|  > \delta \widetilde{\xi}_n  \right \}
\\
& \leq P \left \{  \underset{k_1, k_2}{\max} \Big| \tilde{\bm{\epsilon}}^T \overline{\bm{P}}_{k_1}^T \overline{\bm{P}}_{k_2} \tilde{\bm{\epsilon}} - \text{tr} \left(  \overline{\bm{P}}_{k_1} \bm{\Omega} \overline{\bm{P}}_{k_2}^T \right) \Big| > \delta \widetilde{\xi}_n  \right \}
\\
& \leq \sum_{k_1 = 1}^K \sum_{k_2 = 1}^K P \left \{  \Big| \tilde{\bm{\epsilon}}^T \overline{\bm{P}}_{k_1}^T \overline{\bm{P}}_{k_2} \tilde{\bm{\epsilon}} - \text{tr} \left(  \overline{\bm{P}}_{k_1} \bm{\Omega} \overline{\bm{P}}_{k_2}^T \right) \Big| > \delta \widetilde{\xi}_n \right \}
\\
& \leq \sum_{k_1 = 1}^K \sum_{k_2 = 1}^K \delta^{-2G} \widetilde{\xi}_n^{-2G} E \left[ \tilde{\bm{\epsilon}}^T \overline{\bm{P}}_{k_1}^T \overline{\bm{P}}_{k_2} \tilde{\bm{\epsilon}} - \text{tr} \left(  \overline{\bm{P}}_{k_1} \bm{\Omega} \overline{\bm{P}}_{k_2}^T  \right) \right]^{2G}
\\
& = \sum_{k_1 = 1}^K \sum_{k_2 = 1}^K \delta^{-2G} \widetilde{\xi}_n^{-2G} E \left[ (\bm{\Sigma}^{- \frac{1}{2}} \bm{\varepsilon})^T \bm{\Sigma}^{\frac{1}{2}} (\bm{S}^{-1})^T \overline{\bm{P}}_{k_1}^T \overline{\bm{P}}_{k_2} \bm{S}^{-1} \bm{\Sigma}^{\frac{1}{2}} (\bm{\Sigma}^{- \frac{1}{2}} \bm{\varepsilon}) - \text{tr} \left(  \overline{\bm{P}}_{k_1} \bm{\Omega} \overline{\bm{P}}_{k_2}^T  \right) \right]^{2G}
\\
& \leq \sum_{k_1 = 1}^K \sum_{k_2 = 1}^K \delta^{-2G} \widetilde{\xi}_n^{-2G} C_2 \Big \{ \text{tr} \left( \bm{\Sigma}^{\frac{1}{2}} (\bm{S}^{-1})^T \overline{\bm{P}}_{k_1}^T \overline{\bm{P}}_{k_2} \bm{S}^{-1} \bm{\Sigma}^{\frac{1}{2}} \bm{\Sigma}^{\frac{1}{2}}  \bm{S}^{-T} \overline{\bm{P}}_{k_2}^T \overline{\bm{P}}_{k_1} \bm{S}^{-1} \bm{\Sigma}^{\frac{1}{2}}  \right) \Big \}^G
\\
& \leq \sum_{k_1 = 1}^K \sum_{k_2 = 1}^K \delta^{-2G} \widetilde{\xi}_n^{-2G} C_2 \lambda_{\max}^G (\bm{\Omega}) \lambda_{\max}^{2G} (\overline{\bm{P}}_{k_2}) \text{tr}^G \left( \overline{\bm{P}}_{k_1} \bm{\Omega} \overline{\bm{P}}_{k_1}^T \right)
\\
& \leq  \lambda_{\max}^G (\bm{\Omega})  C_2 K \delta^{-2G} \widetilde{\xi}_n^{-2G} \underset{k}{\sup} \lambda_{\max}^{2G} (\overline{\bm{P}}_k)  \sum_{k=1}^K   (R_k^*)^G
\end{align*}
where $C_2$ is a constant. Then from condition (\ref{C2}), (\ref{C10}) and \ref{barP}, we obtain (\ref{verify2}). The equations (\ref{verify3})-(\ref{verify5}) can be proved similarly.

From condition (\ref{C2}) and (\ref{C12}), we have
\begin{eqnarray*}
&&\underset{\w}{\sup} R^* (\w)^{-1} \Vert ( \widetilde{\bm{P}} (\w) - \overline{\bm{P}} (\w) ) \bm{\mu} \Vert^2 \\
 &\leq &\widetilde{\xi}_n^{-1} \underset{\w}{\sup}  \bm{\mu}^T  \left \{  \sum_{k_1=1}^K \sum_{k_2=1}^K w_{k_1} w_{k_2} (\widetilde{\bm{P}}_{k_1} - \overline{\bm{P}}_{k_1})^T  (\widetilde{\bm{P}}_{k_2} - \overline{\bm{P}}_{k_2}) \right \} \bm{\mu}
\\
 &\leq &\widetilde{\xi}_n^{-1}  \Vert \bm{\mu} \Vert^2 \underset{\w}{\sup} \lambda_{\max} \left \{  \sum_{k_1=1}^K \sum_{k_2=1}^K w_{k_1} w_{k_2} (\widetilde{\bm{P}}_{k_1} - \overline{\bm{P}}_{k_1})^T  (\widetilde{\bm{P}}_{k_2} - \overline{\bm{P}}_{k_2}) \right \}
\\
& \leq &\widetilde{\xi}_n^{-1}  \Vert \bm{\mu} \Vert^2 \underset{k_1, k_2}{\sup} \lambda_{\max} \left \{  (\widetilde{\bm{P}}_{k_1} - \overline{\bm{P}}_{k_1})^T  (\widetilde{\bm{P}}_{k_2} - \overline{\bm{P}}_{k_2}) \right \}
\\
& \leq &\widetilde{\xi}_n^{-1}  \Vert \bm{\mu} \Vert^2 \underset{k}{\sup} \lambda_{\max}^2 (\widetilde{\bm{P}}_k - \overline{\bm{P}}_k) = o_p (1),
\end{eqnarray*}
which is (\ref{verify6}). Similarly, we can obtain (\ref{verify7}).

From condition (\ref{C2}), (\ref{C12}), we have
\begin{eqnarray*}
&&\underset{\w}{\sup} R^* (\w)^{-1} \Big| \text{tr} \left( ( \widetilde{\bm{P}} (\w) - \overline{\bm{P}} (\w))  \widehat{\bm{\Omega}} \right) \Big|\\
& \leq &\widetilde{\xi}_n^{-1} \underset{\w}{\sup} \sum_{k=1}^K w_k \Big| \text{tr} \left( ( \widetilde{\bm{P}}_k - \overline{\bm{P}}_k ) \widehat{\bm{\Omega}} \right) \Big|
\\
& \leq &\widetilde{\xi}_n^{-1} \underset{k}{\sup} \Big| \text{tr} \left( ( \widetilde{\bm{P}}_k - \overline{\bm{P}}_k ) \widehat{\bm{\Omega}} \right) \Big|
\\
& \leq &2pq \widetilde{\xi}_n^{-1} \underset{k}{\sup} \left \{  \lambda_{\max} (\widetilde{\bm{P}}_k - \overline{\bm{P}}_k) \right \} \lambda_{\max} (\widehat{\bm{\Omega}}) = o_p (1),
\end{eqnarray*}
which is (\ref{verify8}).

For (\ref{verify9}), using (\ref{barP}), condition (\ref{C2}) and (\ref{C12}),  we have
\begin{align*}
\underset{\w}{\sup} R^* (\w)^{-1} \Big| \text{tr} \left( \overline{\bm{P}} (\w) (\bm{\Omega} - \widehat{\bm{\Omega}}) \right)  \Big| & \leq \widetilde{\xi}_n^{-1} \underset{\w}{\sup} \sum_{k=1}^K w_k \left| \text{tr} \left( \overline{\bm{P}}_k (\bm{\Omega} - \widehat{\bm{\Omega}}) \right) \right|
\\
& \leq  \widetilde{\xi}_n^{-1} \underset{k}{\sup} \left| \text{tr} \left( \overline{\bm{P}}_k (\bm{\Omega} - \widehat{\bm{\Omega}}) \right) \right|
\\
& \leq pq \widetilde{\xi}_n^{-1} (\lambda_{\max} (\bm{\Omega}) + \lambda_{\max} (\widehat{\bm{\Omega}})) \underset{k}{\sup} \lambda_{\max} (\overline{\bm{P}}_k) = o_p (1).
\end{align*}

Similarly, we can get
\begin{subequations}
\begin{eqnarray*}
&& \underset{\w}{\sup} R^* (\w)^{-1} \Vert \bm{\mu} \Vert \Vert ( \widetilde{\bm{P}} (\w) - \overline{\bm{P}} (\w) )^T \tilde{\bm{\epsilon}} \Vert \leq \widetilde{\xi}_n^{-1} \Vert \bm{\mu} \Vert \underset{k}{\sup} \lambda_{\max} (\widetilde{\bm{P}}_k - \overline{\bm{P}}_k) \Vert \tilde{\bm{\epsilon}} \Vert = o_p (1),
\\
&& \underset{\w}{\sup} R^* (\w)^{-1} \Vert \tilde{\bm{\epsilon}}\Vert \Vert ( \widetilde{\bm{P}} (\w) - \overline{\bm{P}} (\w)  )^T \tilde{\bm{\epsilon}} \Vert \leq \widetilde{\xi}_n^{-1} \Vert \tilde{\bm{\epsilon}} \Vert^2 \underset{k}{\sup} \lambda_{\max} (\widetilde{\bm{P}}_k - \overline{\bm{P}}_k) = o_p (1),
\\
&& \underset{\w}{\sup} R^* (\w)^{-1} \Big| \bm{\mu}^T ( \widetilde{\bm{P}}(\w) - \overline{\bm{P}}(\w))^T \overline{\bm{P}}(\w) \tilde{\bm{\epsilon}} \Big|\\
&&\quad\quad\quad\quad\quad\quad\quad\quad\quad\quad\quad\leq \widetilde{\xi}_n^{-1} \Vert \bm{\mu} \Vert \Vert \tilde{\bm{\epsilon}} \Vert \underset{k}{\sup} \lambda_{\max} (\widetilde{\bm{P}}_k - \overline{\bm{P}}_k) \underset{k}{\sup} \lambda_{\max} (\overline{\bm{P}}_k) = o_p(1),
\\
&& \underset{\w}{\sup} R^* (\w)^{-1}  \Big| \tilde{\bm{\epsilon}}^T ( \widetilde{\bm{P}}(\w) - \overline{\bm{P}}(\w))^T \overline{\bm{P}}(\w) \tilde{\bm{\epsilon}} \Big| \\
&&\quad\quad\quad\quad\quad\quad\quad\quad\quad\quad\quad\leq \widetilde{\xi}_n^{-1} \Vert \tilde{\bm{\epsilon}} \Vert^2 \underset{k}{\sup} \lambda_{\max} (\widetilde{\bm{P}}_k - \overline{\bm{P}}_k) \underset{k}{\sup} \lambda_{\max} (\overline{\bm{P}}_k) = o_p(1),
\\
&&  \underset{\w}{\sup} R^* (\w)^{-1} \Big| \bm{\mu}^T ( \widetilde{\bm{P}}(\w) - \overline{\bm{P}}(\w))^T ( \widetilde{\bm{P}}(\w) - \overline{\bm{P}}(\w)) \tilde{\bm{\epsilon}} \Big|
\\
&& \quad\quad \leq \underset{\w}{\sup} R^* (\w)^{- \frac{1}{2}} \Vert ( \widetilde{\bm{P}}(\w) - \overline{\bm{P}}(\w)) \bm{\mu} \Vert  \underset{\w}{\sup} R^* (\w)^{- \frac{1}{2}} \Vert ( \widetilde{\bm{P}}(\w) - \overline{\bm{P}}(\w)) \tilde{\bm{\epsilon}} \Vert = o_p(1),
\end{eqnarray*}
\end{subequations}
where the last equation holds due to (\ref{verify6}) and (\ref{verify7}).
This proves (\ref{verify10}) - (\ref{verify14}) and finishes the proof.

\end{document}